\documentclass[11pt,en,cite=authoryear]{elegantpaper}
\usepackage{mathtools}
\usepackage{tikz}
\usepackage{eso-pic}
\usepackage{booktabs}
\usepackage{titlesec}
\usepackage{setspace}
\usepackage{rotating}
\usepackage{multirow}
\usepackage{makecell}
\usepackage{url}
\usepackage{graphicx}
\usepackage{float}
\usepackage{framed}
\usepackage{subfigure}
\allowdisplaybreaks[4]

\title{Endogenous Growth Under Multiple Uses of Data\footnote{The authors thank Shota Ichihashi and Liyan Yang for helpful discussions and comments, as well as to the conference participants at the 2021 Conference on Markets and Economies with Information Frictions. Desheng Ma and Ramtin Salamat provided excellent assistance in proofreading the paper. Cong is at SC Johnson College of Business, Cornell University. Email: \url{will.cong@cornell.edu}. Wei is at Institute of Economics, Tsinghua University. Email: \url{wei_wenshi@163.com}. Xie is at Institute of Economics, Tsinghua University. Email: \url{xiedanxia@tsinghua.edu.cn}. Zhang is at School of International Trade and Economics, Central University of Finance and Economics. Email: \url{zhanglongtian@cufe.edu.cn}.}}
\author{Lin William Cong  \and Wenshi Wei  \and Danxia Xie  \and Longtian Zhang}

\date{August 2021}

\usepackage{array}

\begin{document}
\begin{spacing}{1.5}

\maketitle

\begin{abstract}
We model a dynamic data economy with fully endogenous growth where agents generate data from consumption and share them with innovation and production firms. Different from other productive factors such as labor or capital, data are nonrival in their uses across sectors which affect both the level and growth of economic outputs. Despite the vertical nonrivalry, the innovation sector dominates the production sector in data usage and contribution to growth because (i) data are dynamically nonrival and add to knowledge accumulation, and (ii) innovations ``desensitize'' raw data and enter production as knowledge, which allays consumers' privacy concerns. Data uses in both sectors interact to generate spillover of allocative distortion and exhibit an apparent substitutability due to labor's rivalry and complementarity with data. Consequently, growth rates under a social planner and a decentralized equilibrium differ, which is novel in the literature and has policy implications. Specifically, consumers' failure to fully internalize knowledge spillover when bearing privacy costs, combined with firms' market power, underprice data and inefficiently limit their supply, leading to underemployment in the innovation sector and a suboptimal long-run growth. Improving data usage efficiency is ineffective in mitigating the underutilization of data, but interventions in the data market and direct subsidies hold promises.

\end{abstract}

\newpage

\section{Introduction}

With the proliferation of Internet-based businesses and digital platforms, data are widely recognized as an important factor in the productive sectors and the long-run growth of any modern economy.\footnote{According to \citet{Manyika2011}, big data research has saved over 100 billion Euros for Europe, and reduce medical care cost of the United States by 8\% or 300 billion dollars every year.} Because extant studies adopt the framework of semi-endogenous growth and focus on the use of data in predicting and production, one can neither quantify the aggregate level of data usage nor study how multiple uses of data interact to impact economic growth. Yet in practice, data are used both in production and in innovation (e.g., the R\&D sector), and the amount of usage directly drives economic growth. For example, while data-intensive industries process large quantities of data to directly improve their products and services (e.g., self-driving cars and VR technologies), universities, many research institutes in the industry, and open-source initiatives (e.g., Google TensorFlow and GitHub) advance frontiers of fundamental data science that add to our understanding of general purpose technologies such as AI and automation.  

We offer the first analysis on the endogenous growth of a data economy with multiple data uses, explicitly modeling the hallmark characteristics of data usage in innovation such as the endogenous supply of data, the dynamic accumulation of knowledge, and the nonrival data usage in other sectors. Agents in our model produce data through their consumption activities, and can sell data to innovators and production firms, while being averse to potential privacy violations and data abuse. Meanwhile, they also supply labor to both the innovation sector and the production sector. Unlike labor, data are nonrival not only horizontally \citep{Jones2020,Ichihashi2020} or over time \citep{Cong2021}, but also vertically across sectors: Data usage in production does not limit their usage in innovation. That said, data uses across sectors interact in nontrivial ways to affect consumers' data contribution, privacy costs, labor allocation across sectors, and thus the long-run growth of the economy. Moreover, consumers endogenously contribute data, balancing the privacy costs incurred in their uses. We characterize the equilibria on the balanced growth path (BGP) and compare the importance of the multiple uses of data for economic growth. 

The larger quantity of data the consumers share, the more severe the creative destruction is with the emergence of new varieties and firms. Because consumers who are only paid by incumbent firms fail to get fully compensated for the knowledge spillover benefit of their data contribution in the creation of future firms, contemporaneous supply of data is suboptimally low. The growth rate is then endogenously lower in a decentralized economy than that in a social planner's solution. Meanwhile, the suboptimal usage of data lowers the productivity of labor more in the innovation sector than that in the production sector, which distorts the allocation of labor towards the production sector, further slowing economic growth (because knowledge is only accumulated in the innovation sector). Production firms' market power that underprices consumers' data contribution aggravates the inefficiency, leading to underemployment in the innovation sector and overemployment in the production sector.

We further derive the long-run usage of data and decompose the contribution of data of the two sectors. The innovation sector dominates in terms of data usage and the contribution of data to growth for two reasons: (i) Data are dynamically nonrival and add to knowledge creation that is cumulative \citep[also highlighted in][]{Cong2021}. (ii) Innovations have a  ``desensitizing'' effect on data usage in the production sector, i.e., knowledge generated from data usage in the innovation sector can be repeatedly utilized in the future and will not bring any additional privacy costs. When the same data used in the innovation sector are used again in the production sector, consumers' privacy costs will be greater. But, if the data generate knowledge in the innovation sector (by expanding the innovation possibility frontier), which then enters the production sector, the privacy costs do not increase because the accumulated knowledge does not reveal private information. 

Emerging studies on data economy typically focus on how data add to contemporaneous production \citep{Jones2020} and prediction \citep{Farboodi2021} without emphasizing long-run growth. While \citet{Cong2021} introduces data usage in the innovation sector with knowledge accumulation and dynamic data nonrivalry, all these models adopt a semi-endogenous growth framework and do not link the aggregate level of data usage to growth or offer insights on the multiple uses of data. In this paper, we use a fully endogenous model to isolate the impact of data on the economy from that of population growth. Our model generates differential growth rates and highlights the inefficiencies of a decentralized equilibrium relative to a social planner's solution. Unlike \citet{Jones2020} and \citet{Cong2021} that feature overuse of data either in the innovation sector or in the production sector, our model reveals a general underutilization of data. Also, while both our paper and \cite{Farboodi2021} show that the contribution of data to growth is bounded, the latter one relies on an informativeness bound of predicting from data, whereas in our case, it comes from a complementary channel of privacy concerns.

Our paper therefore adds to the large literature of economic growth. After \citet{Romer1990}'s seminal study introducing innovation into the economy to generate long-run growth, \citet{Jones1995} adds the knowledge spillover effect to derive a semi-endogenous result: The long-run growth rate now depends on the growth rate of population, instead of the population level. Subsequent studies such as \citet{Stokey1998} and \citet{Jones2016} also adopt semi-endogenous models, but they leave much to be desired because zero population growth in these models lead to zero growth of the economy, which is counterfactual. Our endogenous growth model circumvents this issue and highlights the distinguishing features of data.

This paper also contributes to the literature on the economics of data, information, and privacy in general. Previous studies such as \citet{Hirshleifer1971}, \citet{Admati1990}, and \citet{Murphy1996} focus on social values, sales, and property rights of information, while recent studies such as \citet{Akcura2005}, \citet{Casadesus2015}, and \citet{Fainmesser2021} strive to connect digital information with privacy issues. Meanwhile, studies such as \citet{Easley2019}, \cite{Jones2020}, and \citet{ichihashi2019competing,Ichihashi2020} highlight the nonrivalry of data horizontally and discuss competitions among data platforms or intermediaries; \cite{Cong2021} emphasizes the dynamic nonrivalry of data and their role in knowledge accumulation. We complement by incorporating the vertical and cross-sector nonrivalry of data, while allowing for data privacy concerns and multiple uses of data.

The remainder of this paper is organized as follows. Section \ref{sec:model} sets up the model and defines the equilibrium. Section \ref{sec:endo_growth} characterizes the social planner's solution, the decentralized economy, and the various usage of data, before drawing policy implications. Section \ref{sec:num} provides numerical results on the contribution of data to growth through their multiple uses, and further discussions misallocation and model robustness under alternative settings. Finally, Section \ref{sec:conclusion} concludes.

\section{The Model}
\label{sec:model}

In this section, we first introduce agents in the economy, i.e., representative consumers, incumbent firms, innovation sector, and data intermediary. Then, we define the equilibrium.

\subsection{Representative Consumers}

Homogeneous agents (consumers) live in a continuous-time economy. We assume that the size of the agent population is a constant $L$ in order to highlight endogenous growth---economic growth that is not driven by population growth. As in \cite{Jones2020} and \cite{Cong2021}, each consumer produces data as a by-product of consumption and chooses the quantity of data to sell for profit, while incurring disutility due to concerns about privacy violations and data breaches.\footnote{For empirical quantification of the value of privacy, see, for example, \cite{tang2019value}.} Each consumer's instantaneous utility is:
\begin{equation}
    u(c(t),d(t))=\ln{c(t)}-\frac{\kappa d(t)^2}{2}, \quad \text{where} \quad c(t) \equiv {\left( \int_{0}^{N(t)} c (v,t) ^{\frac{\gamma -1}{\gamma }} \mathrm{d}v \right)}^{\frac{\gamma }{\gamma -1}} \label{cons:Index}
\end{equation}
is the consumption index, i.e., the CES aggregate of the consumption of the differentiated varieties. $\gamma>1$ is the elasticity of substitution, $c(v,t)$ is the consumption level of variety $v$ at time $t$ (which costs $p(v,t)$), $N(t)$ is the total varieties of products, and $d(t)$ is the quantity of data shared or sold. Note that $\kappa\in(0,1)$ captures the extent of privacy concerns such as expected losses from leakage, extra costs when a platform use consumer data to price discriminate, or hacking of a centralized database, or discomfort under a Big Brother's constant surveillance. The disutility in increasing and convex in the quantity of data sold.  

Each consumer supplies one unit of labor inelastically in exchange for an endogenous wage $w(t)$. They hold assets $a(t)$ that earn returns at an interest rate $r(t)$, and they sell data $d(t)$ at an endogenous price $p_d(t)$. The consumers’ utility maximization problem is then:
\begin{equation}
    \max_{ \left\{ c(v,t),d(t) \right\} } \int_0^\infty e^{-\rho t} u(c(t),d(t)) \mathrm{d} t,
    \label{con:householdProb}
\end{equation}
subject to
\begin{equation}\nonumber
    \dot a(t) = r(t) a(t) + w(t) + p_d(t) d(t) - \int_0^{N(t)} p(v,t) c(v,t) \mathrm{d} v
\end{equation}
and 
\begin{equation}
    d(t) \le g(c(t)).
    \label{dataCons}
\end{equation}
Here, $\rho$ captures the consumers' time preference. Constraint (\ref{dataCons}) dictates that consumers cannot supply more data than what they generate. $g(c(t))$ is an exogenous general function of data generating process. 

Without loss of generality, we normalize the following price index to one:
\begin{equation}
    P(t) \equiv \left( \int_0^{N(t)} p (v,t)^{1-\gamma} \mathrm{d} v \right)^{\frac{1}{\gamma -1}} = 1.
    \label{priceIndex}
\end{equation}
Then, the Euler equations for consumption and data sales from Hamiltonian are simply:
\begin{equation}
    \frac{\dot c (t)}{c(t)} = r(t) - \rho
    \label{Euler:c}
\end{equation}
and 
\begin{equation}\nonumber
    \frac{\dot p_d (t)}{p_d (t)} - \frac{\dot d(t)}{d(t)} = r(t)-\rho.
\end{equation}

\subsection{Incumbent Firms}

Each incumbent firm owns a patent and produces a distinct variety $v\in[0,N(t)]$ through buying data sets $D(v,t)$ from a data intermediary (whom we will introduce shortly) and hiring labor $L_E(v,t)$. Each firm takes the price of data sets and the wage of labor as given. The nonrival nature of data implies that, different from other productive inputs such as labor or capital, the duplication and repeated usage of data incur negligible costs. Consequently, per capita output no longer depends on per capita data usage, but depends on the total quantity of data bought by the firms.

All the incumbent firms constitute the production sector. Similar to the consumption index shown in (\ref{cons:Index}), we define the aggregate output $Y(t)$ as follows:
\begin{equation}
    Y(t) \equiv \left( \int_0^{N(t)} Y (v,t)^{\frac{\gamma -1}{\gamma }} \mathrm{d}v \right)^{\frac{\gamma }{\gamma -1}}.
    \label{output:Index}
\end{equation}
With the FOCs from the consumers' problem and the price index shown in (\ref{priceIndex}), a firm producing variety $v$ takes the following demand function as given: 
\begin{equation}\nonumber
    p(v,t) = \left( \frac{c(t)}{c(v,t)} \right)^{\frac{1}{\gamma }} = \left( \frac{Y(t)}{Y(v,t)} \right)^{\frac{1}{\gamma }}.
\end{equation}
We denote the market value of the incumbent firm $v$ as $V(v,t)$. The profit maximization then becomes:
\begin{equation}
    r(t)V(v,t) = \max_{\left\{L_E(v,t), D(v,t)\right\}}  \left( \frac{Y(t)}{Y(v,t)} \right)^{\frac{1}{\gamma }} Y(v,t) - w(t)L_E(v,t) - q^{prod}_{d}(v,t) D(v,t) +\dot V(v,t),
    \label{firmProb}
\end{equation}
where the price of data sets faced by incumbent firms is denoted as $q^{prod}_{d}(v,t)$. In this optimization problem, the production function capturing the nonrival nature of data in production is defined as follows: 
\begin{equation}
    Y(v,t)= L_E (v,t) D(v,t)^\eta,
    \label{production}
\end{equation}
where parameter $\eta\in(0,1)$ captures the importance of data in production. We assume that data have diminishing marginal returns, consistent with the theoretical foundation in \cite{Farboodi2021} and the empirical evidence or calibration in  \cite{sun2017revisiting} and \cite{Jones2020}.

With firms and products symmetrically entering the economy, FOCs of incumbent firms are:
\begin{equation}
    \left(1-\frac{1}{\gamma } \right)\left( \frac{Y(t)}{Y(v,t)} \right)^{\frac{1}{\gamma }}\frac{Y(v,t)}{L_E(v,t)}=w(t)
    \label{firm:L}
\end{equation}
and
\begin{equation}
    \left(1-\frac{1}{\gamma } \right)\eta \left( \frac{Y(t)}{Y(v,t)} \right)^{\frac{1}{\gamma }}\frac{Y(v,t)}{D(v,t)}=q_{d}^{prod}(v,t).
    \label{firm:D}
\end{equation}
Since the price, outputs, and profits are equal for all incumbent firms, we simplify notations by writing $V(t)=V(v,t)$ and $D(t)=D(v,t)$ for all $v$ and $t$ in the remainder of the paper.

\subsection{Innovation Sector}

Different from \cite{Jones2020}, potential entrants, also known as the innovation sector, can invent new varieties by using data and employing R\&D labor, as in \cite{Cong2021}. Specifically, we set the innovation sector production function and the dynamics of innovation frontior as: 
\begin{equation}
    \dot N (t)= \varepsilon N(t)L_R(t)^{1-\xi}D(t)^\xi,
    \label{newentrants:p}
\end{equation}
where $L_R (t)$ is labor employed in the innovation sector, $\varepsilon>0$ captures the innovation efficiency, and $\xi\in(0,1)$ is the relative importance of data $D(t)$. To generate endogenous growth, we adopt the coefficient for the knowledge spillover effect as in \cite{Romer1990} (exponent of $N$ being 1). Also, we assume a Cobb-Douglas combination of bundles of data and labor in (\ref{newentrants:p}) which captures a competitive innovation process while maintaining free-entry conditions.   

The potential entrants take the price of data, $q^{inno}_{d}(t)$, as given and maximize the expected profit by choosing the quantity of data to buy and the units of labor to employ:
\begin{equation}
    \max_{\left\{L_R(t), D(t)\right\}} \dot N (t)V(t)-w(t)L_R(t)-q^{inno}_{d}(t)D(t),
    \label{newentrants:prob}
\end{equation}
where $D(t)$ denotes the data potential entrants purchase, which reflects that all agents can utilize the same data. The FOCs with respect to labor and data usage are then:
\begin{equation}
    (1-\xi) \varepsilon N(t)L_R(t)^{-\xi}D(t)^{\xi}V(t)=w(t)
    \label{newentrants:L}
\end{equation}
and
\begin{equation}
    \xi \varepsilon N(t)L_R(t)^{1-\xi}D(t)^{\xi-1}V(t)=q_d^{inno}(t).
    \label{newentrants:D}
\end{equation}

\subsection{Data Intermediary}

Following \cite{Jones2020}, we introduce a data intermediary industry with competitive entry. One data intermediary collects data from consumers at a given price $p_d(t)$ and integrates data into a data set $D(t)$. The intermediary maximizes profits by choosing the quantity of data to buy from consumers and the discriminatory price at which it sells data sets to incumbent firms and potential entrants, respectively. The data intermediary makes zero profit in the presence of free entry. We thus characterize data intermediation through a cost minimization problem and a zero profit condition. The intermediary solves:
\begin{equation}
    \min_{\left\{ d(t)\right\}} p_d(t) d(t)L, 
    \label{dataInter:minicost}
\end{equation}
subject to 
\begin{equation}
    D(t) \le L d(t).
    \label{datacomb}
\end{equation}
And the zero profit condition is:
\begin{equation}
    \int_0^{N(t)} q_{d}^{prod}(v,t)D(t)\mathrm{d}v+q_{d}^{inno}(t)D(t)=p_d(t)d(t)L,
    \label{dataInter:zero}
\end{equation}
where $q_{d}^{prod}(v,t)$ is a discriminatory price subject to the demand curve $q_{d}^{prod}(D(t))$ in (\ref{firm:D}), and $q_{d}^{inno}(t)$ is subject to the demand curve $q_{d}^{inno}(D(t))$ in (\ref{newentrants:D}). (\ref{datacomb}) and (\ref{dataInter:zero}) together represent that, with data nonrivalry, the intermediary can profit from combining all consumers’ data and selling them to production firms and entrant innovators simultaneously. 

\subsection{Equilibrium Definition}

An equilibrium in a decentralized economy consists of quantities $\{ c(t), Y(t), N(t), a(t), d(t), D(t), \\ L_R(t)\}$, $\{ c(v,t), Y(v,t), L_E(v,t)\}$, and prices $\{ p_d(t), q_d^{inno}(t), w(t), r(t), V(t)\}$, $\{p(v,t), q_d^{prod}(v,t)\}$, where $v\in[0,N(t)]$, such that:

(i) $\{c(v,t), c(t), a(t), d(t)\}$ maximize consumers' utility in (\ref{con:householdProb}), $\{Y(v,t), Y(t), L_E(v,t), D(t), p(v,t), \\ V(t)\}$ maximize the discounted value of incumbent firms in (\ref{firmProb}), $\{L_R(t), D(t)\}$ maximize the expected return of new entrants in (\ref{newentrants:prob}), and $\{q_d^{prod}(v,t),q_d^{inno}(t)\}$ minimize the cost of data intermediary in (\ref{dataInter:minicost}) and (\ref{dataInter:zero}).

(ii) $\{w(t)\}$ clears the labor market with  $\int_{0}^{N(t)}L_E(v,t)\mathrm{d}v+L_R(t)=L$, $\{r(t)\}$ clears the asset market with $a(t)L=N(t)V(t)$, and $\{p_d(t)\}$ clears the data market with $d(t)L=D(t)$. $\{N(t)\}$ follows R\&D production function in (\ref{newentrants:p}).

\section{Data Uses for Endogenous Growth}
\label{sec:endo_growth}

We characterize the equilibria under the optimal allocation and in a decentralized economy, respectively, to identify inefficiencies due to knowledge accumulation externality and market power. We then contrast multiple uses of data with singular use of data in either the production or innovation sector, before drawing implications for regulation and policy intervention.

\subsection{Optimal Allocation}
\label{OptimalAllocation}

For simplicity, we first discuss the multiple uses of data under the optimal allocation. We denote the fraction of labor hired in the innovation sector by $l_R(t)$ and the fraction in the incumbent firm $v$ by $l_E(v,t)$, then the labor market clearing becomes $\int_{0}^{N(t)}l_E(v,t)\mathrm{d}v+l_R(t)= l_E(t)+l_R(t) = 1$. The planner needs to determine the labor allocation and the quantity of data consumers contribute, solving:
\begin{equation}\nonumber
\max _{\left\{l_E(t), d(t)\right\}} \int_0^\infty e^{-\rho t} L\left(\ln{c(t)}-\frac{\kappa{d(t)^2}}{2} \right) \mathrm{d} t,
\end{equation}
such that
\begin{align}
c(t) &=Y(t) / L,  \nonumber  \\
Y(t) &=N(t)^{\frac{1}{\gamma-1}}l_E(t)L^{1+\eta}d(t)^{\eta}, \nonumber \\
\dot N(t) &=\varepsilon l_R(t)^{1-\xi} L d(t)^{\xi} N(t), \nonumber \\
1 &= l_E(t) + l_R(t), \nonumber \\
d(t) &\le g(c(t)) \equiv \bar{d}(t) \label{s:datacons}.
\end{align}
Note that in (\ref{s:datacons}), we denote the maximum data available by $\bar{d}(t)$. 

The Hamiltonian for the optimal allocation is:
\begin{equation}\nonumber
    \mathcal{H} (l_E(t),d(t),N(t),\lambda (t))=\ln{ \left[ N(t)^{\frac{1}{\gamma-1}}l_E(t)L^{\eta}d^{\eta} \right]}-\frac{\kappa{d(t)}^2}{2}+\varepsilon \lambda (t)l_R^{1-\xi }{L}d (t)^{\xi } N(t).
\end{equation}
Then, the FOC with respect to $d(t)$ is:
\begin{align}
    \frac{\partial \mathcal{H}}{\partial d(t)} &=- \kappa d (t) +\frac{\eta }{d(t)}+\lambda (t)\varepsilon \xi l_R(t)^{1-\xi } L d(t)^{\xi-1 }N(t) \nonumber \\
    &=\underbrace{- \kappa d(t)}_\text{Marginal Disutility}+\underbrace{\frac{\eta }{d(t)}}_\text{Marginal Production}+\underbrace{\lambda (t)\xi \frac{\dot N (t) }{d(t)}}_\text{Marginal Innovation}. \label{s:dMain}
\end{align}
Different from \citet{Jones2020} as well as \citet{Cong2021}, the planner in our model wants to share data since data facilitate both production and innovation. 

Due to the upper limit of data shown by the constraint (\ref{s:datacons}), for some period $t'$, the maximum data available for the social planner $\bar d(t')=g(c(t'))$ may not be large enough to satisfy $\kappa\bar d (t')^2 -\eta >0$, which is the sum of the first two terms in (\ref{s:dMain}). In these corner solutions, the social planner opts to share all the data available, and the quantity of shared data increases synchronously with consumption (according to $\bar d(t) = g(c(t))$). Then, the increasing quantity of data in both the production sector and the innovation sector interact together to fuel economic growth, which is clear when we express $Y(t)$ in terms of growth rate and derive the following results:
\begin{align}
    \frac{\dot Y (t)}{Y(t)}&=\frac{\gamma}{\gamma-1}\frac{\dot N (t)}{N(t)}+\frac{\dot Y (v,t)}{Y(v,t)} \nonumber \\
    &=\underbrace{\frac{1}{\gamma-1}\varepsilon  l_R(t)^{1-\xi}L  d(t)^{\xi}}_{\text{Growth from innovation sector}}+\underbrace{\frac{\dot l_E(t)}{l_E(t)}+\eta\frac{\dot d(t)}{d(t)}}_\text{Growth from production sector}.
    \label{bgpumlimited}
\end{align}
When $\kappa \bar d(t)^2 -\eta>0$, an interior optimal quantity of data $d_s(t)$ balances the marginal disutility and the sum of marginal production and marginal innovation. Along the BGP, the social planner does not share unlimited data, but keeps a constant level of data $d_s$ in the long run. We derive the next proposition in Appendix \ref{App:OptimalAllocation}.

\begin{proposition}
When $\kappa{d^2} -\eta>0$, along the BGP, the optimal balanced growth rate of varieties $g_{Ns}$ and optimal data contribution per capita $d_s$ exist and are uniquely determined by two equations:
\begin{equation}
    g_{Ns}(d) = \frac{(\gamma-1)\rho}{\xi}(\kappa{d}^2 -\eta)
    \label{s:gNd}
\end{equation}
and
\begin{equation}
    g_{Ns}(d) = \varepsilon L \underbrace{\left[{\frac{\frac{1-\xi}{\xi}(\kappa{d}^2 -\eta)}{1+\frac{1-\xi}{\xi}(\kappa{d}^2 -\eta)}}\right]^{1-\xi }}_\text{Changes in labor allocation caused by data}d{^{\xi }}.
    \label{s:gNd2}
\end{equation}
\label{prop1}
\end{proposition}

\begin{figure}[H]
\begin{center}
\includegraphics[width=5in]{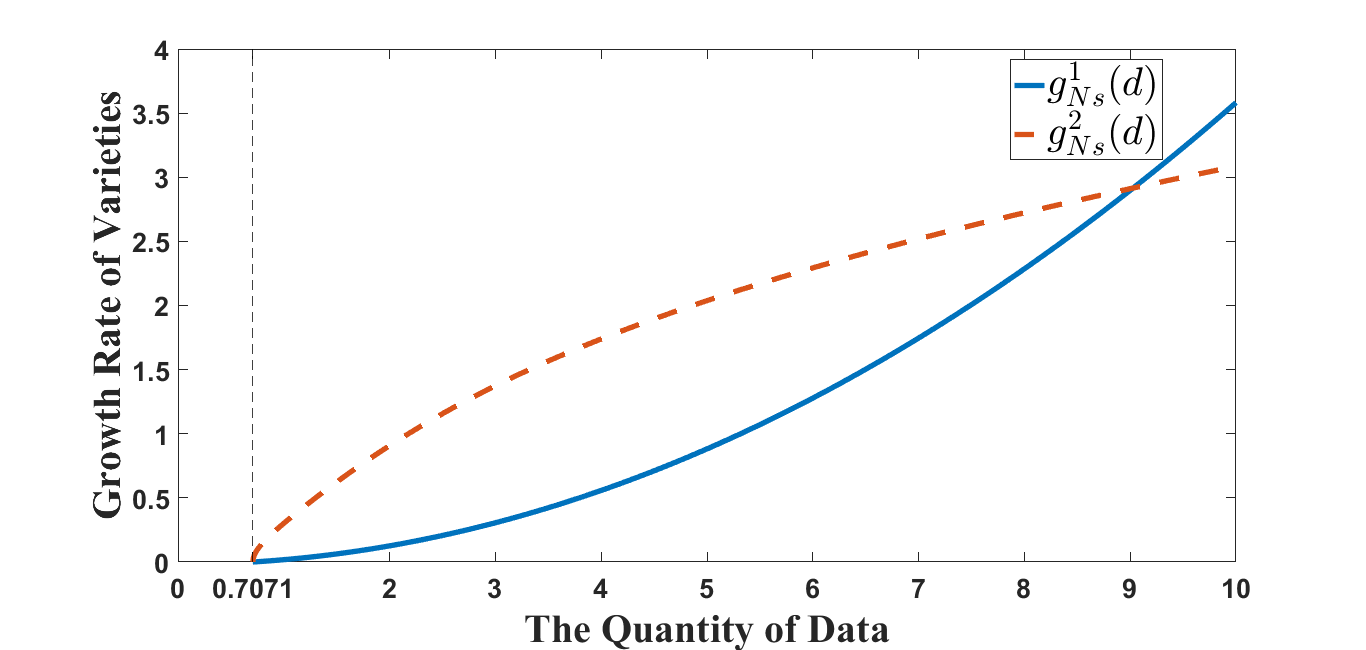}
\caption{Steady states in the optimal allocation in (\ref{s:gNd}) and (\ref{s:gNd2})}
\label{fig:solutionSP}
\end{center}
\qquad Notes: This figure shows the steady states in the optimal allocation. The full line is (\ref{s:gNd}) and the dashed line is (\ref{s:gNd2}). The position marked by the vertical line is the lower bound of the domain of $d$. When the full line exceeds the dashed line, no more data are shared. The intersection is the value of $g_{Ns}$ and $d_s$ along the BGP. The first state where $g_{Ns}=0$ and $d = (\eta / \kappa)^\frac{1}{2}=0.7071$ can be ruled out in the long run, because the FOC with respect to $d$ is ${\partial \mathcal{H}}/{\partial d(t)}>0$ in (\ref{s:dMain}) and the social planner can share data until the quantity of data reaches the intersection where ${\partial \mathcal{H}}/{\partial d(t)}=0$.
\end{figure}

(\ref{s:gNd}) characterizes the relationship between data disutility and growth rate. The social planner requires higher growth rates as the compensation for higher disutility caused by larger usage of data. (\ref{s:gNd2}) reflects a technology constraint, which is the growth rate that can be achieved through each unit of data produced. The two equations together pin down the economic growth and per capita data usage along the BGP, which are shown in Figure \ref{fig:solutionSP}. Importantly, endogenous long-run growth occurs even when the population experiences no growth---a desirable and realistic feature unattainable under semi-endogenous growth models.

Moreover, along the BGP, the growth rate of aggregate output $Y(t)$ is
\begin{equation}
    \frac{\dot Y (t)}{Y(t)}=\frac{\gamma}{\gamma-1}\frac{\dot N (t)}{N(t)}+\frac{\dot Y (v,t)}{Y(v,t)} 
    =\underbrace{\frac{1}{\gamma-1}\varepsilon{l}_{Rs}^{1-\xi}L{d_s^{\xi}}}_{\text{Growth from innovation sector}},
    \label{bgpgY}
\end{equation}
the aggregate output $Y(t)$ at time $t$ is
\begin{equation}
    Y(t)=N(t)^{\frac{1}{\gamma-1}}(1-l_{Rs})d^{\eta}_s{L^{1+\eta}},
    \label{bgpY}
\end{equation}
and the fraction of labor employed in the innovation sector is
\begin{equation}
    l_{Rs}=\frac{\frac{1-\xi}{\xi}(\kappa{d_s^2} -\eta)}{1+\frac{1-\xi}{\xi}(\kappa{d_s^2} -\eta)}.
    \label{s:lRinMain}
\end{equation}

(\ref{bgpgY}) reveals that the growth of aggregate output is determined by the growth of varieties in the long run, and the usage of data in the innovation sector plays an important role in economic growth, whose contribution is denoted by the parameter $\xi$. Compared with (\ref{bgpumlimited}) where the maximum quantity of data available is not large enough, the growth from the production in (\ref{bgpgY}) is zero in the long run. The reason is that increasing privacy costs dominate and the quantity of data input is limited. The growth from the production needs the quantity of data to become larger, but this means that consumers will face larger privacy costs. Because of the diminishing returns of data, the output contribution from directly using data is difficult to counteract the exponential growth of privacy costs. As a result, the quantity of data input is constant, implying that privacy costs are limited and the growth from the production sector alone cannot be sustained.

That said, the innovation sector transforms data into knowledge, creating ``desensitized'' data such as blueprints, algorithm codes or aggregate economic patterns. A firm may innovate incrementally by observing previous data-based knowledge repeatedly without incurring any additional privacy costs. As such, the ``desensitized'' data can be utilized in every future period, and data supplied at a constant level in each period can promote endogenous economic growth in the long run due to the ``dynamic nonrivalry'' in \citet{Cong2021}. As we will see in the decentralized economy in Section \ref{decentralized economy}, knowledge represented by the ``desensitized'' data still suffers from insufficient data sharing, because consumers cannot internalize the benefits of knowledge spillovers. 

The results in \cite{Jones2020} are based on an assumption that privacy costs are measured by the proportion of shared data rather than the total quantity of data. As a result, the corresponding disutility caused by data usage is kept at a constant level while the total quantity of shared data keeps growing. In semi-endogenous growth models, the population growth can impact the dynamics of data usage  because a greater quantity of data are contributed in aggregate without increasing privacy cost per capita. Thus, in the long run, they note that the quantity of data each person shares will decrease while we find it remain fixed in our model. Finally, in both \cite{Cong2021} and our paper, economic growth is sustainable, which is in contrast to \cite{Xie2017} that finds that innovations with disutility stall the endogenous growth of an economy. 

(\ref{bgpY}) reveals that the usage of data in the production sector determines the current period productivity $Y(t)$, and its contribution is denoted by the parameter $\eta$. Then, it is natural that parameter $\eta$ can also affect the growth rate shown in Proposition \ref{prop1}, since $\eta$ affects the quantity of data in the equilibrium, and then further affects the labor allocation and the growth rate through the innovation sector. However, this indirect impact on economic growth is based on the utilization of data in the innovation sector, which is crucial in the analysis here. We extend the model in Section \ref{WithoutInnovation} to further elaborate on this point.

Several comparative statistics reveal that labor allocations under multiple data usage differ significantly from those in studies such as \cite{Cong2021} and \cite{Jones2020}. From (\ref{s:lRinMain}), the fraction of labor employed in the innovation sector is determined by the following three factors: the contribution of data in production $\eta$, the contribution of data in innovation $\xi$, and the quantity of data $d_s$.  Also, from Proposition \ref{prop1}, $d_s$ is determined by $\eta$ and $\xi$ endogenously. As a result, $\xi$ and $\eta$ affect the labor allocation through the following two channels.

The first channel reflects a direct effect. As in \cite{Jones2020}, in the production function (\ref{production}), we let the labor contribution in the production sector to be 1 and assume that labor contribution is independent to data contribution which is denoted by $\eta$, to capture the increasing returns to scale of incumbent firms. Along the BGP, the demand curve for labor from the social planner is
\begin{equation}\nonumber
     \kappa d_s^2 = \eta+\frac{\xi}{1-\xi}\frac{l_R(t)}{l_E(t)}.
\end{equation}
We treat the quantity of data $d_s$ as given, and consider the direct effect of $\eta$. According to the above equation, with the increase of $\eta$, social planner needs less labor in the innovation sector to compensate for privacy costs. Thus, $\eta$ does not cause the substitution of labor, but encourages the  planner to devote more labor to production. Meanwhile, we use the settings in \cite{Cong2021} and assume constant returns to scale in innovation: The contributions of data and labor are $\xi$ and $(1-\xi)$, respectively. Thus, the increase of $\xi$ substitutes labor in the innovation sector, which is similar to automation as discussed in \cite{Aghion2019} and \cite{Acemoglu2018}. The difference in returns to scale helps us capture the differences in data utilization between incumbent of the production sector and potential entrants of the innovation sector. 

Besides the direct effect, an indirect effect can be seen through (\ref{s:dMain}), which can be transformed into:
\begin{align}
    \underbrace{\eta }_\text{Compensation from production}+\underbrace{\xi \lambda (t) \dot N (t)}_\text{Compensation from innovation} =\underbrace{\kappa d_s^2}_\text{Disutility}. 
    \label{s:compensation}
\end{align}
The increases of the two parameters $\eta$ and $\xi$ on the left hand side can increase the quantity of data $d_s$, which means that the planner can tolerate larger privacy costs. Then, in (\ref{s:lRinMain}), $l_{Rs}$ should increase to compensate for the additional privacy costs which are denoted in the term $(\kappa{d{_s}}^2-\eta)$. This indirect effect counteracts the direct effect, rendering the overall impact parameter dependent.

\subsection{Decentralized Economy}
\label{decentralized economy}

In the decentralized economy, the Hamiltonian for consumers' problem is:
\begin{equation}
    \mathcal{H} (c(v,t), d(t),a(t),\mu(t))=\ln{ c(t)}-\frac{ \kappa d(t)^2}{2}+\mu (t)\left[\begin{aligned}
     &r(t)a(t)+w(t) +p_d(t)d(t) \\ &-\int_0^{N(t)} p(v,t)c(v,t)\mathrm{d}v    
    \end{aligned} \right].
    \label{consumerHam}
\end{equation}
With $\mu(t)=1/c(t)$, the FOC for consumers problem with respect to $d(t)$ is:
\begin{equation}
     \frac{\partial \mathcal{H}}{\partial d(t)}=- \kappa d(t)^2+\underbrace{ \left(1-\frac{1}{\gamma}\right)\eta}_\text{Compensation from production sector}+\underbrace{\frac{\xi\dot N (t)V(t)}{N(t)^{\frac{\gamma}{\gamma-1}}Y(v,t)}}_\text{Compensation from innovation sector}=0.
     \label{c:FOCd}
\end{equation}
Here, consumers' privacy costs are compensated by the products from the production sector and assets from the innovation sector. Similar to what we have discussed in the optimal allocation, along the BGP, we have $d(t)\to d_c$ and $g_{Yc} \equiv \dot Y(t)/Y(t) \to g_{Nc}/(\gamma-1)$, so the function of data in productivity determines the output $Y(t)$ and the function in innovation determines the growth rate of varieties $g_{Nc}$. Focusing on the interior solution, i.e., $d_c^2> (1-1/\gamma)\eta/\kappa$, we have:

\begin{proposition}
In decentralized economy along the BGP, the balanced growth rate of varieties $g_{Nc}$ and data per capita $d_c$ are determined by the following two equations:
\begin{equation}\label{c:gNd1}
    g_{Nc}(d)=\rho\frac{ \kappa d^2 - \left(1-\frac{1}{\gamma } \right) \eta}{\xi\Gamma-\left[ \kappa d^2- \left(1-\frac{1}{\gamma } \right)\eta \right]}
\end{equation}
and
\begin{equation}\label{c:gNd2}
    g_{Nc}(d) = \varepsilon L \left[ \frac{\frac{1-\xi}{\xi}\frac{\kappa d^2 -\left(1-\frac{1}{\gamma}\right)\eta}{1-\frac{1}{\gamma}}}{1+\frac{1-\xi}{\xi}\frac{\kappa d^2-\left(1-\frac{1}{\gamma} \right)\eta}{1-\frac{1}{\gamma}}} \right]^{1-\xi} d^\xi,
\end{equation}
where
\begin{equation}\nonumber
    \Gamma \equiv \left[1- \left(1-\frac{1}{\gamma } \right) (1+\eta) \right].
\end{equation}
Moreover, if $1- (1-1/\gamma ) (1+\eta)>0$, then $\kappa d^2-(1-1/\gamma)\eta \in (0,\xi\Gamma)$.
\label{propMis}
\end{proposition}

\begin{figure}[H]
\begin{center}
\includegraphics[width=5in]{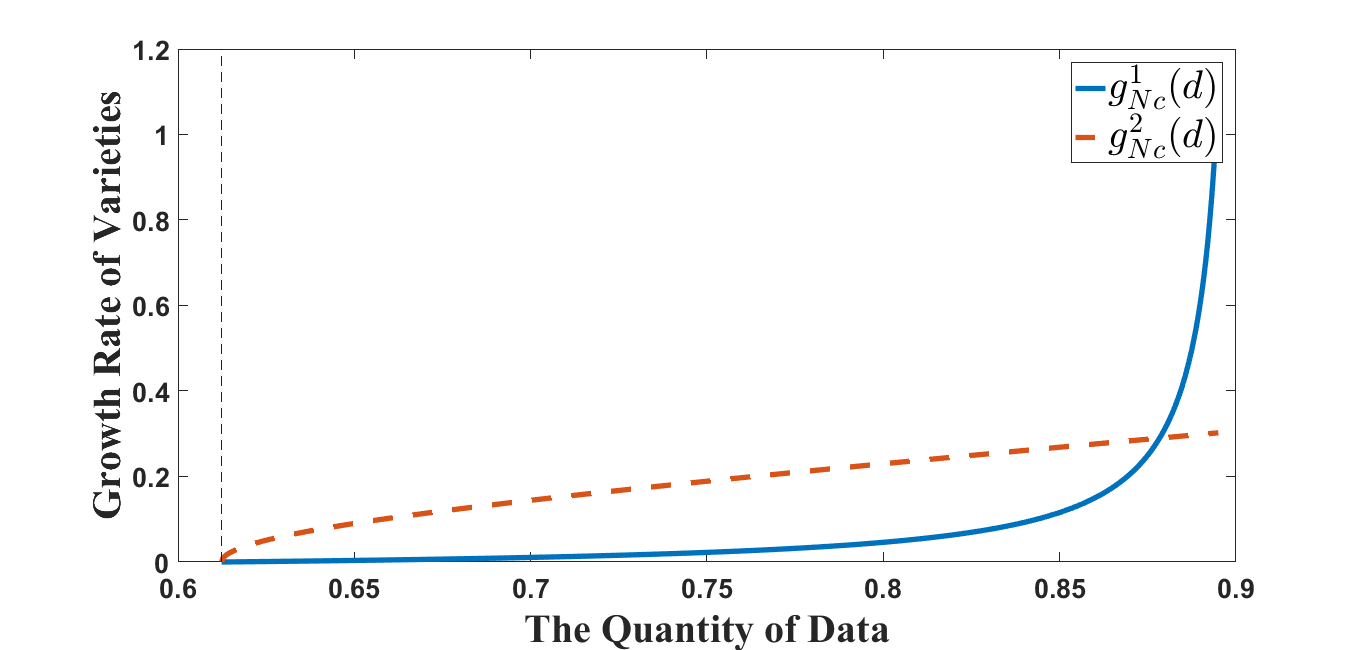}
\caption{Steady states in the decentralized economy in (\ref{c:gNd1}) and (\ref{c:gNd2})}
\label{fig:solutionDC}
\end{center}
\qquad Notes: This figure depicts the steady states in the decentralized economy. The full line is (\ref{c:gNd1}) and the dashed line is (\ref{c:gNd2}). The intersection after the vertical line is the value of $g_{Nc}$ and $d_c$ along the BGP. Because of this inefficiency, (\ref{c:gNd1}) has a larger convexity than (\ref{s:gNd}) in Figure \ref{fig:solutionSP}.
\end{figure}

$\Gamma$ here can be viewed as the share of profit owned by the incumbent firms and we assume $\Gamma>0$. In the above two equations, (\ref{c:gNd1}) describes the relationship between the exponential disutility and required growth from the view of consumers, while (\ref{c:gNd2}) describes the relationship between data demand and achievable growth subject to innovation possibility frontier from the view of potential entrants. Figure \ref{fig:solutionDC} depicts the above two equations and their intersection. When data can be used in both production and innovation, allocations are distorted by the monopoly markup in the production sector, as well as the higher required rate of return compared with the optimal allocation. Figure \ref{fig:solutionDC} depicts the above two equations and their intersection. Generally, we can observe the following three channels of inefficiencies from the above two equations in the decentralized economy. 

The first inefficiency concerns distortions in the price of data and is similar to that discussed in \cite{Jones2020}, which is reflected by the term $(1-1/\gamma)\eta$ in (\ref{c:gNd1}). Compared with the optimal allocation, consumers are compensated by less products from the monopolists, which is equal to $(1-1/\gamma)\eta$. The reason for this is that the price is inefficiently low because the monopolistic incumbent firms require a markup $(1-1/\gamma)$. With the same privacy costs, consumers are compensated by less output. As is shown in (\ref{c:FOCd}), this distortion pushes consumers to share smaller quantity of data while other conditions remain unchanged.  

The second inefficiency comes from the insufficient growth rate to compensate for the disutility of consumers, which is reflected by the term $-[\kappa d^2 - (1-1/\gamma)\eta]$ in (\ref{c:gNd1}). The required growth rate of varieties increases faster than that in optimal allocation (see the denominator in (\ref{s:gNd}) and (\ref{c:gNd1})), which means that with the same privacy concerns, consumers require more growth rate to compensate for the privacy concerns. The reason lies in the insufficient compensations to consumers due to knowledge spillover effects. As a result, the value of new innovations are lower in competitive equilibrium than that in the optimal allocation. Formally, along the BGP and under the optimal allocation, the disutility compensated by new innovations is equal to
\begin{equation}\nonumber
    \kappa d_s^2 -\eta =\lambda(t)\xi\dot N{(t)} 
    =\underbrace{\frac{\xi}{\rho(\gamma-1)}\frac{1}{N(t)}}_\text{Privacy compensation of each time of innovation}\dot{N}(t).
\end{equation}
Meanwhile, in the decentralized economy, the disutility compensated by new innovations is
\begin{align}
    \kappa d_c^2 (t)- \left(1-\frac{1}{\gamma} \right) \eta &=\xi\frac{V(t)}{N(t)^{\frac{1}{\gamma-1}}Y(v,t)}  \frac{\dot N(t)}{N(t)}=\frac{\xi\Gamma}{r(t)-\frac{\dot V(t)}{V(t)}}  \frac{\dot{N}{(t)}}{N(t)} \nonumber \\
    &=\underbrace{\frac{\xi\Gamma}{\rho+g_{Nc}}\frac{1}{N(t)}}_\text{Privacy compensation of each time of innovation}\dot{N}{(t)}.
    \label{c:V}
\end{align}
As seen in (\ref{c:V}), when consumers share more data, the innovation sector can use it to generate dynamically nonrival ``desensitized'' data and increase the growth rate of varieties $g_N$. Through the ``desensitizing'' effect of the innovation sector, consumers' privacy concerns from such dynamically nonrival data are alleviated in future reuse, while at the same time they cannot get the corresponding benefits. This process increases the required rate of return $(r(t)-\dot V(t)/V(t))$ to $(\rho+g_N)$, which diminishes the value of innovation (and the value of each asset, both are denoted by $V(t)$). Therefore, the larger quantity of data consumers share, the lower value of each times of innovation have, and the less compensation consumers can get. This inefficiency pushes consumers share smaller quantity of data under the same growth rate, given other conditions unchanged. 

The last inefficiency comes from different allocations of labor, which is reflected by the term
\begin{equation}\nonumber
    \left[ \frac{\frac{1-\xi}{\xi}\frac{\kappa d^2 -\left(1-\frac{1}{\gamma}\right)\eta}{1-\frac{1}{\gamma}}}{1+\frac{1-\xi}{\xi}\frac{\kappa d^2-\left(1-\frac{1}{\gamma} \right)\eta}{1-\frac{1}{\gamma}}} \right]^{1-\xi}
\end{equation}
in (\ref{c:gNd2}). In the decentralized economy, labor is allocated as
\begin{equation}
    \left(\frac{l_R}{l_E}\right)_c=\frac{1-\xi}{\xi}\frac{\kappa d_c^2 - \left(1-\frac{1}{\gamma} \right)\eta}{1-\frac{1}{\gamma}}.
    \label{c:lRlE}
\end{equation}
Since the monopolistic markup of incumbent firms distorts the equilibrium allocation, the production sector underemploys labor, which is captured by the denominator in (\ref{c:lRlE}). However, when data are used in multiple channels, insufficient data also affect the labor allocation which is captured by the corresponding numerator. This data shortage may trap the decentralized economy in a low-growth regime with underemployment in the innovation sector. Section \ref{sec:num} further investigates $g_{Nc}$ and $d_c$ relative to the optimal allocations numerically.

From (\ref{c:lRlE}), the fraction of labor employed in the innovation sector is also determined by a direct and an indirect effect like what we discussed before. The demand function for labor from incumbent firms can be rewritten as
\begin{equation}\nonumber
    \left(1-\frac{1}{\gamma } \right)N(t)^{\frac{\gamma}{\gamma-1}}(d_{c}(\eta,\xi)L)^\eta=w(t).
\end{equation}
Treating $d_c(\eta,\xi)$ as given, we see the demand for labor from incumbent firms increases with the $\eta$ because of increasing returns to scale, which reflects the direct effect of $\eta$. On the other hand, as for the indirect effect of $\eta$, the demand for labor from incumbent firms increases with $d_c(\eta,\xi)$. Meanwhile, the demand function for labor from the innovation sector is 
\begin{equation}\nonumber
    (1-\xi) \varepsilon N(t)L_R(t)^{-\xi}(d_{c}(\eta,\xi)L)^{\xi}V(t)=w(t).
\end{equation}
When considering the direct effect of $\xi$, we see that the demand for labor from the innovation sector decreases with $\xi$ and increases with $d_c(\eta,\xi)$.

Similar patterns concerning comparative statistics manifest in the optimal allocation as well, but the relative magnitude of direct and indirect effects play a nontrivial role. Section \ref{sec:num} discuss further results through numerical examples. Note that the market power in the production sector should lead to underemployment in production \citep[e.g.,][]{Cong2021}. Yet, we find the opposite.

\subsection{Allocations with Singular Usage of Data}
\label{WithoutInnovation}

We now investigate the two uses of data separately to understand their contribution to economic growth. We first set $\xi=0$, i.e., the social planner can only use data in production sector. Juxtaposing the results with those in Section \ref{OptimalAllocation} reveals the contribution of data in the innovation sector and its complementarity with data usage in production. The social planner problem now becomes: 
\begin{equation}\nonumber
\max_{\left\{l_E(t), d(t)\right\}} \int_0^{\infty} e^{-\rho t} L \left(\ln{c(t)}-\frac{\kappa d(t)^2}{2} \right) \mathrm{d} t,
\end{equation}
subject to
\begin{align}
c(t) &=Y(t) / L, \nonumber \\
Y(t) &=N(t)^{\frac{1}{\gamma-1}}l_E(t)L^{1+\eta}d(t)^{\eta}, \nonumber \\
\dot{N}(t) &=\varepsilon{l_R(t)}LN(t), \nonumber \\
1 &= l_E(t) + l_R(t), \nonumber \\
d(t) &\le g(c(t)) \equiv \bar{d}(t). \nonumber
\end{align}
The Hamiltonian for the optimal allocation is:
\begin{equation}\nonumber
    \mathcal{H} (l_E(t),d(t),N(t),\lambda (t))=\ln{\left[ N(t)^{\frac{1}{\gamma-1}}l_E(t)L^{\eta}d^{\eta} \right]}-\frac{{\kappa d(t)^2}}{2}+\varepsilon \lambda (t) l_R L N(t).
\end{equation}

\begin{proposition}
When data are used only in the production sector, along the BGP in the optimal allocation, per capita usage of data $d'$ is
\begin{equation}\label{dprime}
    d'=\left(\frac{\eta}{\kappa}\right)^{\frac{1}{2}},
\end{equation}
the growth rate of varieties is
\begin{equation}\nonumber
    g'_{N}=\varepsilon L-(\gamma-1)\rho,
\end{equation}
the aggregate output at time $t$ is 
\begin{equation}\nonumber
    Y'(t)=\frac{\rho (\gamma -1)}{\varepsilon }\left(\frac{\eta}{\kappa}\right)^{\frac{\eta}{2}} L^{1+\eta}  N'(t)^{\frac{1}{\gamma -1}},
\end{equation}
and the fraction of labor employed in the R\&D is
\begin{equation}\nonumber
    l'_{R}=1-\frac{\rho (\gamma -1)}{\varepsilon L}.
\end{equation}
\label{propWI}
\end{proposition}

Appendix \ref{App:Only} contains the derivations. Shutting down the data usage in innovation helps us understand better why the quantity of data shared becomes constant in the long run. Consider the data usage shown in (\ref{dprime}), it is a constant. Due to diminishing marginal returns ($\eta<1$), the social planner does not share all the data available and instead chooses a fixed quantity. Also, comparing the results with Section \ref{OptimalAllocation}, we get:

\begin{corollary}
When data are used only in production, data usage (e.g., captured by $\kappa$ and $\eta$) does not affect the growth rate of economy in the long run, but only affects the level of GDP $Y'(t)$ in each period. Moreover, per capita usage of data is less than that when data can be used in innovation.
\label{propd}
\end{corollary}

We next set $\eta=0$ to analyze the case in which the social planner can only use data in the innovation sector. The social planner problem now becomes:
\begin{equation}\nonumber
\max_{\left\{l_E(t), d(t)\right\}} \int_0^{\infty} e^{-\rho t} L \left(\ln{c(t)} - \frac{\kappa d(t)^2}{2} \right) \mathrm{d} t,
\end{equation}
subject to
\begin{align}
c(t) &=Y(t) / L, \nonumber \\
Y(t) &=N(t)^{\frac{1}{\gamma-1}}l_E(t)L, \nonumber \\
\dot{N}(t) &=\varepsilon{l_R(t)}^{1-\xi}L d(t)^{\xi}N(t), \nonumber \\
1 &= l_E(t) + l_R(t), \nonumber \\
d(t) &\le g(c(t)) \equiv \bar{d}(t). \nonumber
\end{align}
The Hamiltonian for the optimal allocation is then:
\begin{equation}\nonumber
    \mathcal{H} (l_E(t),d(t),N(t),\lambda (t))=\ln{\left[ N(t)^{\frac{1}{\gamma-1}}l_E(t)\right]}-\frac{\kappa d(t)^2}{2}+\varepsilon \lambda (t) l_R^{1-\xi} d^{\xi} L N(t).
\end{equation}

\begin{proposition}
When data are used only in the production sector, along the BGP in optimal allocation, the balanced growth rate of varieties $g''_N$ and data per capita $d''$ are determined by the following two equations: 
\begin{equation}
    (g^{1}_N)''(d)=\frac{(\gamma-1)\rho}{\xi}\kappa{d}^2,
    \label{onlyI:gN1}
\end{equation}
and
\begin{equation}
    (g^2_N)''(d)=\varepsilon L \left[\frac{\frac{1-\xi}{\xi}\kappa d^2}{1+\frac{1-\xi}{\xi}\kappa d^2}\right]^{1-\xi}d^{\xi}.
    \label{onlyI:gN2}
\end{equation}
Also, the fraction of labor employed in the innovation sector is
\begin{equation}\nonumber
    l''_R =\frac{\frac{1-\xi}{\xi}\kappa (d'')^2}{1+\frac{1-\xi}{\xi}\kappa (d'')^2}.
\end{equation}
The aggregate output at time $t$ is:
\begin{equation}\nonumber
    Y''(t)=N(t)^{\frac{1}{\gamma-1}}(1-l''_{R})L.
\end{equation}
\label{propIN}
\end{proposition}

\begin{figure}[H]
\begin{center}
\includegraphics[width=5in]{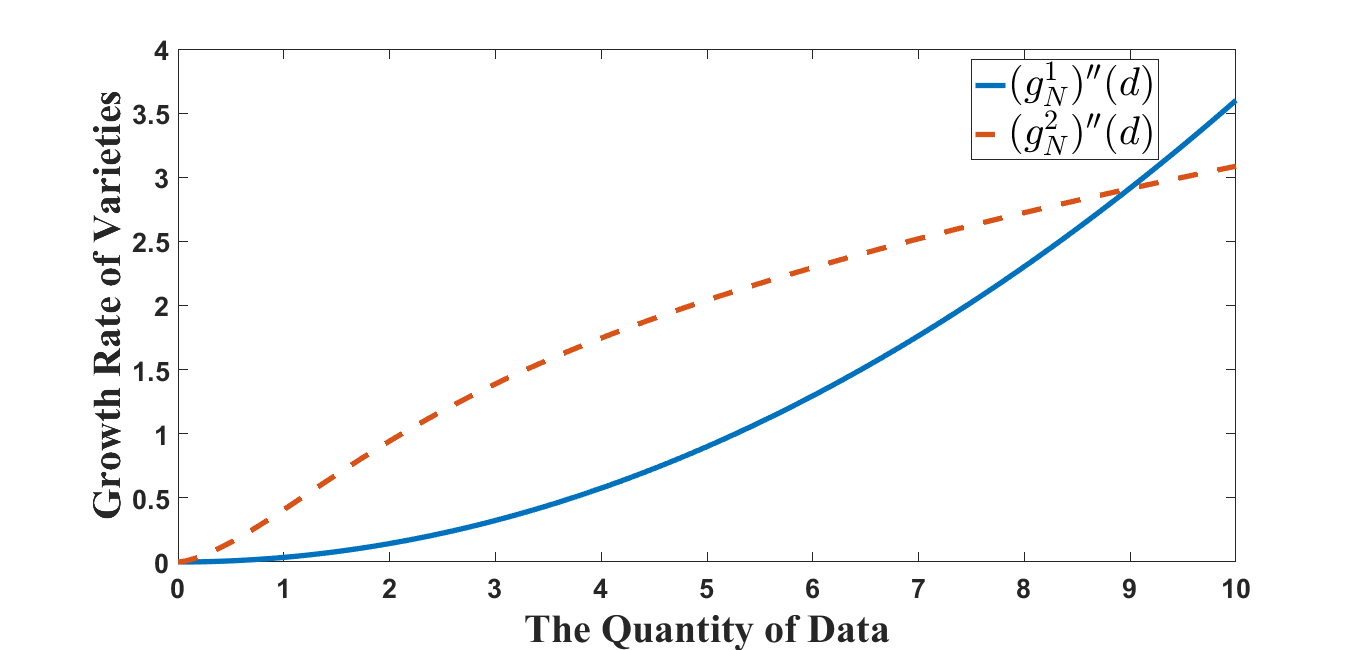}
\caption{Steady states when data only enter innovation sector in (\ref{onlyI:gN1}) and (\ref{onlyI:gN2})}
\label{fig:solutionOnlyI}
\end{center}
\qquad Notes: This figure depicts the steady states when data only enter to the innovation sector. The full line is (\ref{onlyI:gN1}) and the dashed line is (\ref{onlyI:gN2}). The intersection is the value of ${{g}''_{N}}$ and $d{''}$ along the BGP.
\end{figure}

Appendix \ref{App:OnlyInno} contains the derivations and Figure \ref{fig:solutionOnlyI} depicts the above two equations and their intersection. When $\eta$ is relatively small \citep[e.g., 0.03 to 0.10, as estimated in][]{Jones2020}, the growth rates of varieties and the quantity of data are similar to those in Proposition \ref{prop1}. 

Comparing the two singular uses of data, we see that the role of data is much more important in the innovation sector because of the ``desensitizing'' effect and the accumulation of knowledge, which are absent in the production sector. However, when the social planner uses data only in the innovation sector, the economy suffers losses in production. This is intuitive because a nonrival use of data in the production sector (without requiring additional data contributed by consumers) can further increase production outputs.

\subsection{Policy Interventions and Further Discussion}

In the decentralized economy, the output level and the value of innovation are lower than those in the optimal allocations. In other words, consumers choose to share a socially inefficient quantity of data. In order to push the decentralized allocations to the optimal level, we propose the following policy interventions to alleviate the distortions in the levels of data usage, growth rates, and labor allocations, which is shown in Proposition \ref{policy:prop}.

\begin{proposition}\label{policy:prop}

To restore social welfare efficiency, a government can apply revenue subsidies in the production and the innovation sectors with the following rates: 
\begin{equation}\nonumber
    1+s_p=\frac{\gamma}{\gamma-1} \quad \textit{and} \quad  1+s_{n}=\frac{\rho+g_{Ns}}{\rho\Gamma(\gamma-1)(1+s_p)}.
\end{equation}
Here, $s_p$ is the production subsidy, which is the conventional markup correction to increase output of the monopolists, and $s_n$ is the innovation subsidy to encourage innovation.

Alternatively, the government can also subsidize firms for purchasing data and employing labors to push the economy to the optimal level. The required subsidy rates are:
\begin{equation}\nonumber
    1-s_{d1}=1-\frac{1}{\gamma}, \qquad 1-s_{d2}=\frac{\rho(\gamma-1)\Gamma}{\rho+g_{Ns}},
\end{equation}
and 
\begin{equation}\nonumber
    1-s_l=\frac{1-s_{d2}}{1-\frac{1}{\gamma}}.
\end{equation}
Here, $s_{d1}$ and $s_{d2}$ are the data subsidies for the production sector and the innovation sector, respectively, and $s_l$ is the labor subsidy for the innovation sector.

\end{proposition}

\section{Numerical Analysis}
\label{sec:num}

In this section, we simulate the growth of our data economy to quantify the misallocation in the decentralized economy and to compare the various uses of data. Furthermore, we explore alternative specifications of the model to demonstrate the robustness of our findings. 

\subsection{Contribution of Multiple Uses of Data to Growth}
\label{subsec:multi_use_num}

We first build on Section \ref{sec:endo_growth} to examine the relative contribution of different uses of data. For simplicity, we focus on the optimal allocations. In Section \ref{OptimalAllocation}, data are used in both the production sector and the innovation sector, while in Section \ref{WithoutInnovation}, data are used either in the production sector or in the innovation sector but not both. We compare the results derived in the these three subsections. Table \ref{tab:parameterValues} summarizes the parameters.\footnote{Some parameters related to data cannot be determined from existing literature, e.g., $\eta$, $\kappa$, and $\xi$. As a result, we illustrate our theoretical framework within a rational range of values for these parameters and choose standard values for the other parameters.} Our numerical analyses mainly focus on the following three key variables: growth rate of varieties $g_N$, quantity of data being shared $d$, and fraction of labor employed in the innovation sector $l_R$. We do not discuss the time-varying output level $Y(t)$ (which is affected by the variety level $N(t)$) here, since it will make our analyses confusing.

Table \ref{tab:Baseline Numerical} briefly summarizes the baseline numerical results. The values for the three key variables shown in the table differ dramatically across different scenarios. The optimal allocation with multiple uses of data realizes the highest growth rate of varieties with the largest quantity of data usage. Compared with only used in the innovation sector, sharing data with the production sector may increase the quantity of data and the growth rate. This is because when the production sector can use data simultaneously, consumers can get additional compensation from selling data to the production sector captured by $\eta$. The additional compensation increases the quantity of data in the economy. Thus, through vertical nonrivalry, the innovation sector can use more data, and thus the growth rate increases. This reflects that the two sectors are complementary in the growth of the economy, where one sector’s data usage can benefit the other’s. In addition, the quantity of data when data are only used in innovation is higher than that when data are only used in production. The reason lies in that when data enter the innovation sector, they get ``desensitized'' as knowledge accumulates, and thus alleviate privacy concerns. This ``desensitization'' makes innovation uses of data more economical compared with the production use and bring higher growth rate.

\begin{figure}[H]
\begin{center}  
\subfigure[Difference in growth rate of varieties]{
\includegraphics[scale=0.18]{{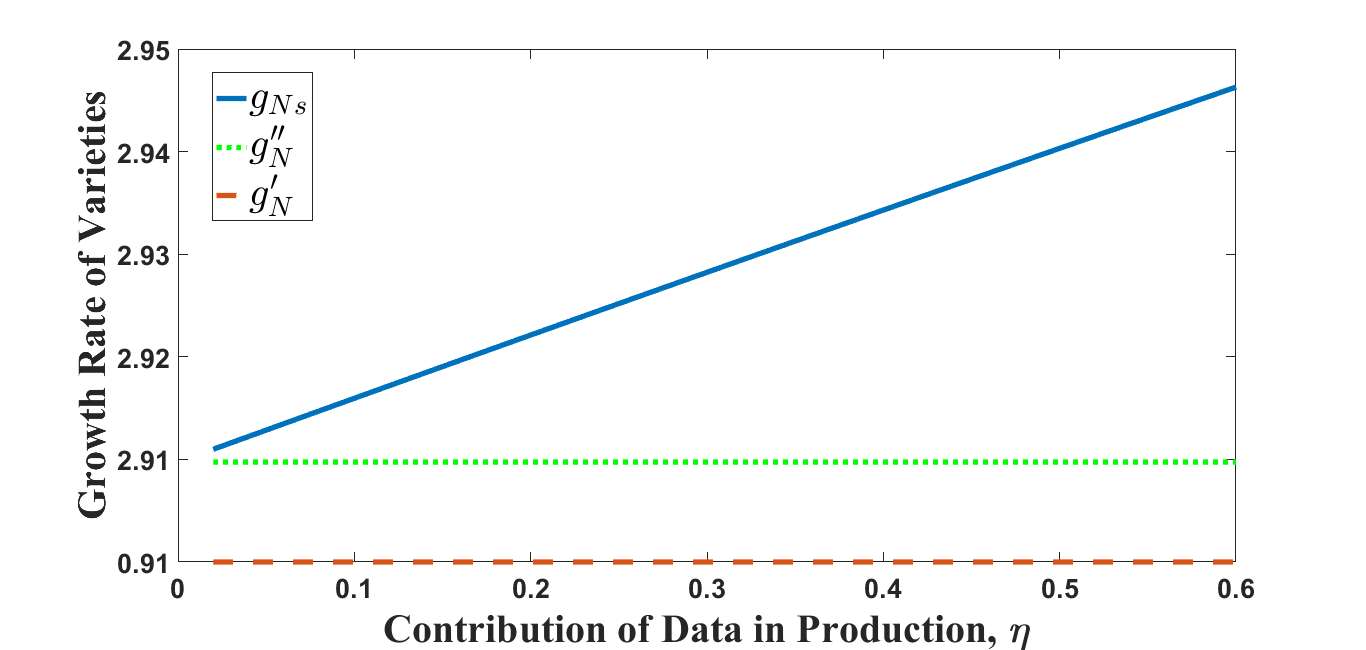}}\label{fig:case_eta:a}}
\subfigure[Difference in the quantity of data]{
\includegraphics[scale=0.18]{{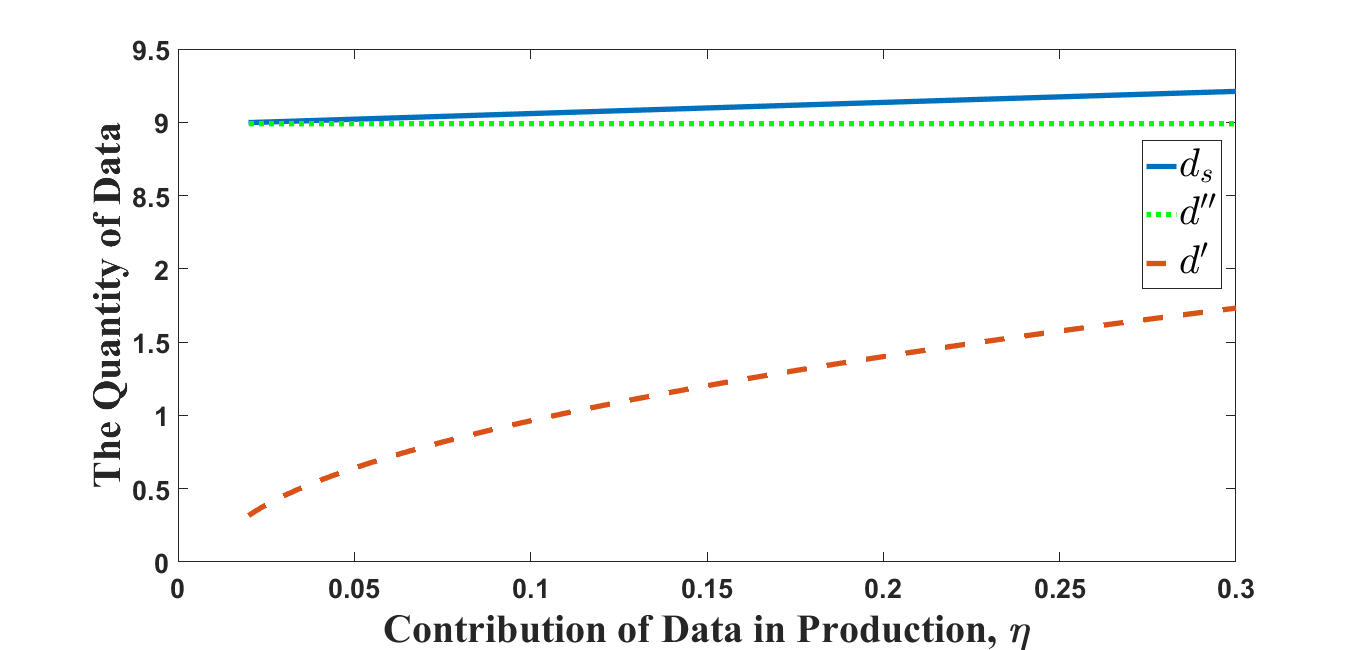}}\label{fig:case_eta:b}}
\subfigure[Difference in labor allocation]{
\includegraphics[scale=0.18]{{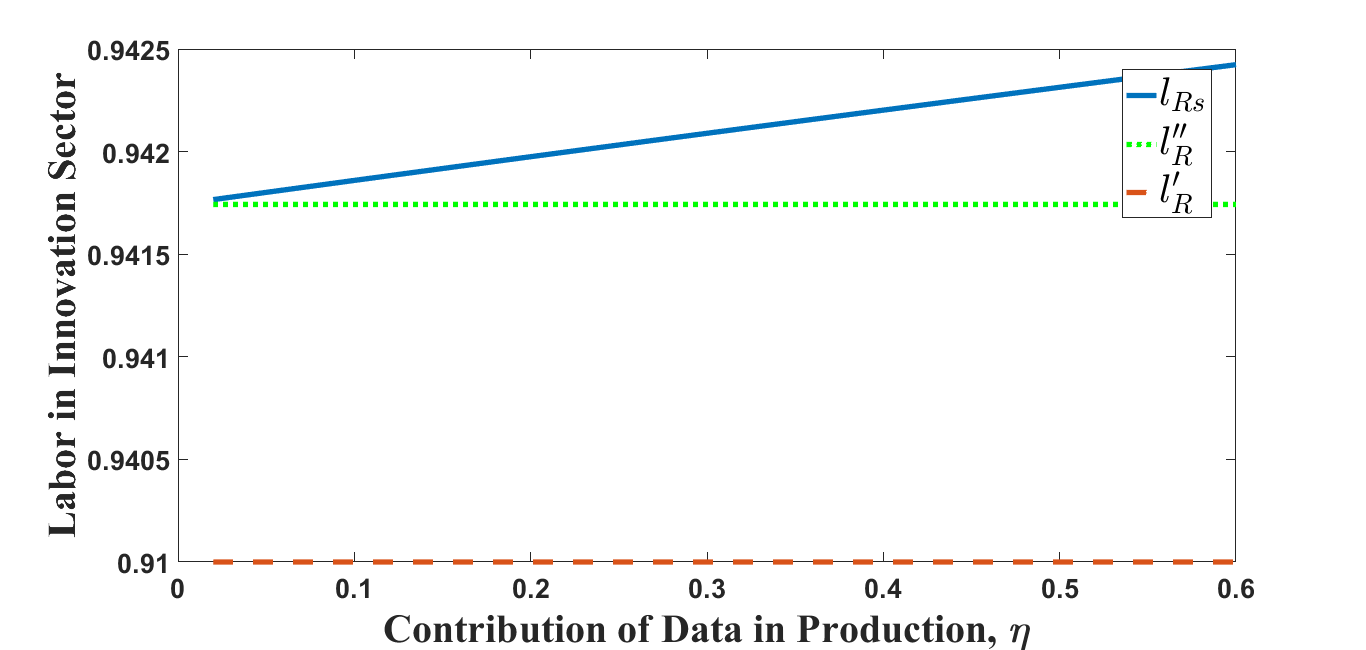}}\label{fig:case_eta:c}}
\caption{Difference allocation among three cases with different $\eta$}
\label{fig:case_eta}
\end{center}
\qquad Notes: The figure depicts different allocations with different $\eta$. For legibility, the $y$-axis is compressed. The full line represents the optimal allocation, the dashed line represents the economy where data only enter the production sector, and the dotted line represents the economy where data only enter the innovation sector. $\eta$ does not impact $g'_N$, $l'_R$ and all variables where data only enter the innovation sector. Other parameters are given in Table \ref{tab:parameterValues}. This color scheme is the same in Figures \ref{fig:case_eta}-\ref{fig:case_kappa}.
\end{figure}

Figure \ref{fig:case_eta} displays the allocations under different values of $\eta$. Against the backdrop of multiple data uses, the quantity of data and the growth rate are strictly larger than those when data are only used either in production or in innovation. Here, the higher contribution of data in production, which is captured by $\eta$, can increase the growth rate of varieties $g_N$, while other parameters remain unchanged. Intuitively, the planner is willing to share larger quantity of data with more compensation from the production sector. With vertical nonrivalry, the additional data can also be used in the innovation sector, so the growth rate increases simultaneously. Recalling the discussion of (\ref{s:lRinMain}) and (\ref{s:compensation}), if the quantity of data is sufficient and increases faster than $\eta$, the planner will choose to devote more labor to the innovation sector, which is shown in Figure \ref{fig:case_eta} and \ref{fig:case_eta:c}, respectively. 

Figure \ref{fig:case_xi} presents the allocations under different values of $\xi$. Unlike the role of $\eta$ shown in Figure \ref{fig:case_eta}, higher value of $\xi$ has more significant influences on the increasing of growth rate and quantity of data shared. Meanwhile, the changing patterns of the quantity of data and labor employment are similar both in the case when data have multiple uses and in that when data only have innovation use. However, when data only have production use, the value of $\xi$ does not influence the three key variables significantly compared with the other two cases.

\begin{figure}[H]
\begin{center}  
\subfigure[Difference in growth rate of varieties]{
\includegraphics[scale=0.16]{{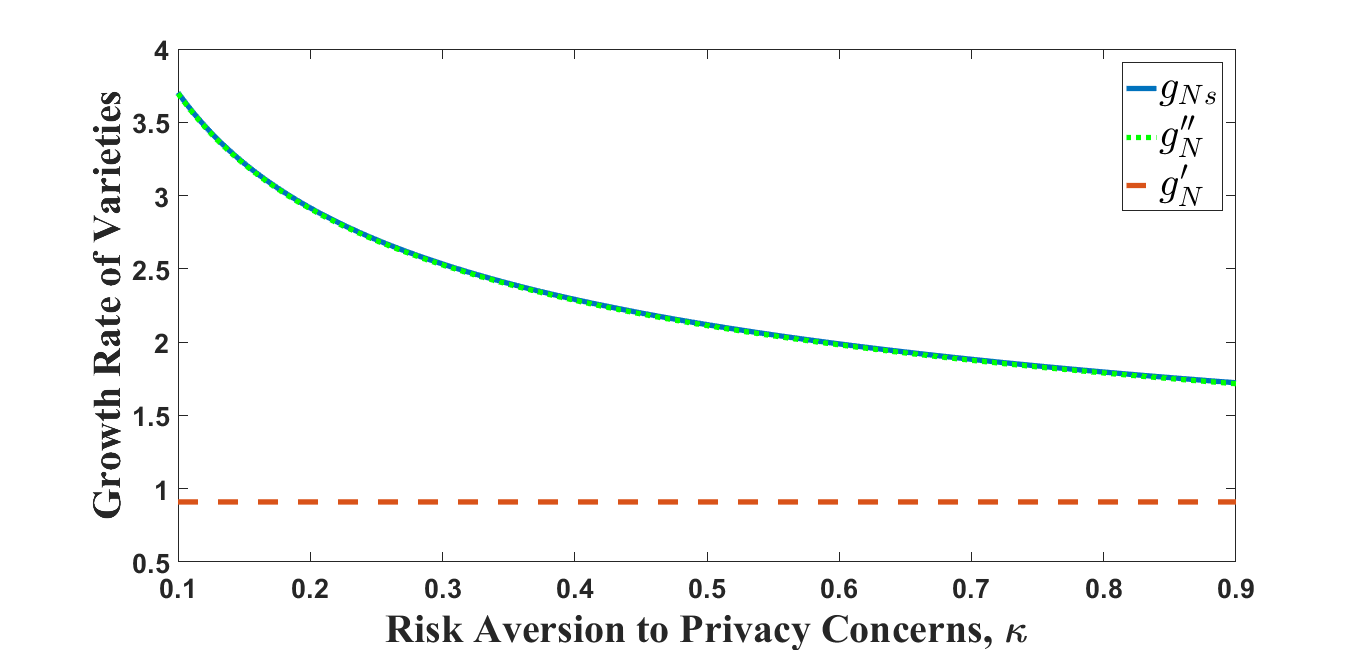}}\label{fig:case_kappa:a}}
\subfigure[Difference in the quantity of data]{
\includegraphics[scale=0.16]{{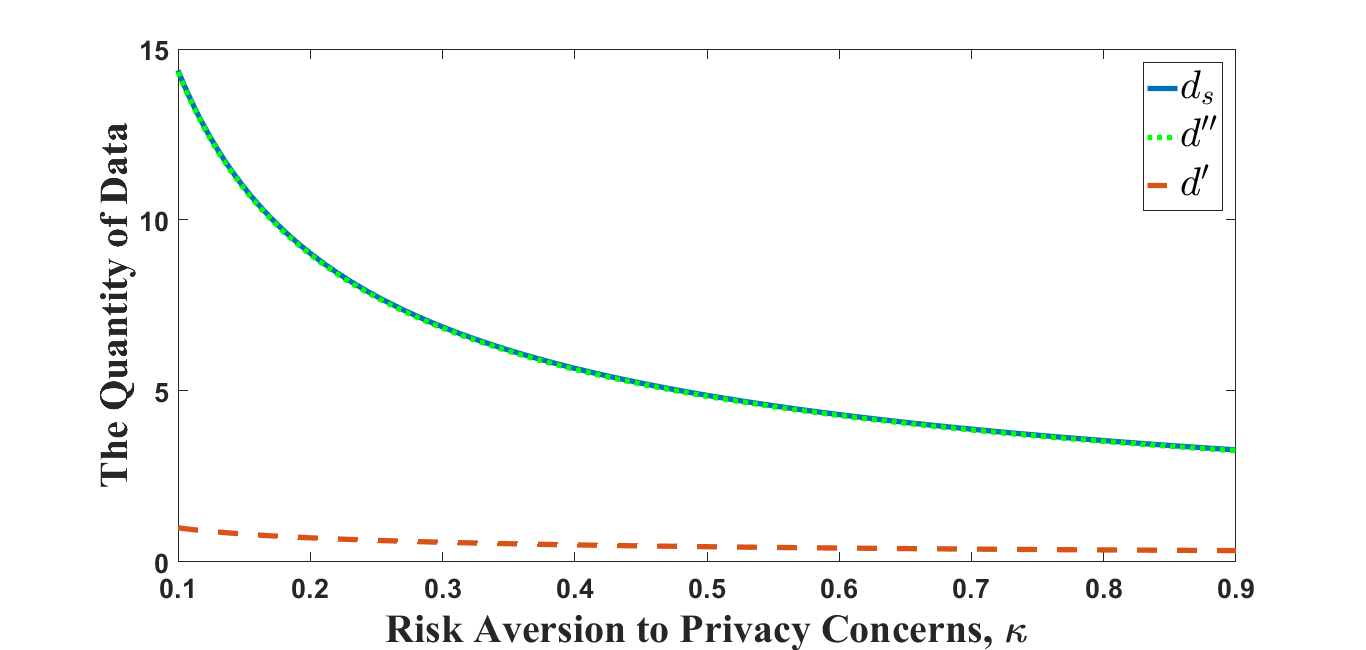}}}
\subfigure[Difference in labor allocation]{
\includegraphics[scale=0.16]{{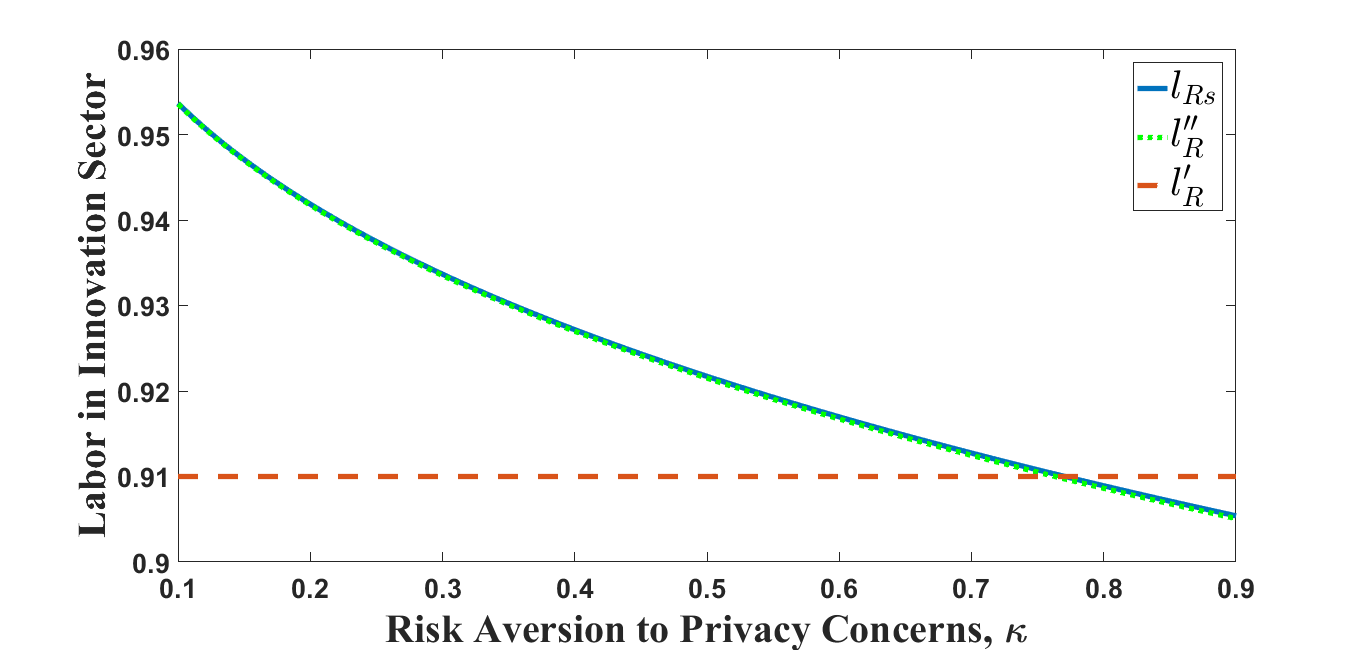}}}
\caption{Difference allocation among three cases with different $\kappa$ }
\label{fig:case_kappa}
\end{center}
\qquad Notes: The figure depicts different allocations with different $\kappa$. When data only enter the production sector, increasing $\kappa$ reduces $d'$ but have no impact on economic growth $g'_N$. In optimal allocation which is depicts by the full line, $\kappa$ affects three variables. 
\end{figure}

Figure \ref{fig:case_kappa} depicts the allocations under different values of $\kappa$. When data are used only in production, increasing $\kappa$ reduces the quantity of data, but has no impact on economic growth. On the contrary, $\kappa$ has a broader impact when data have multiple uses and when data only have innovation use. Under these two cases, as the household's risk aversion to privacy concerns becomes higher, the planner will reduce the extent of data sharing and the fraction of employment in the innovation sector, and thus further affect the growth rate. 

\subsection{Misallocation in Equilibrium}

\begin{figure}[H]
\begin{center}  
\subfigure[Difference in growth rate of varieties]{
\includegraphics[scale=0.18]{{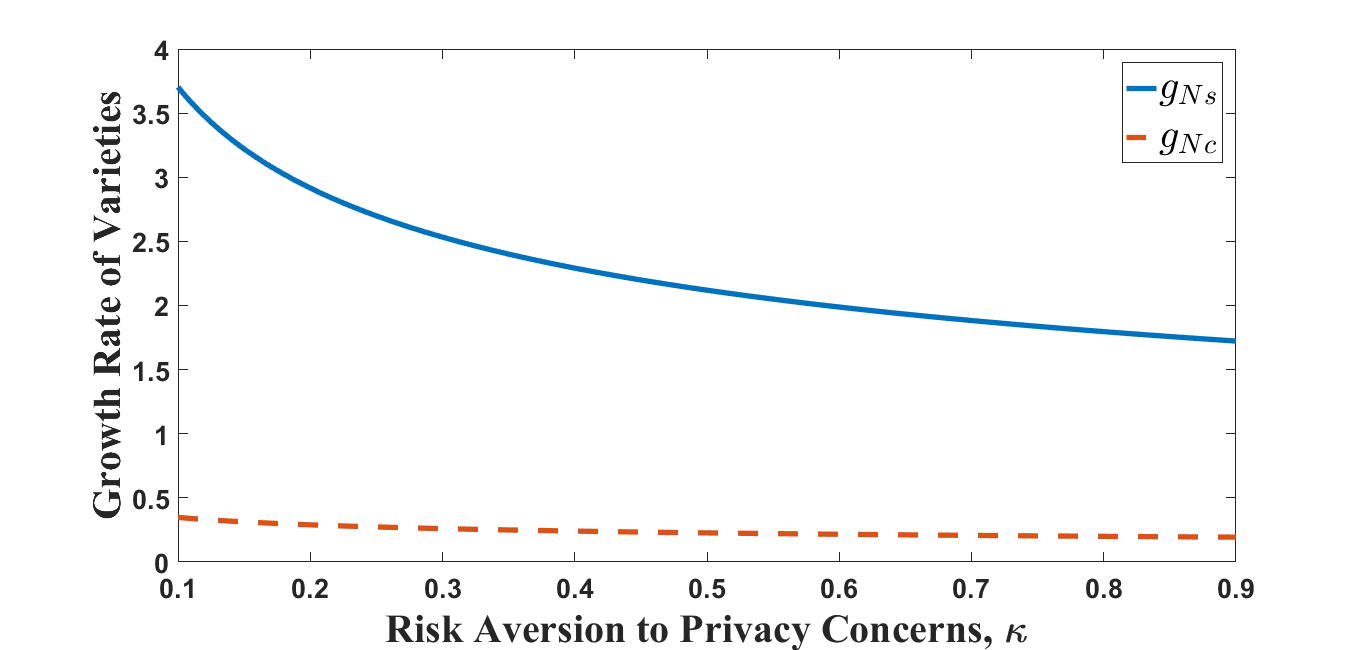}}}
\subfigure[Difference in the quantity of data]{
\includegraphics[scale=0.18]{{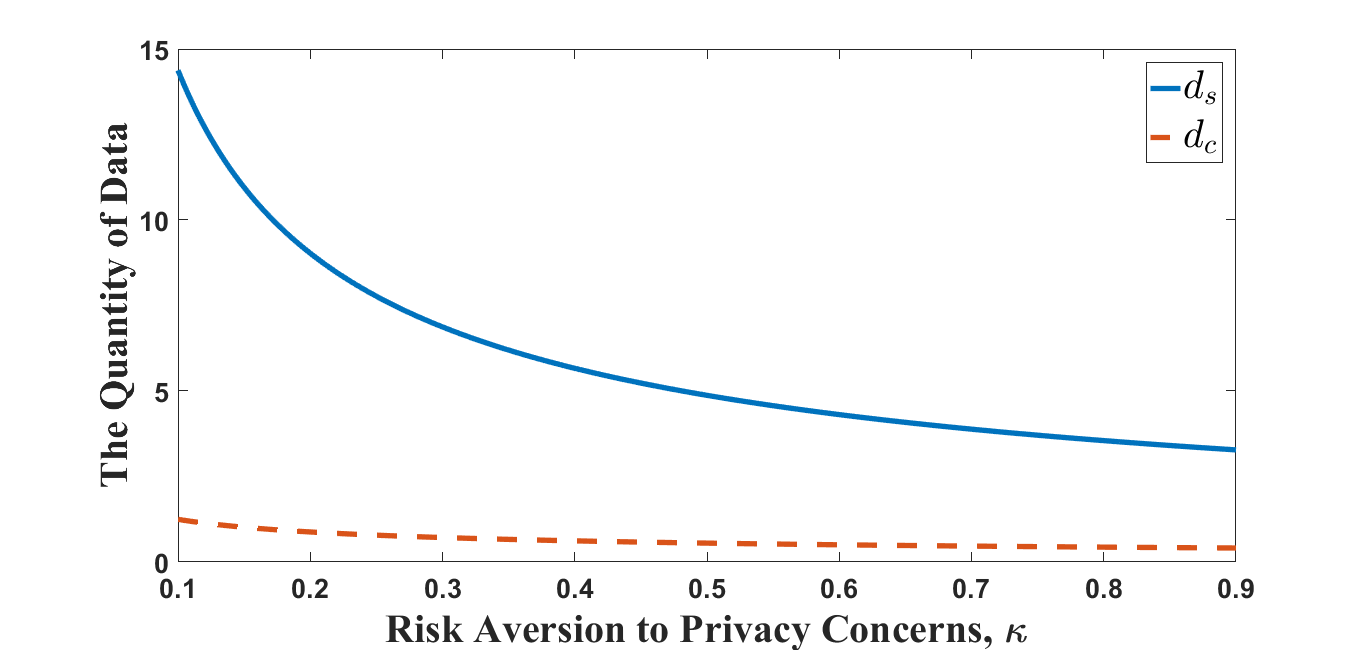}}}
\subfigure[Difference in labor allocation]{
\includegraphics[scale=0.18]{{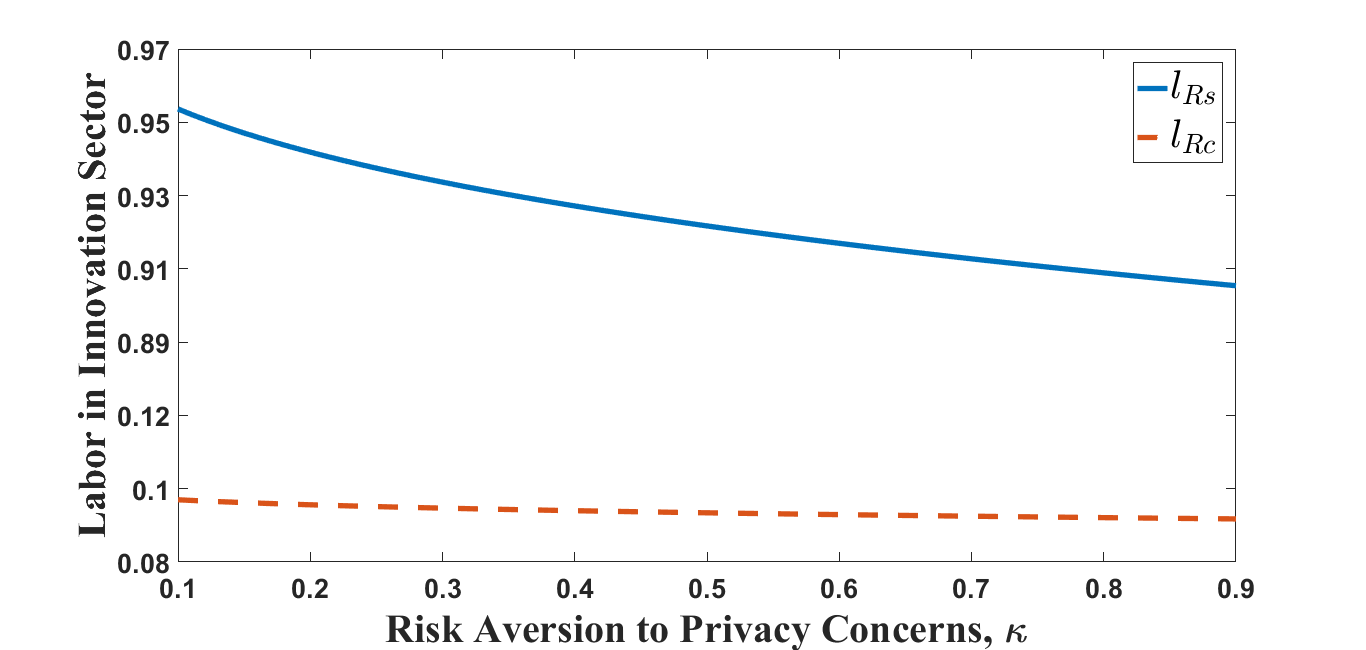}}\label{fig:diff_dc_kappa:c}}
\caption{Different allocation between optimal allocation and decentralized economy with different $\kappa$ }
\label{fig:diff_dc_kappa}
\end{center}
\qquad Notes: Orange dashed lines depict three key variables with different $\kappa$ in the decentralized economy, and blue lines depict variables in the optimal allocation. $d_c$, $g_{Nc}$, $l_{Rc}$ display the negative correlation with $\kappa$.
\end{figure}

We next discuss allocations in the four different models in Section \ref{sec:endo_growth} to characterize misallocation in the decentralized economy. According to Table \ref{tab:Baseline Numerical}, in the decentralized economy, due to market distortions, the quantity of data usage is only slightly higher than that when data are only used in production. Compared with the optimal allocation, the underutilization of data is obvious. In the long run, the decentralized economy suffers from a growth trap of insufficient varieties and innovation. Figure \ref{fig:diff_dc_kappa} reveals that this growth trap becomes more serious as $\kappa$ increases.

\begin{figure}[H]
\begin{center} 
\subfigure[Difference in growth rate of varieties]{
\includegraphics[scale=0.16]{{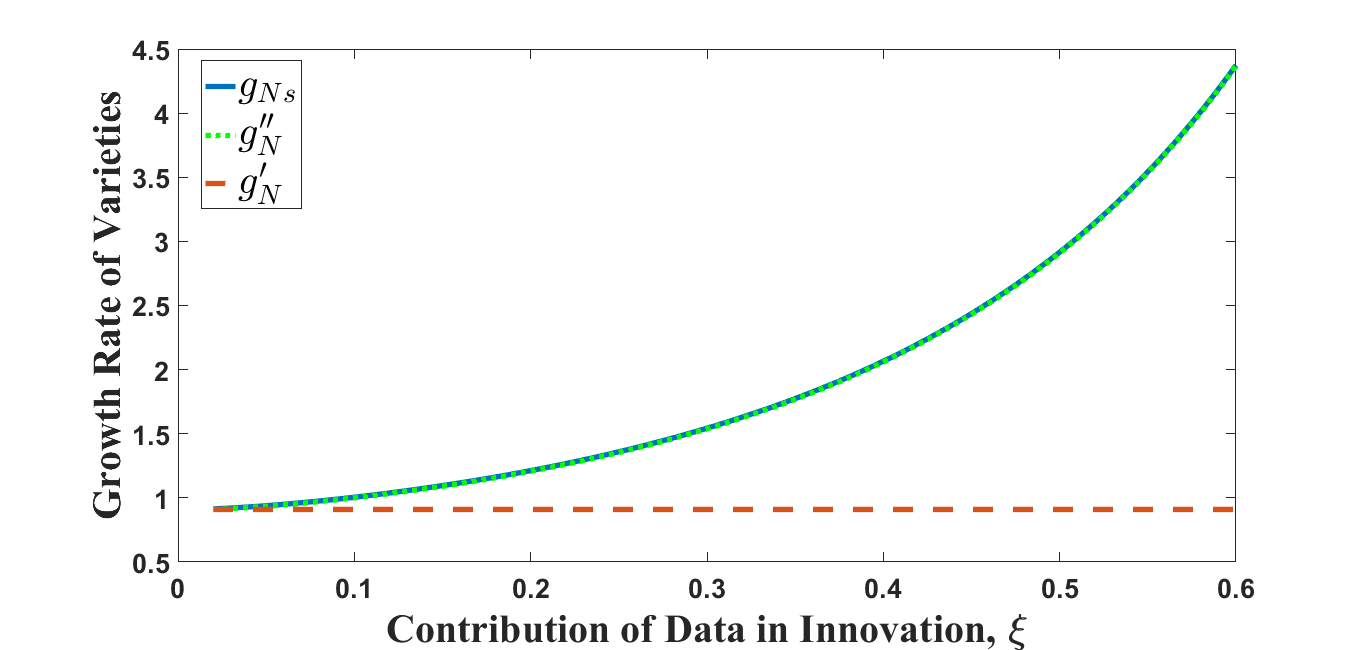}}\label{fig:case_xi:a}}
\subfigure[Difference in the quantity of data]{
\includegraphics[scale=0.16]{{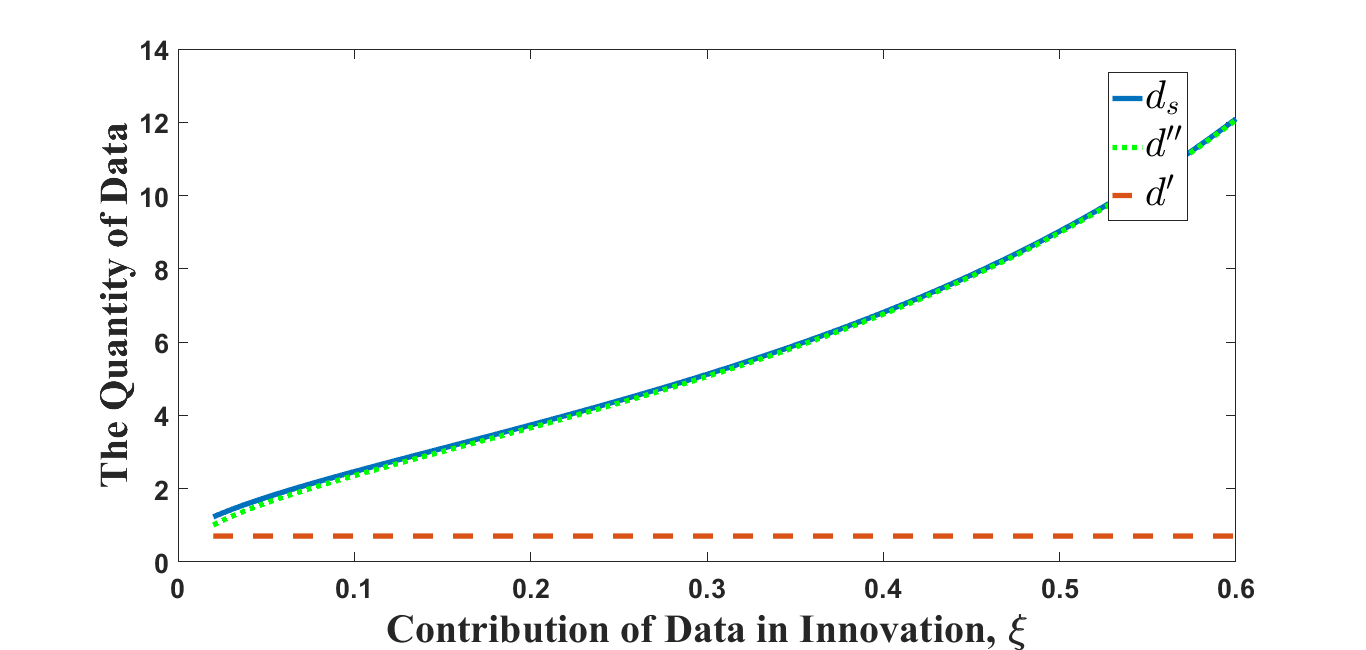}}\label{fig:case_xi:b}}
\subfigure[Difference in labor allocation]{
\includegraphics[scale=0.16]{{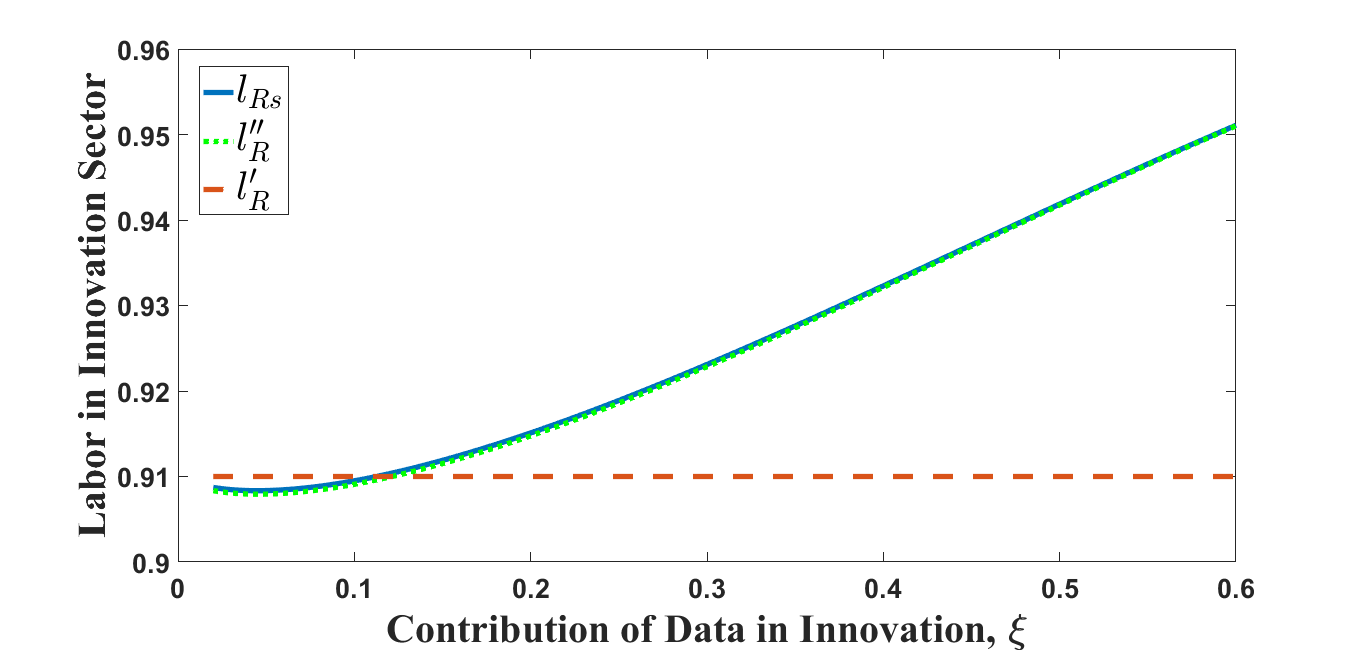}}\label{fig:case_xi:c}}
\caption{Difference allocation among three cases with different $\xi$}
\label{fig:case_xi}
\end{center}
\qquad Notes: The figure depicts different allocations with different $\xi$. The impact of $\xi$ on the multiple uses and only innovation use is similar. For example, $\xi$ significantly increases $d_s$ and $g_{Ns}$, but it also causes the outflow of labor from the production sector.
\end{figure}

\begin{figure}[H]
\begin{center}  
\subfigure[Difference in growth rate of varieties]{
\includegraphics[scale=0.18]{{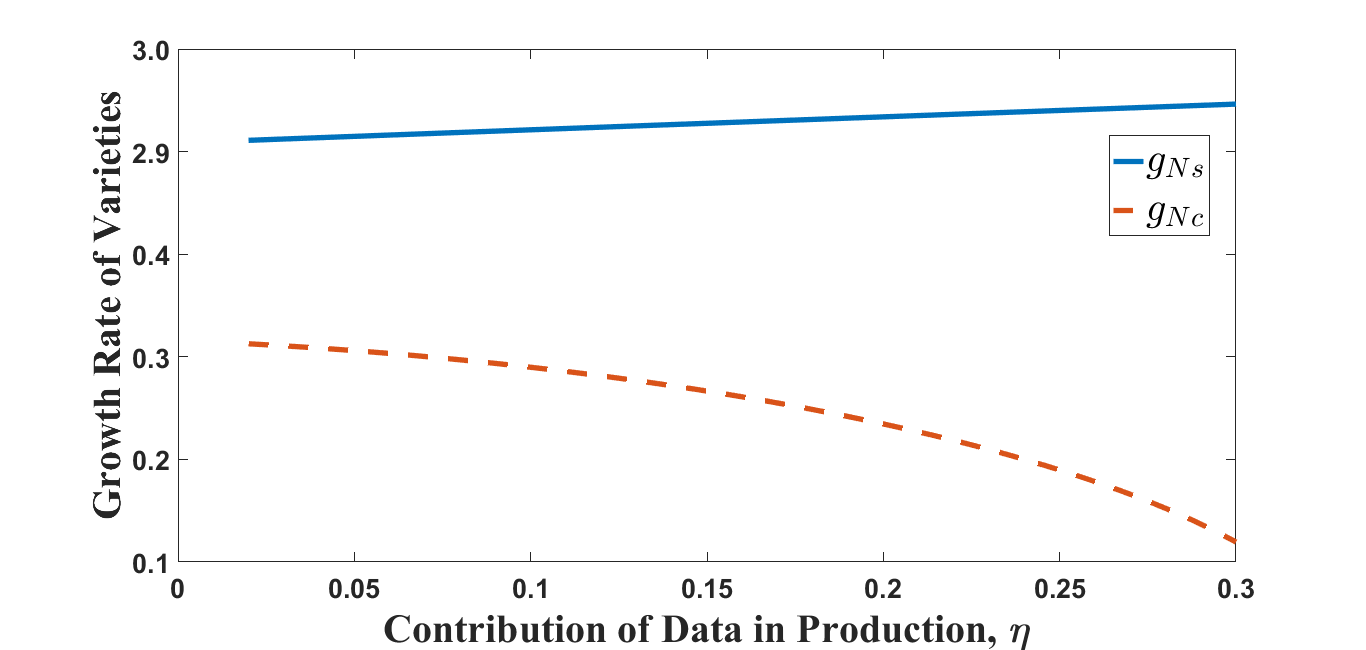}}}
\subfigure[Difference in the quantity of data]{
\includegraphics[scale=0.18]{{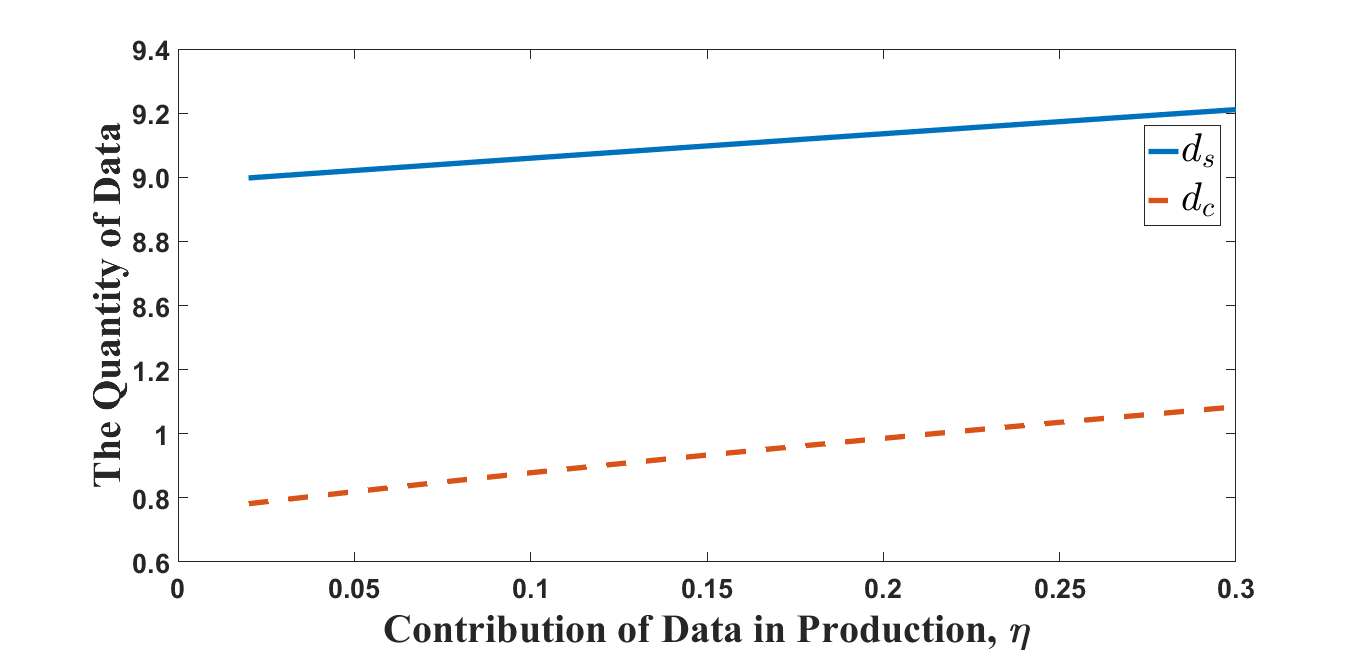}}}
\subfigure[Difference in labor allocation]{
\includegraphics[scale=0.18]{{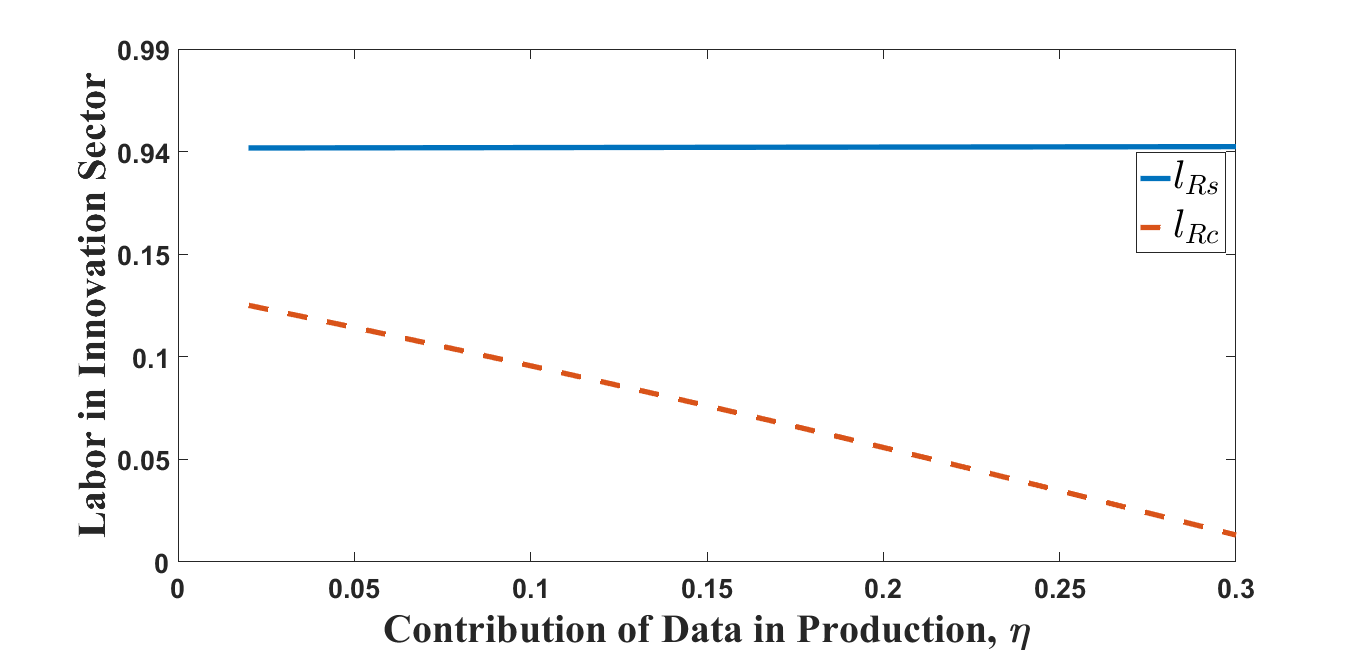}}\label{fig:diff_dc_eta:c}}
\caption{Different allocation between optimal allocation and decentralized economy with different $\eta$ }
\label{fig:diff_dc_eta}
\end{center}
\qquad Notes: The orange dashed lines depict three key variables with different $\eta$ in the decentralized economy. The directions of $g_{Nc}$ and $l_{Rc}$ are opposite to $g_{Ns}$ and $l_{Rs}$ in Figure \ref{fig:case_eta:a} and \ref{fig:case_eta:c}.   
\end{figure}

\begin{figure}[H]
\begin{center}  
\subfigure[Difference in growth rate of varieties]{
\includegraphics[scale=0.18]{{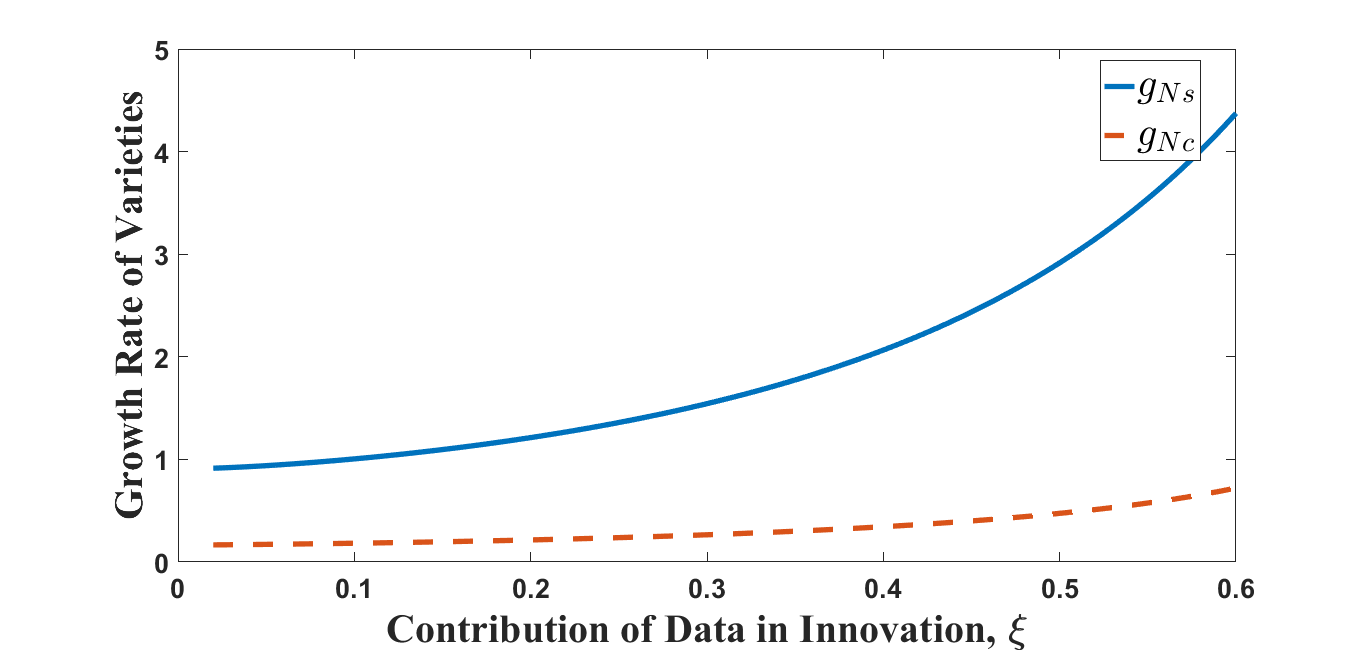}}}
\subfigure[Difference in the quantity of data]{
\includegraphics[scale=0.18]{{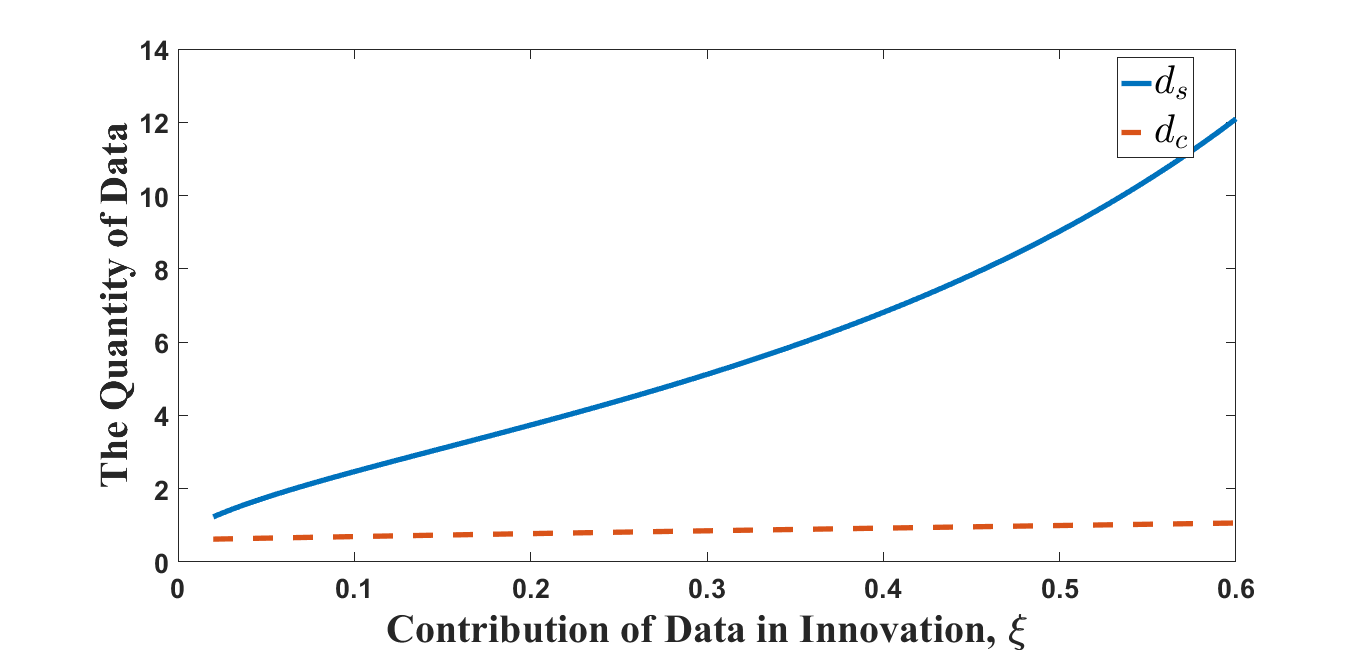}}}
\subfigure[Difference in labor allocation]{
\includegraphics[scale=0.18]{{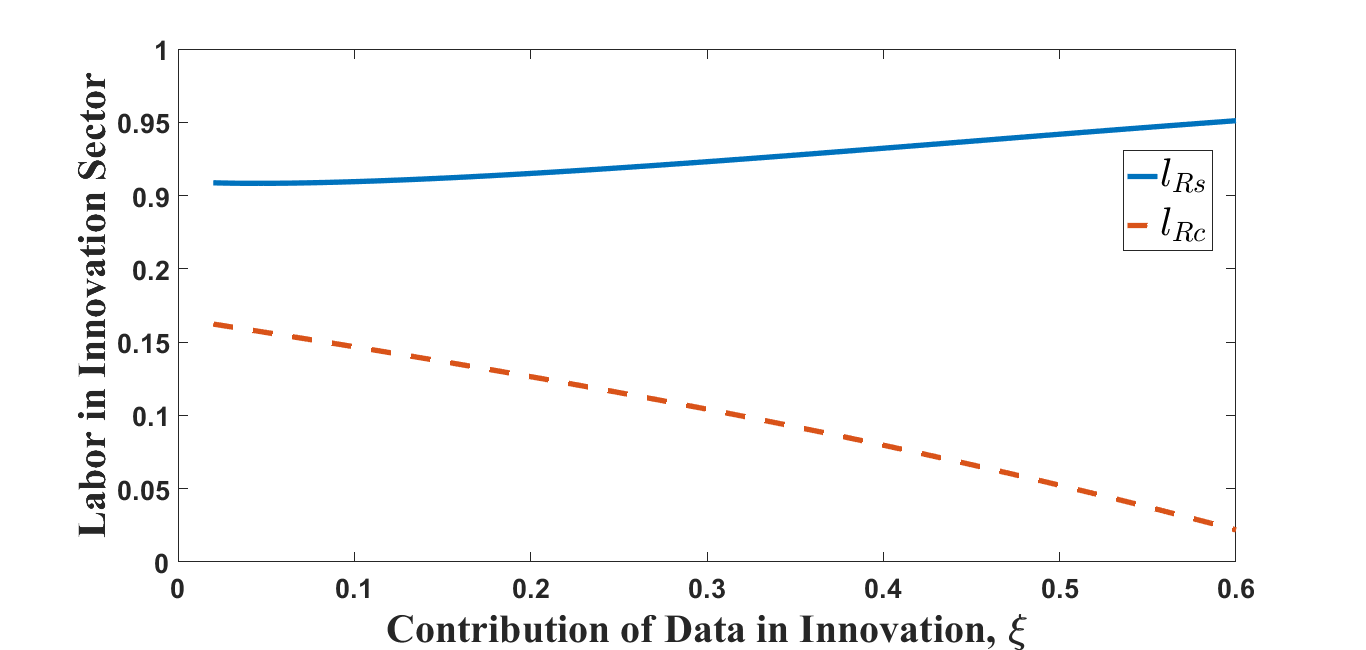}}\label{fig:diff_dc_xi:c}}
\caption{Different allocation between optimal allocation and decentralized economy with different $\xi$ }
\label{fig:diff_dc_xi}
\end{center}
\qquad Notes: Orange dashed lines depict three key variables with different $\xi$ in the decentralized economy,and blue lines depict variables in the optimal allocation. The direction of $l_{Rc}$ is opposite to $l_{Rs}$ in Figure \ref{fig:case_eta:c}.
\end{figure}

Moreover, we simulate across a wide range values for $\eta$ and $\xi$ to find out whether the inefficiency can be mitigated by the improvement in data utilization efficiency, say, $\eta$ and $\xi$. In Figure \ref{fig:diff_dc_eta}, with the increasing of $\eta$, the fraction of labor employed in the innovation sector and the growth rate of varieties are decreasing. Although the quantity of data has increased, it is still too limited to attract labor to the innovation sector. The increase in $\eta$ shows an effect of attracting labor to the production sector. This process of labor allocation goes opposite compared with the optimal allocation shown in Figure \ref{fig:case_eta:c}, that is, the increase in the quantity of data has little effect on economic growth here. Figure \ref{fig:diff_dc_xi} reveals a substitution effect of $\xi$ on labor, which differs from Figure \ref{fig:case_xi:c}, since the quantity of data has not increased significantly (see (\ref{c:lRlE})). In addition, in Figure \ref{fig:diff_dc_xi}, the maximum value of $\xi$ is only 0.9. There is still a big gap from the optimal distribution because of an insufficient compensation for data sharing (in Section \ref{decentralized economy}) and the loss of labor in the innovation sector.


\subsection{Alternative Specifications}

Following both \cite{Jones2020} and \cite{Cong2021}, we set the production sector to have increasing returns and the innovation sector, constant returns. However, one may question if the production sector should also have constant returns to scale, e.g.,  $Y(v,t)=L_E(v,t)^{1-\theta}D(t)^\theta$, where $L_E(v,t)$ and $\theta$ are the labor employed and the data contributed in production. Here both the direct and indirect effects of $\theta$ increase labor employment in the innovation sector. Nonetheless, we show in Appendix \ref{APP:constant} and Figure \ref{fig:constant_theta}-\ref{fig:constant_kappa} that our key findings remain robust.

One may suspect that privacy concerns become greater as the number of firms using data increases, because the repeated use among different firms increase the likelihood of leakage and hacking. We can allay this concern by allowing the disutility term to be $\int_0^{N(t)} [\kappa d(v,t)^2/2] \mathrm{d}v$. Along the BGP, the quantity of data decreases and the growth rate of varieties gradually declines towards zero, as shown in Appendix \ref{cumulativeAPP}. In this case, innovation leads to the increasing number of firms, while it causes larger privacy costs which impedes further data sharing. Meanwhile, the decrease of the quantity of data sharing in turn pushes innovation to become insufficient over time. In reality, with multiple uses, it is difficult for consumers or the government to distinguish the specific use of data. Also, the corresponding extends of privacy breaches are also hard to separate, which further slows down the growth of data economy. This alternative setting complements the discussion about the growth in data economy, such as \citet{Cong2021} and \citet{Farboodi2021}, from the perspective of increasing privacy costs. 

Similarly, privacy costs may increase with multiple uses of data in multiple sectors. Specifically, each consumer's instantaneous utility function is now adjusted as
\begin{equation}\nonumber
\max _{\left\{l_E(t), d(t)\right\}}  \int_0^\infty e^{-\rho t} L\left[\ln{c(t)}- (1+\alpha)\frac{\kappa d(t)^2}{2} \right] \mathrm{d} t, \end{equation}
where $\alpha\geq 0$ denotes the additional privacy concerns due to multiple uses. In Appendix \ref{additionalPrivacy}, Figure \ref{fig:additional privacy} and Table \ref{tab:additional privacy}, we show the equilibrium and the comparative statistics with respect to $\alpha$ (or equivalently, with respect to $\kappa$). When $\alpha$ is close to zero, the model reduces to the baseline with multiple uses of data as discussed in Section \ref{OptimalAllocation}. The multiple uses bring higher compensation for privacy costs, and consequently a higher quantity of data usage as we discuss in Section \ref{subsec:multi_use_num}. When $\alpha$ increases, sharing data with the production sector may decrease the quantity of data and the growth rate because the production sector cannot provide economic growth and additional compensation coming from $\eta$ are limited, making sharing the same quantity of data as the innovation sector to the production sector may be accompanied by huge privacy costs. With vertical nonrivalry, higher privacy costs lead to a reduction of the quantity of data and the growth rate, and the results shown in Table \ref{tab:Baseline Numerical} will become reversed. In this case, the quantity of data used in the production sector should be limited, in order to impede the increase of additional privacy costs.

\section{Conclusion}
\label{sec:conclusion}

The nonrival nature of data allows them to be employed in both innovation and production simultaneously. We develop an endogenous growth model to understand the interactions and differences of data uses in these sectors. We show that consumers' privacy concerns bound the growth in the overall quantity of data used. Moreover, the usage of data in the innovation sector plays a more important role in economic growth than that in the production sector, because that (i) data are dynamically nonrival and add to knowledge accumulation, and that (ii) innovations ``desensitize'' data, reducing consumers' privacy costs when knowledge enters the production sector instead of raw data. 

Data uses in both sectors interact to generate spillover of allocative distortion and exhibit an apparent substitutability due to labor's rivalry and complementarity with data. Consequently, growth rates under a social planner and a decentralized equilibrium differ, which is a novel result obtainable only in a model with fully endogenous growth. In an optimal allocation, the majority of the labor is employed in innovation. But in a decentralized equilibrium, consumers' failure to fully internalize knowledge spillover in the face of privacy concerns, combined with firms' market power, underprices data and inefficiently limits their supply, leading to underemployment in the innovation sector and a substantially lower data utilization and growth in the long run. Also, we discuss interventions in the data market and potential direct subsidies to mitigate the aforementioned inefficiencies.

We are at the dawn of research on data economy and growth. To offer clear insights in a tractable way, we have necessarily left interesting aspects of the economics of data for future research. For example, a lower $\kappa$, the exogenous parameter for consumers' disutility, facilitates faster growth. It can be endogenously influenced by technological innovations such as privacy-preserving computation \citep{cao2019financial,hastings2020privacy} and distributed ledgers \citep{cong2019blockchain,chen2021brief}. It is equally interesting to understand how regulatory policies such as GDPR affect $\kappa$ in the long run. Furthermore, consumers' cost of data contribution is modeled in reduced form here; an active literature starts to provide microfoundations for privacy costs \citep[e.g.,][]{ichihashi2021economics,Ichihashi2020,liu2020data}, including possibly negative data externalities exacerbating privacy concerns \citep{Acemoglu2020,ichihashi2021economics}. 

\bibliography{refs.bib}

\newpage

\begin{center}
{\Large \textbf{Figures and Tables}}
\end{center}

\begin{table}[H]
\caption{Parameters Values in Baseline}
\begin{center}
\begin{tabular*}{\hsize}{@{\extracolsep{\fill}}c l c c}
    \toprule
    Parameter & Description & Value & Source \\
    \midrule
    $\eta$ & Contribution of data in production sector & 0.1 & Standard \\
    $\xi$ & Contribution of data in innovation sector& 0.5 & Discretionary \\
    $\kappa$ & Risk aversion to privacy concerns & 0.2 & Discretionary \\
    $\gamma$ & Elasticity of substitution between varieties & 4 & Standard \\
    $\rho$ & Rate of time preference & 0.03 & Standard \\
    $L$ & Population level & 1 & Standard  \\
    $\varepsilon$ & Innovation efficiency & 1 & Standard  \\
    \bottomrule
    \label{tab:parameterValues}
\end{tabular*}
\end{center}
\qquad Notes: Baseline parameter values for the numerical analysis. 
\end{table}


\begin{table}[H]
\caption{The Baseline Numerical Example}
\begin{center}
\begin{tabular*}{\hsize}{@{\extracolsep{\fill}}l c c c}
    \toprule
    Model & $g_N$ & $d$ & $l_R$ \\
    \midrule
    Social Planner & 2.9160 & 9.0277 & 0.9419 \\
    Decentralized Economy & 0.2897 & 0.8783 & 0.0956  \\
    Only in Production (SP) & 0.9100 & 0.7071 & 0.9100  \\
    Only in Innovation (SP) & 2.9097 & 8.9903 & 0.9417  \\
    \bottomrule
    \label{tab:Baseline Numerical}   
    \end{tabular*} 
\end{center}
\qquad Notes: The table reports three key variables for different allocations using the parameter values in Table \ref{tab:parameterValues}. $g_N$ is BGP level of the growth rate of varieties, $d$ is BGP level of the quantity of data and $l_R$ is BGP level of research employment. Compared with single use in last two lines, multiple uses in the social planner have higher $g_N$, and the quantity of data is not a simple addition of the former. 
\end{table}

\end{spacing}
\pagebreak
		
\titleformat{\section}{\normalfont\Large\bfseries}{Appendix~\Alph{section}}{11pt}{\Large}
\setcounter{equation}{0}
\setcounter{figure}{0}
\setcounter{table}{0}  
\renewcommand{\theequation}{A.\arabic{equation}}
\renewcommand{\thefigure}{A.\arabic{figure}}
\renewcommand{\thetable}{A.\arabic{table}}
\begin{appendices}
\renewcommand{\thesection}{\Alph{section}}

\begin{center}
    {\large{\textbf{Online Appendix}}}
    
    \bigskip
    
    {\LARGE{\textbf{Endogenous Growth Under Multiple Uses of Data}}}
\end{center}

\section{Endogenous Growth}

\subsection{BGP in Optimal Allocation}
\label{App:OptimalAllocation}

The social planner problem is
\begin{equation}\nonumber
\max_{\left\{l_E(t), d(t)\right\}} \int_0^\infty e^{-\rho t} L\left(\ln{c(t)}-\frac{\kappa d(t)^2}{2} \right) \mathrm{d}t,
\end{equation}
s.t.,
\begin{align}
c(t) &=Y(t) / L, \nonumber \\
Y(t) &=N(t)^{\frac{1}{\gamma-1}}l_E(t)L^{1+\eta}d(t)^{\eta}, \nonumber \\
\dot{N}(t) &=\varepsilon l_R(t)^{1-\xi} L d(t)^{\xi} N(t), \nonumber \\
1 &= l_E(t) + l_R(t), \nonumber \\
d(t) &\le g(c(t)). \nonumber
\end{align}

Define the following Hamiltonian:
\begin{equation}\nonumber
    \mathcal{H} (l_E(t),d(t),N(t),\lambda (t))=\ln{\left[ N(t)^{\frac{1}{\gamma-1}} l_E(t) L^{\eta} d(t)^{\eta} \right]}-\frac{\kappa d(t)^2}{2}+\varepsilon \lambda (t) l_R(t)^{1-\xi} L d(t)^\xi N(t).
\end{equation}

We then have following FOC:
\begin{equation}
    \frac{\partial \mathcal{H}}{\partial l_E(t)}=\frac{1}{l_E(t)}-\lambda (t)\varepsilon (1-\xi) l_R(t)^{-\xi} L d(t)^\xi N(t)=0,
    \label{s:HlE}
\end{equation}
\begin{equation}
    \frac{\partial \mathcal{H}}{\partial d(t)}=\frac{\eta }{d(t)}- \kappa d(t)+\lambda (t)\varepsilon \xi l_R(t)^{1-\xi } L d(t)^{\xi-1}N(t)=0,
    \label{s:Hd}
\end{equation}
\begin{equation}
    \frac{\partial \mathcal{H}}{\partial N(t)}=-\dot \lambda (t)+\rho \lambda (t)=\frac{1}{\gamma -1}\frac{1}{N(t)}+\varepsilon \lambda (t){l_R(t)^{1-\xi}} L d(t)^{\xi},
    \label{s:HN}
\end{equation}

Consider (\ref{s:HlE}), we have
\begin{equation}
    \frac{l_R(t)}{l_E(t)}=(1-\xi)\lambda(t)\dot N(t).
    \label{s:lRlE}
\end{equation}

Consider (\ref{newentrants:p}), we have
\begin{equation}
    \frac{\dot N(t)}{N(t)}=\varepsilon{l_R(t)^{1-\xi}}L d(t)^{\xi}.
    \label{s:gN}
\end{equation}

Next, plug (\ref{s:lRlE}) into (\ref{s:Hd}), we have
\begin{align}
    \kappa d(t)^2 &= \eta+\xi\lambda(t)\dot N(t) \nonumber \\
    &= \eta+\frac{\xi}{1-\xi}\frac{l_R(t)}{l_E(t)}.
    \label{s:dataUstility}
\end{align}

Then, we have the fraction of labor employed in potential entrants or the R\&D sector:
\begin{equation}
    l_R = \frac{\frac{1-\xi}{\xi}(\kappa d^2 -\eta)}{1+\frac{1-\xi}{\xi}(\kappa d^2 -\eta)}.
    \label{s:lRd}
\end{equation} 

From (\ref{s:lRlE}), we have $g_\lambda=-g_N$ along BGP. Multiplying (\ref{s:HN}) by $N(t)$, we have
\begin{align}
    N(t)\frac{\partial \mathcal{H}}{\partial N(t)}=-\dot \lambda (t)N(t)+\rho \lambda (t)N(t) &= \frac{1}{\gamma -1}+\varepsilon \lambda (t) l_R(t)^{1-\xi} L d(t)^{\xi} N(t) \nonumber \\
    \Rightarrow \quad g_N \lambda(t) N(t) + \rho\lambda(t) N(t) &= \frac{1}{\gamma -1} +g_N \lambda(t) N(t) \nonumber \\
    \Rightarrow \quad \lambda(t) N(t) &= \lambda N = \frac{1}{\gamma-1} \frac{1}{\rho}. \label{s:lambdagN}
\end{align}

Plug (\ref{s:lambdagN}) into (\ref{s:dataUstility}), we have
\begin{equation}\nonumber
    g_N^1(d) = \frac{(\gamma-1)\rho}{\xi}(\kappa d^2 -\eta).
\end{equation}

Plug $\lambda{N}$ in (\ref{s:lambdagN}) and $l_R$ in (\ref{s:lRd}) into (\ref{s:gN}), respectively, we have (\ref{s:gNd2}):
\begin{equation}\nonumber
    g_N^2(d) = \varepsilon L \left[{\frac{\frac{1-\xi}{\xi}(\kappa d^2 -\eta)}{1+\frac{1-\xi}{\xi}(\kappa d^2 -\eta)}}\right]^{1-\xi }d^{\xi }.
\end{equation}

Note that $g_N^2(d)$ increases with $d$ and is smaller than $\varepsilon L d^{\xi}$. Define $F(d)=g_N^1(d)-g_N^2(d)$, we have
\begin{align}
    F(d) &= \frac{(\gamma-1)\rho}{\xi}(\kappa d^2 -\eta)-\varepsilon L \left[\frac{\frac{1-\xi}{\xi}(\kappa d^2 -\eta)}{1+\frac{1-\xi}{\xi}(\kappa d^2 -\eta)}\right]^{1-\xi }d^{\xi} \nonumber \\
    &>\frac{(\gamma-1)\rho}{\xi}(\kappa d^2 -\eta)-\varepsilon L d^{\xi}. \nonumber
\end{align}
Obviously, with $0<\xi<1$, there exists an $d_0$, where $F(d)>0$ for all $d>d_0$.

Next, because $F(\left( \eta/\kappa\right)^\frac{1}{2})=0$ and $F(d)$ is continuous, if $F'(\left(\eta/\kappa \right)^\frac{1}{2})<0$, there exists a $d_1$ to satisfy $F(d_1)<0$. The first-order derivative of $F(d)$ is:
\begin{align}
    F'(d) =& \frac{(\gamma-1)\rho}{\xi} 2 \kappa d 
    -\varepsilon L \xi d^{\xi-1}\left[\frac{\frac{1-\xi}{\xi}(\kappa d^2 -\eta)}{1+\frac{1-\xi}{\xi}(\kappa d^2 -\eta)}\right]^{1-\xi }-... \nonumber \\
    &\varepsilon L (1-\xi) d^\xi \left[ \frac{\frac{1-\xi}{\xi}(\kappa d^2 -\eta)}{1+\frac{1-\xi}{\xi}(\kappa d^2 -\eta)} \right]^{-\xi }
    \left[\frac{\frac{1-\xi}{\xi} 2 \kappa d }{(1+\frac{1-\xi}{\xi}(\kappa d^2 -\eta))^2}\right]. \nonumber
\end{align}

Then, we have $F'(\left(\eta/\kappa\right)^\frac{1}{2}) \to -\infty$. Combining with $F(d)>0$ for all $d>d_0$, the solution of $g_N^1(d)$ and $g_N^2(d)$ exists. Also, combining the concavity and convexity, the solution is unique.

\subsection{Data Intermediary Problem}

Divide (\ref{firm:D}) by (\ref{newentrants:D}) and substitute $Y(t)$ by definition (\ref{output:Index}), with symmetry we have
\begin{equation}
   \frac{(1-\frac{1}{\gamma })\eta N(t)^{\frac{1}{\gamma -1}} Y(v,t)}{\xi\dot N(t)V(t)}=\frac{q_d^{prod} (v,t)}{q_d^{inno} (t)}.
   \label{qq}
\end{equation}

Similarly, divide (\ref{firm:L}) by (\ref{newentrants:L}), we have
\begin{equation}
   \frac{(1-\frac{1}{\gamma }) N(t)^{\frac{1}{\gamma -1}} Y(v,t)}{(1-\xi) \dot N(t)V(t)}=\frac{L_E (v,t)}{L_R (t)}.
   \label{ww}
\end{equation}

Plug $L_E(v,t)N(t)=L_E(t)$ into (\ref{ww}), then
\begin{equation}
   \frac{(1-\frac{1}{\gamma })N (t)^{\frac{1}{\gamma -1}}Y(v,t)N(t)}{(1-\xi)\dot N(t)V(t)}=\frac{l_E(t)}{l_R(t)}.
   \label{ww2}
\end{equation}

Plug (\ref{ww2}) into (\ref{qq}), we get the data price ratio as
\begin{equation}
    \frac{\eta(1-\xi) }{\xi }\frac{l_E(t)}{l_R(t)}q_d^{inno}(t)=q_d^{prod}(v,t)N(t).
    \label{qqrelation}
\end{equation}

Plug (\ref{qqrelation}) into the zero condition (\ref{dataInter:zero}), we have
\begin{equation}\nonumber
    q_d^{inno}(t)+\frac{\eta(1-\xi)}{\xi }\frac{l_E(t)}{l_R(t)}q_d^{inno}(t)=p_d(t).
\end{equation}

\subsection{BGP in Decentralized Model}
\label{APP:Decentralized}

The problem for consumers is defined by Hamiltonian (\ref{consumerHam}), the FOC for consumption $\partial \mathcal{H} /\partial c(t)$ is
\begin{equation}\nonumber
    \frac{1}{\left( \int_0^{N(t)} c(v,t)^{\frac{\gamma -1}{\gamma }} \mathrm{d}v \right)^{\frac{\gamma }{\gamma -1}}} \left( \int_0^{N(t)} c(v,t)^{\frac{\gamma -1}{\gamma }} \mathrm{d}v \right)^{\frac{1}{\gamma -1}} c(v,t)^{\frac{-1}{\gamma }} - \mu  (t) p(v,t)=0.
\end{equation}
Then, we have
\begin{equation}\nonumber
    \mu (t)=\frac{1}{c(t) \left[ \int_0^{N(t)} p(v,t)^{1-\gamma }\mathrm{d}v \right]^{\frac{1}{1-\gamma }}}.
\end{equation}
We normalize the price index $P(t) \equiv \left( \int_0^{N(t)} p(v,t)^{1-\gamma} \mathrm{d}v \right)^{\frac{1 }{\gamma -1}} = 1$ in (\ref{priceIndex}), which means $\mu(t) =1/c(t)$.

Plug $\mu(t) =1/c(t)$ into $\partial \mathcal{H}/ \partial d(t)$, we have
\begin{equation}
    \frac{\partial \mathcal{H}}{\partial d(t)} = \kappa d(t)^2 - \frac{p_d(t) d(t)}{c(t)}=0,
    \label{d=pdc}
\end{equation}
where $d(t)^2>(1-1/\gamma)\eta/\kappa$.

Plug (\ref{firm:D}) and (\ref{newentrants:D}) into the zero condition of the data intermediary (\ref{dataInter:zero}), then
\begin{equation}\nonumber
    \left(1-\frac{1}{\gamma }\right)\eta N(t)^{\frac{\gamma }{\gamma -1}}Y(v,t) + \xi\dot N(t)V(t)=p_d(t)d(t)L.
\end{equation}
Then, plug (\ref{d=pdc}) into the above equation, we have
\begin{align}
    \left(1-\frac{1}{\gamma }\right) \eta N(t)^{\frac{\gamma }{\gamma -1}} Y(v,t) + \xi\dot N(t)V(t) =& \kappa d(t)^2 c(t) L \nonumber \\
    \Rightarrow \quad \left( 1-\frac{1}{\gamma }\right) \eta +\xi\frac{\dot N(t)V(t)}{N(t)^{\frac{\gamma}{\gamma-1}} Y(v,t) } =& \kappa d(t)^2 \nonumber \\
    \Rightarrow \quad \frac{\dot N(t)V(t)}{N(t)^{\frac{\gamma }{\gamma -1}}Y(v,t)} =& \frac{ \kappa d(t)^2-\left(1-\frac{1}{\gamma }\right)\eta }{\xi}.     \label{NV/NY}
\end{align}
The second line is a transformation of (\ref{c:FOCd}). At time $t$, the value of the monopolist can be written as
\begin{equation}\nonumber
    V(t)=\int_0^{\infty} \exp{\left(-\int_t^s r(s') \mathrm{d}s' \right)} \left\{ \left[ 1- \left(1-\frac{1}{\gamma } \right)-\left(1-\frac{1}{\gamma }\right) \eta \right] N(s)^{\frac{1}{\gamma -1}} Y(v,s) \right\} \mathrm{d}s.
\end{equation}

If $l_E(t)=0$, the value $V(t)$ will be zero with $Y(v,t)=0$, then we have $d(t)$ is zero in (\ref{NV/NY}). In order to keep $d(t)>0$, there must exist $l_E(t)>0$ to keep $V(t)>0$. Then, we discuss two situations. First, if $l_E(t)=1$, we will get a constant $d$ because $\dot N{(t)=0}$. Second, if $l_E(t)\in(0,1)$, we plug (\ref{ww2}) into the second line of (\ref{c:FOCd}), then we get
\begin{equation}\nonumber
    \left(1-\frac{1}{\gamma } \right) \eta+\frac{1-\gamma}{1-\xi} \frac{l_R(t)}{l_E(t)}= \kappa d (t)^2.
\end{equation}
Because the left side is less than $(1-1/\gamma )\eta+(1-\gamma)/(1-\xi)$, data cannot keep increasing. The economic insight of this discussion is that the diminishing return cannot compensate the exponential disutility forever. Along the BGP, when $l_R(t)/l_E(t)$ is constant, $d$ and $g_N$ are constant, too.

Change the R\&D production function (\ref{newentrants:p}) into form of growth rate, along the BGP there is
\begin{equation}\nonumber
    g_N \equiv \frac{\dot N(t)}{N(t)} = \varepsilon l_R^{1-\xi} L d^\xi.
\end{equation}

Plug (\ref{NV/NY}) into (\ref{ww2}), then we have expression of the labor fraction:
\begin{equation}
    \frac{l_R(t)}{1-l_R(t)}=\frac{1-\xi}{\xi} \frac{\kappa d(t)^2 - \left(1-\frac{1}{\gamma}\right)\eta}{1-\frac{1}{\gamma}}.
    \label{c:lR}
\end{equation}

Substitute $l_R$ with (\ref{c:lR}) to get (\ref{c:gNd2}):
\begin{equation}\nonumber
    g^2_{Nc}(d) = \varepsilon L \left[ \frac{\frac{1-\xi}{\xi} \frac{\kappa d^2 - \left(1-\frac{1}{\gamma}\right)\eta}{1-\frac{1}{\gamma}}}{1+\frac{1-\xi}{\xi} \frac{\kappa d^2- \left(1-\frac{1}{\gamma} \right)\eta}{1-\frac{1}{\gamma}}} \right]^{1-\xi } d^\xi.
\end{equation}

Recall the problem the incumbent firms in (\ref{firmProb}), plug FOC (\ref{firm:D}) and (\ref{firm:L}). With symmetry we have:
\begin{equation}
    r(t)-\frac{\dot V(t)}{V(t)}=\frac{ \left[1- \left(1-\frac{1}{\gamma } \right)- \left(1-\frac{1}{\gamma } \right)\eta \right] N(t)^{\frac{1}{\gamma -1}} Y(v,t)}{V(t)}.
    \label{NY/V}
\end{equation}
Along the BGP, we have:
\begin{equation}\nonumber
    \frac{\dot V(t)}{V(t)}=\frac{1}{\gamma-1}g_N + \frac{\dot Y(v,t)}{Y(v,t)}.
\end{equation}

Change the production function (\ref{production}) into the form of growth rate, along the BGP we have:
\begin{equation}\nonumber
    \frac{\dot Y(v,t)}{Y(v,t)}=-g_N,
\end{equation}
and plug $\dot Y(v,t)/Y(v,t)$ into $\dot V(t)/V(t)$, then:
\begin{equation}\nonumber
    \frac{\dot V(t)}{V(t)}=\frac{1}{\gamma-1}g_N-g_N.
\end{equation}

Change the consumption index (\ref{cons:Index}) into the form of growth rate, along the BGP we have:
\begin{align}
    \frac{\dot c(t)}{c(t)} &= \frac{\gamma}{\gamma-1}g_N - \frac{\dot c(v,t)}{c(v,t)} \nonumber \\
    &=\frac{\gamma}{\gamma-1}g_N-\frac{\dot Y(v,t)}{Y(v,t)} \nonumber \\
    &=\frac{\gamma}{\gamma-1}g_N-g_N=\frac{\dot Y(t)}{Y(t)} \equiv g_Y. \nonumber
\end{align}
Recall the Euler equations for consumption (\ref{Euler:c}), then:
\begin{equation}\nonumber
    r(t)=\rho+\frac{\dot c(t)}{c(t)}=\rho+\frac{1}{\gamma-1}g_N.
\end{equation}

Combine $r(t)$ and $\dot V(t)/V(t)$, and plug them into (\ref{NY/V}), then: 
\begin{equation}\nonumber
    \rho+g_N=\frac{ \left[1-\left(1-\frac{1}{\gamma }\right)-\left(1-\frac{1}{\gamma }\right)\eta \right]N(t)^{\frac{1}{\gamma -1}} Y(v,t)}{V(t)}.
\end{equation}
We define
\begin{equation}\nonumber
    \Gamma \equiv \left[ 1- \left(1-\frac{1}{\gamma } \right)-\left(1-\frac{1}{\gamma } \right)\eta \right].
\end{equation}

With $d(t)=d$ and plugging the above equation into (\ref{NV/NY}), we have
\begin{align}
    \frac{\dot N(t)V(t)}{N(t)^{\frac{\gamma }{\gamma -1}}Y(v,t)} =& \frac{ \kappa d^2- \left(1-\frac{1}{\gamma }\right)\eta }{\xi} \nonumber \\
    \Rightarrow  \quad \frac{g_N V(t)}{N(t)^{\frac{1}{\gamma -1}} Y(v,t)} =& \frac{ \kappa d^2-\left(1-\frac{1}{\gamma }\right)\eta }{\xi} \nonumber \\
    \Rightarrow  \quad \frac{ \Gamma g_N}{\rho+g_N} =& \frac{\kappa d^2- \left(1-\frac{1}{\gamma }\right)\eta }{\xi}. \nonumber
\end{align}
Rearrange the above to obtain (\ref{c:gNd1}):
\begin{equation}\nonumber
     g_{Nc}^1(d)=\rho\frac{\kappa d^2- \left(1-\frac{1}{\gamma } \right)\eta}{\xi\Gamma- \left[\kappa d^2-\left(1-\frac{1}{\gamma }\right)\eta \right]}.
\end{equation} 

To satisfy $g_{Nc}>0$, the additional balance privacy concerns $[\kappa d^2 - (1-1/\gamma )\eta ]$ in decentralized economy must smaller than $\xi\Gamma$. Both $g_{Nc}^1$ and $g_{Nc}^2$ are monotone increasing functions. Obviously, when $g_{Nc}^1>0$, there exist $g_{Nc}^1>\rho/(\xi\Gamma)[ \kappa d^2 - (1-1/\gamma )\eta]$, according to the proof in the optimal allocation, with increasing $d$, $g_{Nc}^1(d)$ will ultimately exceed $g_{Nc}^2(d)$. Next, we need to prove whether there exists $d_2$ that satisfies $g_{Nc}^2(d_2)>g_{Nc}^1(d_2)$.

The first-order derivative of $g_{Nc}^1(d)$ is
\begin{align}
    (g_{Nc}^1)'(d) &= \frac{\partial g_{Nc}^1(d)}{\partial \left[\kappa d^2- \left(1-\frac{1}{\gamma } \right)\eta \right]} \frac{\partial \left[ \kappa d^2 - \left(1-\frac{1}{\gamma }\right)\eta \right]}{\partial d} \nonumber \\
    &=\rho \frac{\xi\Gamma}{\left[\xi\Gamma-\left(\kappa d^2-\left(1-\frac{1}{\gamma }\right)\eta\right) \right]^2} \frac{\partial \left(\kappa d^2 - \left(1-\frac{1}{\gamma }\right)\eta \right)}{\partial d}. \nonumber
\end{align}
Then, we know that
\begin{equation}\nonumber
    (g_{Nc}^1)' \left( \left[\frac{ \left(1-\frac{1}{\gamma} \right)\eta}{\kappa}\right]^{\frac{1}{2}} \right)
\end{equation}
is finite.

Also, we have
\begin{equation}\nonumber
    g_{Nc}^2(d) > \varepsilon L \left[ \frac{\frac{1-\xi}{\xi} \left[ \kappa d^2-\left(1-\frac{1}{\gamma}\right)\eta\right]}{1+\frac{1-\xi}{\xi} \left[\kappa d^2- \left(1-\frac{1}{\gamma} \right)\eta \right]} \right]^{1-\xi } d^\xi \equiv \Psi(d).
\end{equation}
We then prove whether
\begin{equation}\label{Psig1cond}
    \Psi' \left( \left[\frac{ \left(1-\frac{1}{\gamma} \right)\eta}{\kappa} \right]^{\frac{1}{2}}\right) > (g_{Nc}^1)' \left( \left[\frac{ \left(1-\frac{1}{\gamma} \right)\eta}{\kappa} \right]^{\frac{1}{2}}\right).
\end{equation}
Because both $\Psi(d)$ and $g_{Nc}^2(d)$ are continuous and equal to zero when
\begin{equation}\nonumber
    d= \left[\frac{ \left(1-\frac{1}{\gamma} \right)\eta}{\kappa} \right]^{\frac{1}{2}}.
\end{equation}
If (\ref{Psig1cond}) exists, there would exist a $d_2$ such that $\Psi(d_2)>g_{Nc}^1(d_2)$. Then, we have $g_{Nc}^1(d_2)<g_{Nc}^2(d_2)$.

The first-order derivative of $\Psi(d)$ is
\begin{align}
    \Psi'(d) =& \varepsilon L \xi d^{\xi-1} \left[ \frac{\frac{1-\xi}{\xi} \left[ \kappa d^2- \left(1-\frac{1}{\gamma} \right)\eta \right]}{1+\frac{1-\xi}{\xi} \left[\kappa d^2 - \left(1-\frac{1}{\gamma} \right)\eta \right]} \right]^{1-\xi }... \nonumber \\
    &+\varepsilon L d^\xi (1-\xi) \left[ \frac{\frac{1-\xi}{\xi} \left[\kappa d^2 - \left(1-\frac{1}{\gamma} \right)\eta\right]}{1+\frac{1-\xi}{\xi} \left[\kappa d^2 - \left(1-\frac{1}{\gamma} \right)\eta \right]} \right]^{-\xi }
    \left[ \frac{\frac{1-\xi}{\xi}2\kappa d}{ \left( 1+\frac{1-\xi}{\xi} \left[\kappa d^2 - \left(1-\frac{1}{\gamma} \right)\eta \right] \right)^2} \right]. \nonumber
\end{align}
Then, we get
\begin{equation}\nonumber
    \Psi' \left( \left[\frac{ \left(1-\frac{1}{\gamma} \right)\eta}{\kappa} \right]^{\frac{1}{2}} \right) \to +\infty.
\end{equation}
Combining with earlier discussion, there exists a $d^*_c$ where $g_{Nc}^1(d^*_c) = g_{Nc}^2(d^*_c)$. Combining the concavity and convexity, the solution is unique.

\subsection{BGP When Data Only Enter Production Sector}
\label{App:Only}

The social planner problem is
\begin{equation} \nonumber
\max_{\left\{l_E(t), d(t)\right\}} \int_0^\infty e^{-\rho t} L \left(\ln{c(t)} - \frac{\kappa d(t)^2}{2} \right) \mathrm{d} t,
\end{equation}
s.t.,
\begin{align}
c(t) &=Y(t) / L, \nonumber \\
Y(t) &=N(t)^{\frac{1}{\gamma-1}}l_E(t)L^{1+\eta}d(t)^\eta, \nonumber \\
\dot N (t) &=\varepsilon l_R(t) L N(t), \nonumber \\
1 &= l_E(t) + l_R(t), \nonumber \\
d(t) &\le g(c(t)) \equiv \bar d (t). \nonumber
\end{align}

Define the following Hamiltonian as
\begin{equation}\nonumber
    \mathcal{H} (l_E (t),d(t),N(t),\lambda (t))= \ln{\left[ N(t)^{\frac{1}{\gamma-1}} l_E(t) L^{\eta} d^\eta \right]} - \frac{ \kappa d(t)^2}{2}+\varepsilon \lambda (t) l_R L N(t).
\end{equation}

We then have following FOCs:
\begin{equation}\label{dataonlyFOClE}
    \frac{\partial \mathcal{H}}{\partial l_E(t)} = \frac{1}{l_E (t)}-\lambda (t)\varepsilon L N(t)=0,
\end{equation}
\begin{equation}\label{dataonlyFOCd}
    \frac{\partial \mathcal{H}}{\partial d(t)} = \frac{\eta }{d(t)}- \kappa d(t)=0,
\end{equation}
\begin{equation}\label{dataonlyFOCN}
    \frac{\partial \mathcal{H}}{\partial N(t)} = -\dot \lambda (t)+\rho \lambda (t)=\frac{1}{\gamma -1} \frac{1}{N(t)} + \varepsilon \lambda (t) l_R(t) L,
\end{equation}

From (\ref{dataonlyFOClE}), we have
\begin{equation}
    \frac{l_R(t)}{l_E (t)} = \lambda (t)\dot N (t).
    \label{lRlEO}
\end{equation}
Along the BGP, then we have:
\begin{equation}\nonumber
    g'_N=-\frac{\dot \lambda (t)}{\lambda(t)}.
\end{equation}

From (\ref{dataonlyFOCd}), we have
\begin{equation}\nonumber
    d'=\left(\frac{\eta}{\kappa}\right)^\frac{1}{2}.
\end{equation}

From (\ref{dataonlyFOCN}), we have
\begin{align}
    -\dot \lambda (t)+\rho \lambda (t) &= \frac{1}{\gamma -1} \frac{1}{N(t)} + \varepsilon \lambda (t) l_R(t) L \nonumber \\
    \Rightarrow \quad -\dot \lambda(t)N(t)+\rho \lambda (t)N(t) &= \frac{1}{\gamma -1}+\varepsilon \lambda (t) l_R(t) L N(t) \nonumber \\
    \Rightarrow \quad \lambda (t)N(t) &= \frac{1}{\gamma -1}\frac{1}{\rho }. \label{lambdaNO}
\end{align}

Plug (\ref{lambdaNO}) into (\ref{lRlEO}), then we have
\begin{align}
    \frac{l_R(t)}{1-l_R(t)} &= \frac{1}{\gamma -1}\frac{1}{\rho } \varepsilon l_R(t) L \nonumber \\
    \Rightarrow \quad l'_R &= 1-\frac{(\gamma-1)\rho}{\varepsilon L}. \nonumber
\end{align}

Combining with $l'_R$, then we get $g'_N$ as
\begin{equation}\nonumber
    g'_N=\varepsilon L-(\gamma-1)\rho.
\end{equation}

\subsection{BGP When Data Only Enter Innovation Sector}
\label{App:OnlyInno}

The social planner problem is
\begin{equation}\nonumber
\max_{\left\{l_E(t), d(t)\right\}} \int_0^\infty e^{-\rho t} L\left(\ln{c(t)}-\frac{\kappa d(t)^2}{2} \right) \mathrm{d} t,
\end{equation}
s.t.,
\begin{align}
c(t) &=Y(t) / L, \nonumber \\
Y(t) &=N(t)^{\frac{1}{\gamma-1}}l_E(t)L, \nonumber \\
\dot N(t) &=\varepsilon l_R(t)^{1-\xi} L d(t)^\xi N(t), \nonumber \\
1 &= l_E(t) + l_R(t), \nonumber \\
d(t) &\le g(c(t)) \equiv \bar d(t). \nonumber
\end{align}

Define the following Hamiltonian as
\begin{equation}\nonumber
    \mathcal{H} (l_E(t),d(t),N(t),\lambda (t))=\ln{ \left[ N(t)^{\frac{1}{\gamma-1}} l_E(t) \right]} - \frac{ \kappa d(t)^2}{2} + \varepsilon \lambda(t) l_R(t)^{1-\xi} L d(t)^\xi N(t).
\end{equation}

We then have following FOCs:
\begin{equation}\label{OnlyInnoFOClE}
    \frac{\partial \mathcal{H}}{\partial l_E(t)} = \frac{1}{l_E(t)} - \varepsilon(1-\xi)\lambda(t)l_R(t)^{-\xi}d(t)^{\xi} L N(t)=0,
\end{equation}
\begin{equation}\label{OnlyInnoFOCd}
    \frac{\partial \mathcal{H}}{\partial d(t)}=\varepsilon \xi \lambda(t)l_R(t)^{1-\xi} L d(t)^{\xi-1}N(t)- \kappa d(t) =0,
\end{equation}
\begin{equation}\label{OnlyInnoFOCN}
    \frac{\partial \mathcal{H}}{\partial N(t)}=-\dot \lambda(t)+\rho \lambda (t)=\frac{1}{\gamma -1}\frac{1}{N(t)}+\varepsilon\lambda(t) l_R(t)^{1-\xi} L d(t)^\xi,
\end{equation}

From (\ref{OnlyInnoFOClE}), we have
\begin{equation}\nonumber
    \frac{l_R(t)}{l_E(t)} = (1-\xi)\lambda (t)\dot N(t).
\end{equation}
Along the BGP, we have
\begin{equation}\nonumber
    g''_N=-\frac{\dot \lambda (t)}{\lambda(t)}.
\end{equation}

From (\ref{OnlyInnoFOCd}), we have
\begin{align}
    \kappa d(t)^2 &= \xi\lambda(t) \dot N(t) \nonumber \\
    &= \frac{\xi}{1-\xi} \frac{l_R(t)}{l_E (t)}. \nonumber
\end{align}
Then, we get
\begin{equation}
    l_R''(t)=\frac{\frac{1-\xi}{\xi}\kappa d(t)^2}{1+\frac{1-\xi}{\xi}\kappa d(t)^2}.
    \label{lRonlyInno}
\end{equation}

From (\ref{OnlyInnoFOCN}), we have
\begin{align}
    -\dot \lambda (t)+\rho \lambda (t) &= \frac{1}{\gamma -1}\frac{1}{N(t)}+\varepsilon\lambda(t) l_R(t)^{1-\xi} L d(t)^\xi \nonumber \\
    \Rightarrow \quad -\dot \lambda(t)N(t)+\rho \lambda (t)N(t) &= \frac{1}{\gamma -1}+\lambda(t)\dot N(t) \nonumber \\
    \Rightarrow \quad \lambda (t)N(t) &= \frac{1}{\gamma -1}\frac{1}{\rho }. \label{lambdaNonlyInno}
\end{align}

Plug (\ref{lambdaNonlyInno}) into (\ref{lRlEO}), we have
\begin{equation}\nonumber
    (g^{1}_N)''(d) = \frac{(\gamma-1)\rho}{\xi} \kappa d^2.
\end{equation}
Plug (\ref{lRonlyInno}) into (\ref{newentrants:p}), we have
\begin{equation}\nonumber
    (g^2_N)''(d)=\varepsilon L \left[\frac{\frac{1-\xi}{\xi}\kappa d^2}{1+\frac{1-\xi}{\xi}\kappa d^2}\right]^{1-\xi}d^\xi.
\end{equation}
Remember that $(g^2_N)''(d)$ increases with $d$ and is smaller than $\varepsilon L d^\xi$. Define $F(d)=(g^1_N)''(d)-(g^2_N)''(d)$, we have
\begin{align}
    F(d) &= \frac{(\gamma-1)\rho}{\xi} \kappa d^2
    -\varepsilon L \left[ \frac{\frac{1-\xi}{\xi}\kappa d^2}{1+\frac{1-\xi}{\xi}\kappa d^2}\right]^{1-\xi}d^\xi \nonumber \\
    &> \frac{(\gamma-1)\rho}{\xi}\kappa d^2
    -\varepsilon L d^\xi . \nonumber
\end{align}
With a slight abuse of notation, we note the equation by $F(d)$ like before. Obviously, with $0<\xi<1$, there exists an $d'_0$, where $F(d)>0$ for all $d>d'_0$.

Next, the first-order derivative of $F(d)$ is
\begin{align}
    F'(d) =& \frac{(\gamma-1)\rho}{\xi}2\kappa d 
    -\varepsilon L \xi d^{1-\xi} \left[{\frac{\frac{1-\xi}{\xi}\kappa}{1+\frac{1-\xi}{\xi}\kappa d^2}} \right]^{1-\xi }-... \nonumber \\
    & \varepsilon L (1-\xi) d^{1-\xi} \left[\frac{\frac{1-\xi}{\xi}\kappa}{1+\frac{1-\xi}{\xi}\kappa d^2}\right]^{-\xi } \left[ \frac{\frac{1-\xi}{\xi}2\kappa}{\left(1+\frac{1-\xi}{\xi}\kappa d^2 \right)^2} \right] \nonumber \\
    =& \frac{(\gamma-1)\rho}{\xi}2\kappa d    - \varepsilon L {\xi}d^{1-\xi} \left( \frac{1-\xi}{\xi} \kappa \right)^{1-\xi} \left( 1+\frac{1-\xi}{\xi} \kappa d^2 \right)^{-1+\xi }- ...\nonumber \\
    & \varepsilon L (1-\xi) d^{1-\xi} \left(\frac{1-\xi}{\xi}\kappa \right)^{-\xi} \left( \frac{1-\xi}{\xi}2\kappa \right) \left(1+\frac{1-\xi}{\xi} \kappa d^2 \right)^{\xi-2}. \nonumber
\end{align}
We get $F{'}(0)=0$ and second-order derivative of $F(d)$ is:
\begin{align}
    F''(d) =& \frac{(\gamma-1)\rho}{\xi}2\kappa
    -\varepsilon L \xi (1-\xi) d^{-\xi} \left( \frac{1-\xi}{\xi}\kappa \right)^{1-\xi} \left( 1+\frac{1-\xi}{\xi} \kappa d^2 \right)^{-1+\xi }- ...\nonumber \\
    & \varepsilon L \xi (\xi-1) d^{1-\xi} \left( \frac{1-\xi}{\xi}\kappa \right)^{1-\xi} \left( 1+\frac{1-\xi}{\xi}\kappa d^2 \right)^{-2+\xi } \left(\frac{1-\xi}{\xi}2\kappa d \right)-... \nonumber \\
    & \varepsilon L (1-\xi)^2 d^{-\xi} \left( \frac{1-\xi}{\xi}\kappa \right)^{-\xi} \left( \frac{1-\xi}{\xi} 2 \kappa \right) \left( 1+\frac{1-\xi}{\xi} \kappa d^2 \right)^{\xi-2}- ...\nonumber \\
    & \varepsilon L (1-\xi) (\xi-2) d^{1-\xi} \left(\frac{1-\xi}{\xi}\kappa \right)^{-\xi} \left(\frac{1-\xi}{\xi}2\kappa \right) \left(1+\frac{1-\xi}{\xi} \kappa d^2 \right)^{\xi-3} \left(\frac{1-\xi}{\xi} 2 \kappa d \right). \nonumber
\end{align}

Then we have $F''(0) \to -\infty$. Considering that $F'(0)=0$ and $F'(d)$ is continuous, there exists a $d_3$, where $F'(d)<0$ for all $d<d_3$. Also, considering $F(0)=0$ and $F(d)$ is continuous, the solution of $(g^1_N)''(d)$ and $(g^2_N)''(d)$ exists. Combining the concavity and convexity, the solution is unique.

\subsection{Policy Implications on BGP}

\subsubsection{A Production Subsidy and An Innovation Subsidy}
\label{APP:Policy}

With a production subsidy, the incumbent firm's problem is
\begin{equation}
    r(t)V(t) = \max_{\left\{ L_E(v,t), D(t) \right\}} (1+s_p) \left( \frac{Y(t)}{Y(v,t)}\right)^{\frac{1}{\gamma }} Y(v,t) - w(t) L_E(v,t) - q^{prod}_d (v,t) D(t) + \dot V(t),
    \label{sfirmProb}
\end{equation}
subject to the following production function:
\begin{equation}\nonumber
    Y(v,t)=L_E (v,t) D(t)^\eta.
\end{equation}

The first order conditions of the incumbent firms are:
\begin{equation}
    (1+s_p) \left(1-\frac{1}{\gamma } \right) \left( \frac{Y(t)}{Y(v,t)} \right)^{\frac{1}{\gamma }} \frac{Y(v,t)}{L_E (v,t)} = w(t)
    \label{sfirm:L}
\end{equation}
and
\begin{equation}
    (1+s_p) \left(1-\frac{1}{\gamma } \right)\eta \left( \frac{Y(t)}{Y(v,t)} \right)^{\frac{1}{\gamma } } \frac{Y(v,t)}{D(t)} = q_d^{prod} (v,t).
    \label{sfirm:D}
\end{equation}

With an innovation subsidy, the potential entrants' problem is
\begin{equation}\nonumber
    \max_{\left\{L_R(t), D(t)\right\}} (1+s_n)\dot N(t)V(t)-w(t)L_R(t)-q^{inno}_d(t) D(t),
\end{equation}
subject to the innovation sector production function:
\begin{equation}\nonumber
    \dot N (t) = \varepsilon N(t) L_R(t)^{1-\xi} D(t)^\xi.
\end{equation}

The first order conditions of innovators are
\begin{equation}
    (1+s_n) (1-\xi) \varepsilon N(t) L_R(t)^{-\xi} D(t)^\xi V(t)=w(t)
    \label{snewentrants:L}
\end{equation}
and
\begin{equation}\nonumber
    (1+s_n) \xi \varepsilon N(t) L_R(t)^{1-\xi} D(t)^{\xi-1} V(t) = q_d^{inno}(t).
\end{equation}

Then, the zero condition of the data intermediary is
\begin{equation}\nonumber
    (1+s_p) \left(1-\frac{1}{\gamma } \right) \eta N(t)^{\frac{\gamma }{\gamma -1}} Y(v,t) + (1+s_n) \xi \dot N(t)V(t) = p_d(t) d(t) L.
\end{equation}
Plug (\ref{d=pdc}) into the above equation, we have
\begin{align}
    &(1+s_p) \left(1-\frac{1}{\gamma } \right)\eta N(t)^{\frac{\gamma }{\gamma -1}} Y(v,t) + (1+s_n) \xi \dot N(t)V(t) = \kappa d(t)^2 c(t) L \nonumber \\
    \Rightarrow \quad & (1+s_p) \left(1-\frac{1}{\gamma } \right)\eta +\frac{\xi(1+s_n)\dot N(t)V(t)}{N(t)^{\frac{\gamma}{\gamma-1}}Y(v,t) } = \kappa d(t)^2 \nonumber \\
    \Rightarrow \quad & \frac{\dot N(t)V(t)}{N(t)^{\frac{\gamma }{\gamma -1}} Y(v,t)} = \frac{\kappa d(t)^2 - (1+s_p) \left(1-\frac{1}{\gamma } \right)\eta }{\xi(1+s_n)}. \label{sNV/NY}
\end{align}

Divide (\ref{sfirm:L}) by (\ref{snewentrants:L}) and plug (\ref{sww}), then we have
\begin{align}
   \frac{(1+s_n){(1-\xi)} \dot N (t)V(t)}{(1+s_p)\left(1-\frac{1}{\gamma } \right) N(t)^{\frac{\gamma}{\gamma-1}} Y(v,t)} &= \frac{l_R(t)}{1-l_R(t)} \nonumber \\
   \Rightarrow \quad \frac{l_R(t)}{1-l_R (t)} &= \frac{1-\xi}{\xi} \frac{\kappa d(t)^2 -(1+s_p)\left(1-\frac{1}{\gamma } \right)\eta}{(1+s_{p})(1-\frac{1}{\gamma})}.    \label{sww} 
\end{align}

Then we get
\begin{equation}\nonumber 
    \tilde g^2_{Nc}(d) = \varepsilon L \left[ \frac{\frac{1-\xi}{\xi} \frac{ \kappa d(t)^2 - (1+s_p) \left(1-\frac{1}{\gamma } \right)\eta}{(1+s_{p}) \left(1-\frac{1}{\gamma} \right)}}{1+\frac{1-\xi}{\xi} \frac{ \kappa d(t)^2 - (1+s_p) \left(1-\frac{1}{\gamma } \right)\eta}{(1+s_p) \left(1-\frac{1}{\gamma} \right)}} \right]^{1-\xi } d(t)^\xi.
\end{equation}

Recall the problem the incumbent firms in (\ref{sfirmProb}), plug FOC (\ref{sfirm:L}) and (\ref{sfirm:D}). With symmetry we have
\begin{equation}
    r(t)-\frac{\dot V (t)}{V(t)} = \frac{ \left[1- \left(1-\frac{1}{\gamma } \right)- \left(1-\frac{1}{\gamma } \right)\eta \right] (1+s_p) N(t)^{\frac{1}{\gamma -1}}Y(v,t)}{V(t)}.
    \label{sNY/V}
\end{equation}
Along BGP, we have
\begin{equation}\label{policy_dotVV}
    \frac{\dot V(t)}{V(t)} = \frac{1}{\gamma-1}g_N-g_N.
\end{equation}

Recall the Euler equations for consumption (\ref{Euler:c}):
\begin{equation}\nonumber
    r(t) = \rho+\frac{\dot c(t)}{c(t)}= \rho+\frac{1}{\gamma-1}g_N.
\end{equation}

Combine $r(t)$ and (\ref{policy_dotVV}), and plug them into (\ref{sNY/V}), then: 
\begin{equation}\nonumber
    \rho+g_N = \frac{\left[1- \left(1-\frac{1}{\gamma } \right)- \left(1-\frac{1}{\gamma } \right)\eta \right] (1+s_p) N(t)^{\frac{1}{\gamma -1}} Y(v,t)}{V(t)}.
\end{equation}
We define
\begin{equation}\nonumber
    \Gamma \equiv \left[1- \left(1-\frac{1}{\gamma } \right)-\left(1-\frac{1}{\gamma } \right)\eta \right] >0.
\end{equation}

Plugging the above equation into (\ref{sNV/NY}), we have
\begin{align}
    \frac{\dot N (t)V(t)}{N(t)^{\frac{\gamma }{\gamma -1}}Y(v,t)} =& \frac{\kappa d(t)^2 - (1+s_p) \left(1-\frac{1}{\gamma } \right)\eta }{\xi(1+s_n)} \nonumber \\
    \Rightarrow \quad  \frac{\Gamma (1+s_p)g_N}{\rho+g_N} =& \frac{\kappa d(t)^2-(1+s_p) \left(1-\frac{1}{\gamma } \right)\eta }{\xi(1+s_n)}. \nonumber
\end{align}
Rearrange the above equation, we have
\begin{equation}\nonumber
    \tilde g_{Nc}^1(d) = \rho \frac{ \kappa d(t)^2 - (1+s_p) \left(1-\frac{1}{\gamma } \right)\eta}{ \Gamma \xi(1+s_p)(1+s_n)- \left[ \kappa d(t)^2 - (1+s_p) \left(1-\frac{1}{\gamma } \right)\eta \right]},
\end{equation} 
or 
\begin{equation}\nonumber
    \frac{\Gamma(1+s_p) \tilde g_{Nc}^1(d)}{\rho+\tilde g_{Nc}^1(d)} = \frac{\kappa d(t)^2 -(1+s_p) \left(1-\frac{1}{\gamma } \right)\eta }{\xi(1+s_n)}.
\end{equation}
Recall in optimal allocation, the optimal balanced growth rate of varieties $g_{Ns}$ and optimal data contribution per capita $d_s$ are determined by
\begin{equation}\nonumber
    g_{Ns}^1(d) = \frac{(\gamma-1)\rho}{\xi}(\kappa d^2 -\eta),
\end{equation}
and
\begin{equation}\nonumber
    g_{Ns}^2(d) = \varepsilon L \left[ \frac{\frac{1-\xi}{\xi}(\kappa d^2 -\eta)}{1+\frac{1-\xi}{\xi}(\kappa d^2 -\eta)} \right]^{1-\xi } d^\xi.
\end{equation} 

We subsidize to make the growth rate and the quantity of data of the decentralized economy the same as the optimal allocation. Then we get a production subsidy to correct monopolistic distortion:
\begin{equation}\nonumber
    1+s_p=\frac{\gamma}{\gamma-1}
\end{equation}
and
\begin{equation}\nonumber
    1+s_n=\frac{\rho+g_{Ns}}{\rho\Gamma(\gamma-1)(1+s_p)}
\end{equation}

\subsubsection{Subsidies for Labor and Data}
\label{APP:Policy2}

With a data subsidy, the incumbent firm's problem is:
\begin{equation}\nonumber
    r(t)V(t) = \max_{\left\{L_E(v,t), D(t)\right\}} \left( \frac{Y(t)}{Y(v,t)} \right)^{\frac{1}{\gamma }} Y(v,t)-w(t)L_E(v,t)-(1-s_{d1}) q^{prod}_d(v,t)D(t) + \dot V(t),
\end{equation}
subject to the following production function:
\begin{equation}\nonumber
    Y(v,t)=L_E (v,t)D(t)^\eta.
\end{equation}

The first order conditions of the incumbent firms are:
\begin{equation}
    \left(1-\frac{1}{\gamma } \right) \left( \frac{Y(t)}{Y(v,t)} \right)^{\frac{1}{\gamma }} \frac{Y(v,t)}{L_E(v,t)} = w(t)
    \label{s2firm:L}
\end{equation}
and
\begin{equation}\nonumber
    \left(1-\frac{1}{\gamma } \right)\eta \left( \frac{Y(t)}{Y(v,t)} \right)^{\frac{1}{\gamma }} \frac{Y(v,t)}{D(t)}= (1-s_{d1}) q_d^{prod}(v,t).
\end{equation}

With an innovation subsidy and a labor subsidy, the potential entrants' problem is 
\begin{equation}\nonumber
    \max_{\left\{L_R(t), D(t)\right\}} \dot N(t)V(t)-(1-s_l)w(t)L_R(t)-(1-s_{d2})q^{inno}_d(t)D(t),
\end{equation}
subject to the innovation sector production function:
\begin{equation}\nonumber
    \dot N(t)=\varepsilon N(t) L_R(t)^{1-\xi} D(t)^\xi.
\end{equation}

The first order conditions of innovators are
\begin{equation}
    (1-\xi) \varepsilon N(t)L_R(t)^{-\xi}D(t)^\xi V(t) = (1-s_l)w(t)
    \label{s2newentrants:L}
\end{equation}
and
\begin{equation}\nonumber
    \xi \varepsilon N(t) L_R(t)^{1-\xi} D(t)^{\xi-1} V(t) = (1-s_{d2}) q_d^{inno}(t).
\end{equation}

Then, we have:
\begin{align}
    \frac{1}{1-s_{d1}} \left(1-\frac{1}{\gamma } \right) \eta N(t)^{\frac{\gamma }{\gamma -1}} Y(v,t) + \frac{1}{1-s_{d2}} \xi \dot N(t)V(t) &= \kappa d(t)^2 c(t) L \nonumber \\
    \Rightarrow \quad \frac{1}{1-s_{d1}} \left(1-\frac{1}{\gamma } \right)\eta +\frac{1}{1-s_{d2}} \frac{\xi\dot N(t)V(t)}{N(t)^{\frac{\gamma}{\gamma-1}}Y(v,t) } &= \kappa d(t)^2 \nonumber \\
    \Rightarrow \quad \frac{\kappa d(t)^2 - \frac{1}{1-s_{d1}} \left(1-\frac{1}{\gamma } \right)\eta }{\xi}(1-s_{d2}) &= \frac{\dot N(t)V(t)}{N(t)^{\frac{\gamma }{\gamma -1}}Y(v,t)}.
        \label{s2NV/NY}
\end{align}

Divide (\ref{s2firm:L}) by (\ref{s2newentrants:L}) and plug (\ref{sww}), then we have
\begin{align}
    \frac{(1-s_l)l_R(t)}{1-l_R(t)} &= \frac{(1-\xi) \dot N(t) V(t)}{\left(1-\frac{1}{\gamma } \right) N(t)^{\frac{\gamma}{\gamma-1}}Y(v,t)} \nonumber \\
   \Rightarrow \quad \frac{l_R (t)}{1-l_R(t)} &= \frac{1-\xi}{\xi} \frac{\kappa d(t)^2 - \frac{1}{1-s_{d1}} \left(1-\frac{1}{\gamma } \right)\eta}{1-\frac{1}{\gamma}}\frac{1-s_{d2}}{1-s_{l}}. \nonumber
\end{align}
From the problem the incumbent firms, we have
\begin{equation}\nonumber
    \rho+g_N=\frac{\Gamma N(t)^{\frac{1}{\gamma -1}}Y(v,t)}{V(t)}.
\end{equation}
Plugging the above equation into (\ref{s2NV/NY}), we have
\begin{align}
  \frac{\dot N(t)V(t)}{N(t)^{\frac{\gamma }{\gamma -1}}Y(v,t)} &=\frac{\kappa d(t)^2-\frac{1}{1-s_{d1}}\left(1-\frac{1}{\gamma } \right)\eta }{\xi}(1-s_{d2}) \nonumber \\
  \Rightarrow  \quad \frac{\Gamma \hat g_{Nc}}{\rho+\hat g_{Nc}} &= \frac{\kappa d(t)^2-\frac{1}{1-s_{d1}} \left(1-\frac{1}{\gamma } \right)\eta }{\xi}(1-s_{d2}). \nonumber 
\end{align}
Then we get a production subsidy to correct monopolistic distortion:
\begin{equation}\nonumber
    1-s_{d1}=1-\frac{1}{\gamma},
\end{equation}
\begin{equation}\nonumber
    1-s_{d2}=\frac{\rho(\gamma-1)\Gamma}{\rho+g_{Ns}},
\end{equation}
and 
\begin{equation}\nonumber
    1-s_l=\frac{1-s_{d2}}{1-\frac{1}{\gamma}}=\frac{\rho\gamma\Gamma}{\rho+g_{Ns}}.
\end{equation}
Because we assume the share of profit owned by the incumbent firms is $\Gamma>0$, then we have 
\begin{equation}\nonumber
    \gamma\Gamma=\gamma[1-(1-\frac{1}{\gamma })-(1-\frac{1}{\gamma })\eta ]=1-(\gamma-1)\eta<1.
\end{equation}
When $\gamma>1$, we have $\rho\gamma\Gamma<\rho$ and $\rho(\gamma-1)\Gamma<\rho$.Then we have $1-s_{d2}\in(0,1)$ and $1-s_l\in(0,1)$.

\subsection{Constant Return of Production}
\label{APP:constant}
\subsubsection{Decentralized Economy}

The incumbent firm's problem is
\begin{equation}\nonumber
    r(t)V(t) = \max_{\left\{L_E(v,t), D(t)\right\}} \left( \frac{Y(t)}{Y(v,t)} \right)^{\frac{1}{\gamma }}Y(v,t)-w(t)L_E(v,t)-q^{prod}_d(v,t)D(t) + \dot V(t),
\end{equation}
subject to the following production function:
\begin{equation}\nonumber
    Y(v,t)=L_E(v,t)^{1-\theta} D(t)^\theta.
\end{equation}

The first order conditions of the incumbent firms are
\begin{equation}
    \left(1-\frac{1}{\gamma } \right) (1-\theta) \left( \frac{Y(t)}{Y(v,t)} \right)^{\frac{1}{\gamma }} \frac{Y(v,t)}{L_E(v,t)}=w(t)
    \label{s3firm:L}
\end{equation}
and
\begin{equation}\nonumber
    \left(1-\frac{1}{\gamma } \right)\theta \left( \frac{Y(t)}{Y(v,t)} \right)^{\frac{1}{\gamma }} \frac{Y(v,t)}{D(t)} = q_d^{prod}(v,t).
\end{equation}
The zero condition of the data intermediary is
\begin{equation}\nonumber
    \left(1-\frac{1}{\gamma }\right)\theta N(t)^{\frac{\gamma }{\gamma -1}} Y(v,t)+\xi\dot N(t)V(t)=p_d(t) d(t)L.
\end{equation}
Plugging (\ref{d=pdc}), we have
\begin{equation}
    \frac{\dot N (t)V(t)}{N(t)^{\frac{\gamma }{\gamma -1}}Y(v,t)} = \frac{\kappa d(t)^2-\left(1-\frac{1}{\gamma }\right)\theta }{\xi}.
        \label{NV/NY2}
\end{equation}
Divide (\ref{newentrants:L}) by (\ref{s3firm:L}), then
\begin{equation}\nonumber
    \frac{l_R(t)}{1-l_R(t)} = \frac{1-\xi}{\xi} \frac{\kappa d(t)^2-\left(1-\frac{1}{\gamma } \right)\theta}{\left(1-\frac{1}{\gamma}\right)(1-\theta)}.
\end{equation}
Then, we have
\begin{equation}\nonumber
    \tilde g_{Nc}^2(d) = \varepsilon L \left[ \frac{\frac{1-\xi}{\xi} \frac{\kappa d(t)^2-\left(1-\frac{1}{\gamma } \right)\theta}{\left(1-\frac{1}{\gamma}\right)(1-\theta)}}{1+\frac{1-\xi}{\xi} \frac{ \kappa d(t)^2 -\left(1-\frac{1}{\gamma } \right)\theta}{\left(1-\frac{1}{\gamma}\right)(1-\theta)}}\right]^{1-\xi }d^\xi.
\end{equation}
Along BGP, we have
\begin{equation}\nonumber
    \frac{\dot V(t)}{V(t)}=\frac{1}{\gamma-1}g_N-(1-\theta)g_N.
\end{equation}
From (\ref{NV/NY2}), we have:
\begin{equation}\nonumber
     g_{Nc}^1(d)=\rho\frac{\kappa d^2-\left(1-\frac{1}{\gamma }\right)\theta}{\frac{\xi}{\gamma}-\left[ \kappa d^2-\left(1-\frac{1}{\gamma } \right)\theta \right]}.
\end{equation} 

\subsubsection{Optimal Allocation}

\begin{equation}\nonumber
\max _{\left\{l_E(t), d(t)\right\}}  \int_0^\infty e^{-\rho t} L\left(\ln{c(t)}-\frac{\kappa d(t)^2}{2} \right) \mathrm{d} t, 
\end{equation}
s.t.,
\begin{align}
c(t) &=Y(t) / L, \nonumber \\
Y(t) &=N(t)^{\frac{\gamma}{\gamma-1}}\left(\frac{l_E(t)}{N(t)}\right)^{1-\theta}L d(t)^{\theta}, \nonumber \\
\dot N(t) &=\varepsilon{l_R(t)^{1-\xi}}L{d(t)^{\xi}}N(t), \nonumber \\
1 &= l_E(t) + l_R(t), \nonumber \\
d(t) &\le g(c(t)). \nonumber
\end{align}

Define following Hamiltonian:
\begin{equation}\nonumber
    \mathcal{H}(l_E(t),d(t),N(t),\lambda (t))=\ln{\left[ N(t)^{\frac{1}{\gamma-1}+\theta}l_E(t)^{1-\theta} d(t)^{\theta} \right]} - \frac{\kappa d(t)^2}{2}+\varepsilon \lambda (t){l_R(t)^{1-\xi}}L d(t)^{\xi} N(t).
\end{equation}

We then have following FOCs:
\begin{equation}\label{constantFOClE}
    \frac{\partial \mathcal{H}}{\partial l_E(t)} = \frac{1-\theta}{l_E(t)}-\lambda (t)\varepsilon (1-\xi) l_R(t)^{-\xi}{L}d(t)^{\xi }N(t)=0,
\end{equation}
\begin{equation}\nonumber
    \frac{\partial \mathcal{H}}{\partial d(t)}=\frac{\theta }{d(t)}- \kappa d(t)+\lambda (t)\varepsilon \xi l_R(t)^{1-\xi } L d(t)^{\xi-1 } N(t)=0,
\end{equation}
\begin{equation}\label{constantFOCN}
    \frac{\partial \mathcal{H}}{\partial N(t)}= -\dot \lambda (t)+\rho \lambda (t)= \left(\frac{1}{\gamma -1}+\theta \right)\frac{1}{N(t)}+\varepsilon \lambda (t){l_R(t)^{1-\xi}}L d(t)^{\xi},
\end{equation}

Consider (\ref{constantFOClE}), we have
\begin{equation}\nonumber
    (1-\theta)\frac{l_R(t)}{l_E(t)}=(1-\xi)\lambda(t)\dot N(t).
\end{equation}

Consider (\ref{constantFOClE}), we have
\begin{equation}\nonumber
    \theta+\xi\lambda (t)\dot N (t)= \kappa d(t)^2.
\end{equation}

Consider (\ref{constantFOCN}), we have
\begin{equation}\nonumber
     \lambda(t)N(t)=\frac{\frac{1}{\gamma -1}+\theta}{\rho}.
\end{equation}

\begin{figure}[ht]
\begin{center}  
\subfigure[Difference in growth rate of varieties]{
\includegraphics[scale=0.22]{{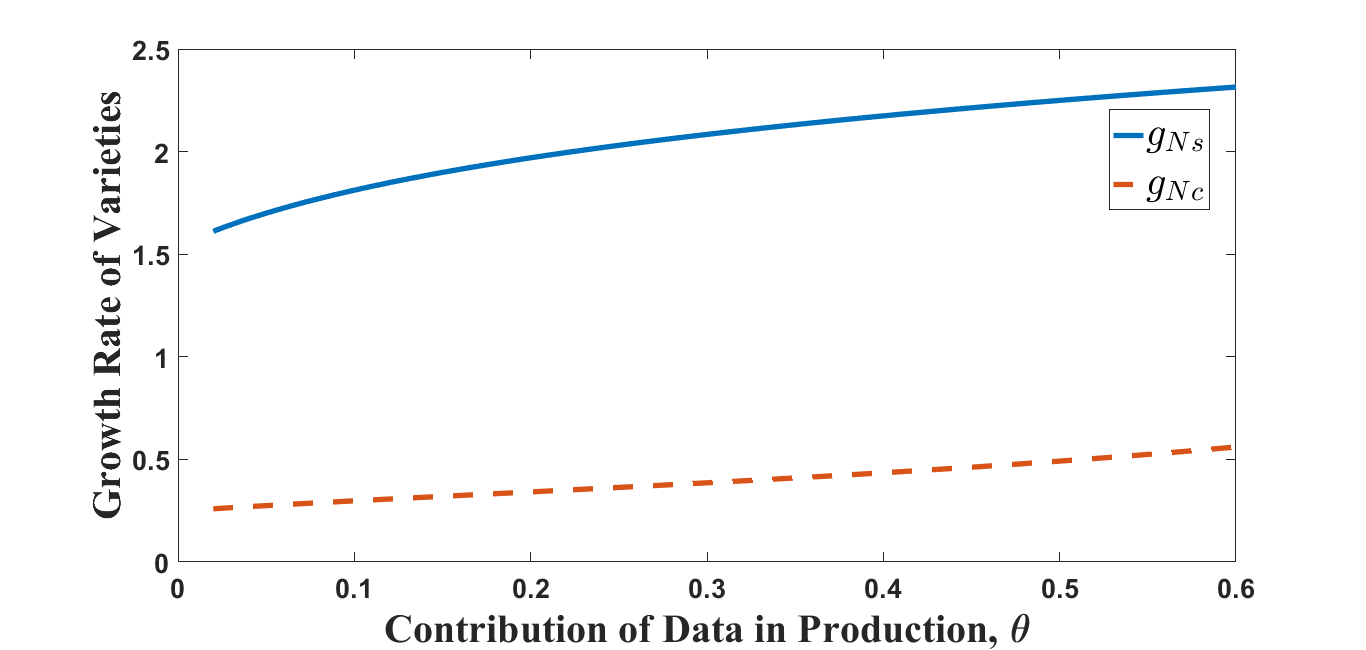}}}
\subfigure[Difference in the quantity of data]{
\includegraphics[scale=0.22]{{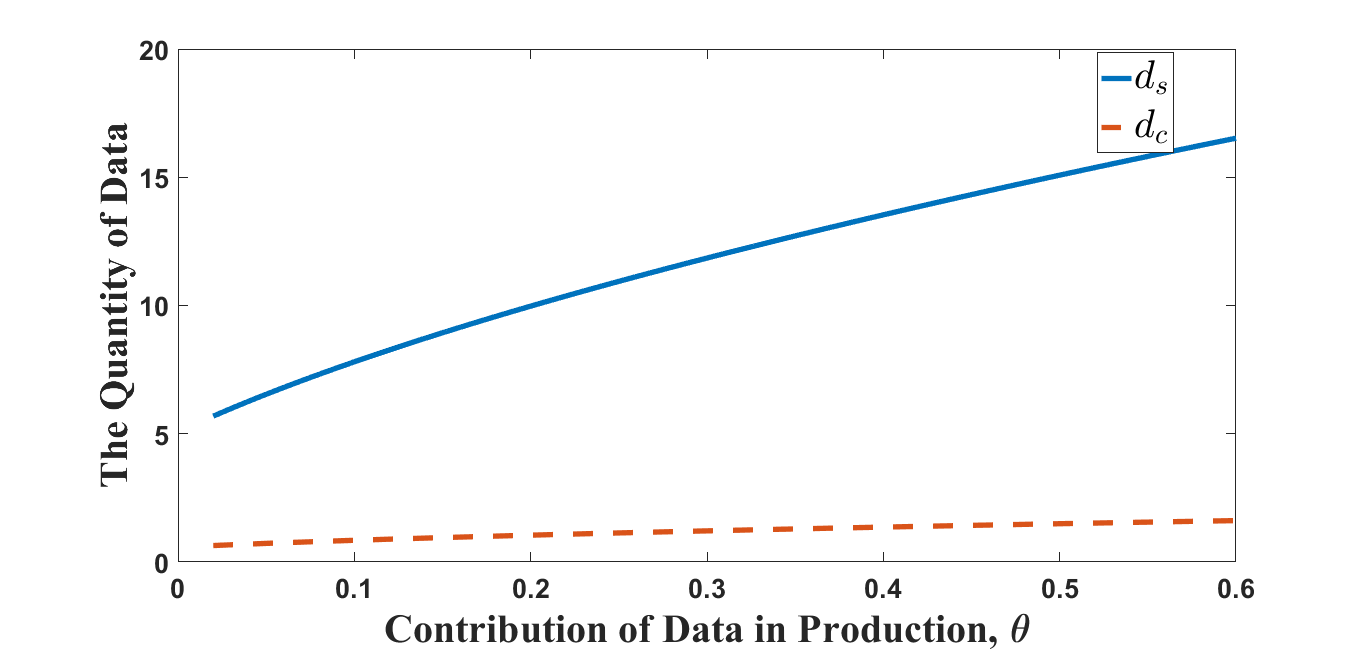}}}
\subfigure[Difference in labor allocation]{
\includegraphics[scale=0.22]{{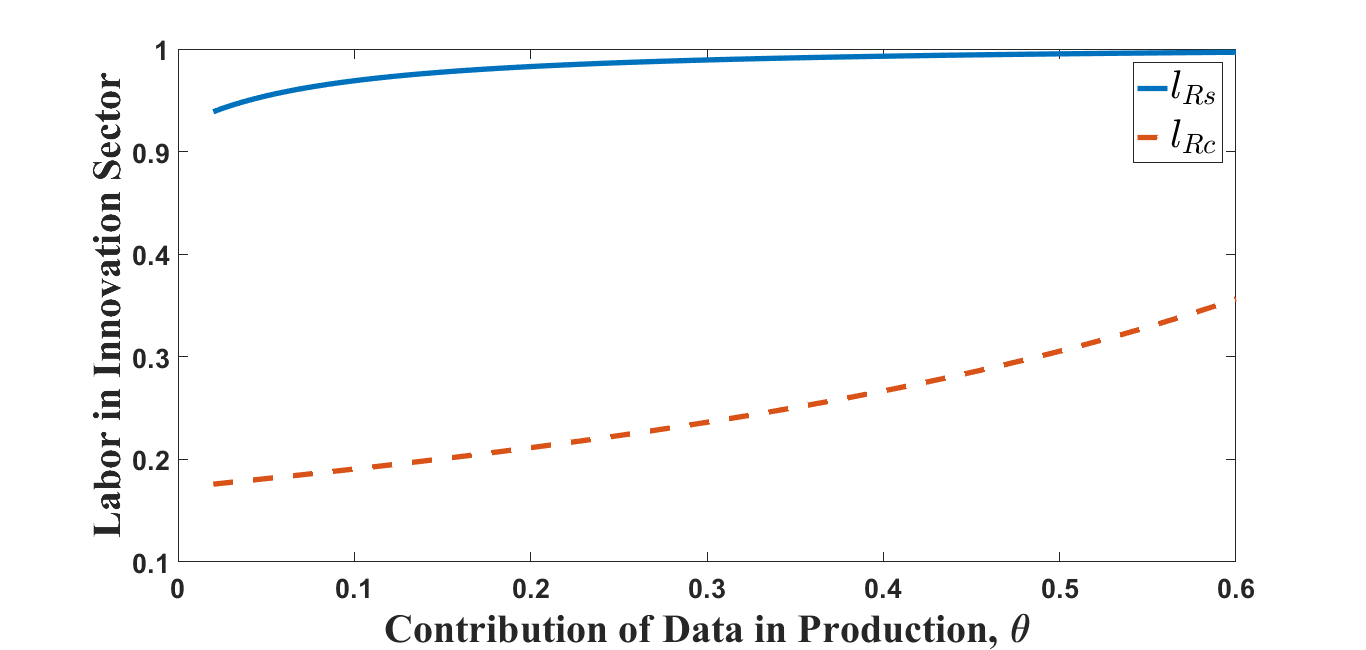}}\label{fig:constant_theta:b}}
\caption{\textbf{Different allocation with constant return of production and  different $\theta$ }}
\label{fig:constant_theta}
\end{center}
\qquad Notes: The figure depicts three key variables with different $\theta$ when the production is constant return, $Y(v,t)=L_E(v,t)^{1-\theta}D(t)^\theta$. The fraction of labor in the innovation sector $l_{Rc}$ in Figure \ref{fig:constant_theta:b} is different with that in Figure \ref{fig:diff_dc_eta:c}, because the parameter $\theta$ causes the substitution of labor.
\end{figure}

\begin{figure}[p]
\begin{center}  
\subfigure[Difference in growth rate of varieties]{
\includegraphics[scale=0.18]{{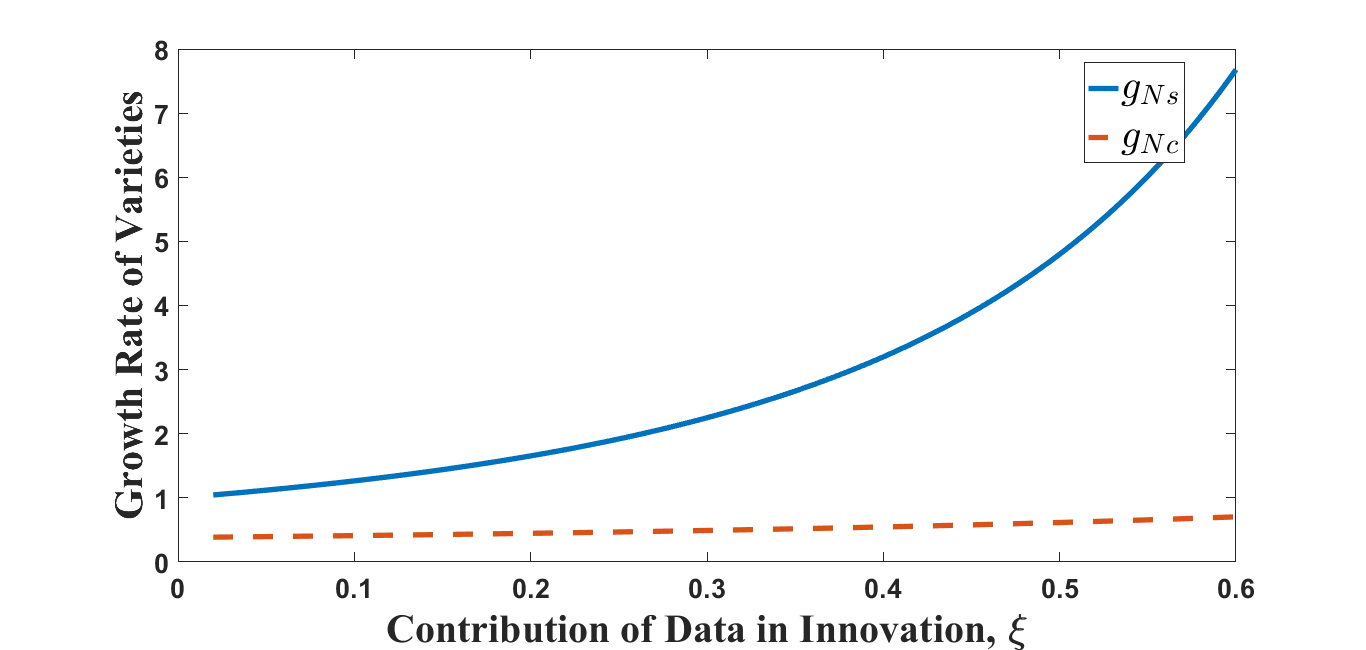}}}
\subfigure[Difference in the quantity of data]{
\includegraphics[scale=0.18]{{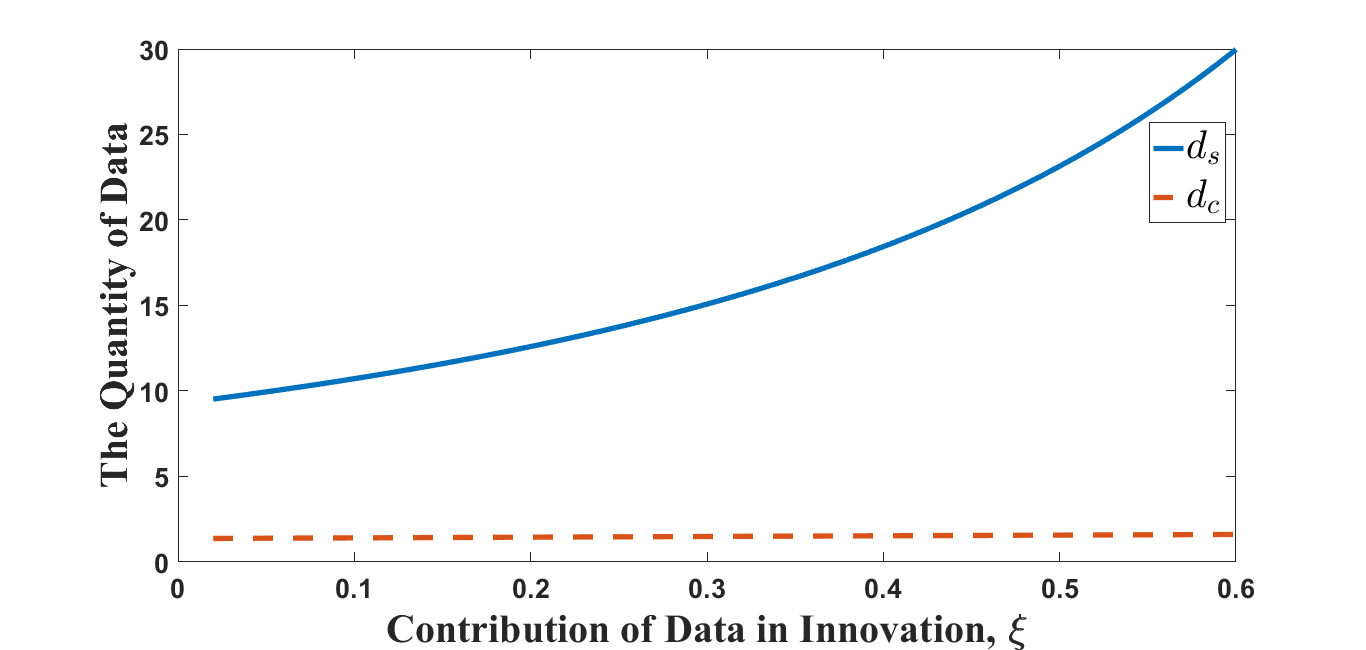}}}
\subfigure[Difference in labor allocation]{
\includegraphics[scale=0.18]{{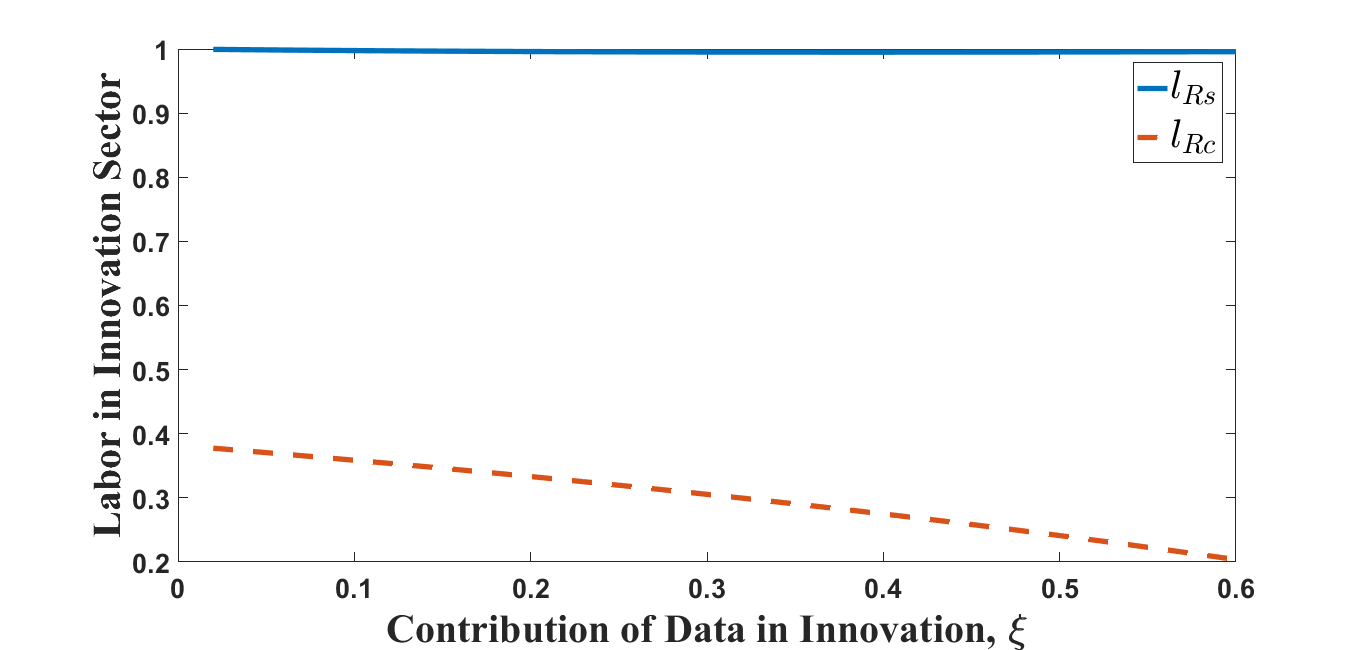}}}
\caption{\textbf{Different allocation with constant return of production and different $\xi$ }}
\label{fig:constant_xi}
\end{center}
\qquad Notes: The figure depicts three key variables with different $\xi$ when the production is constant return. The result is similar to Figure \ref{fig:diff_dc_xi}.
\end{figure}

\begin{figure}[p]
\begin{center}  
\subfigure[Difference in growth rate of varieties]{
\includegraphics[scale=0.18]{{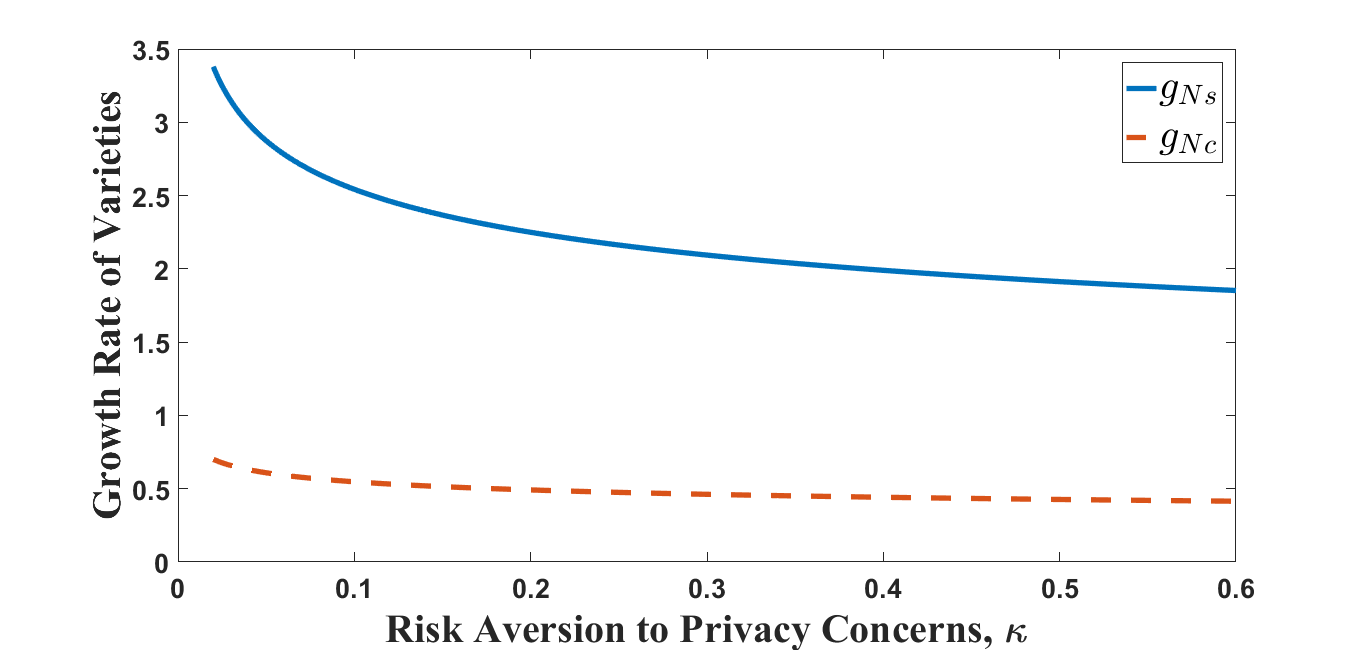}}}
\subfigure[Difference in the quantity of data]{
\includegraphics[scale=0.18]{{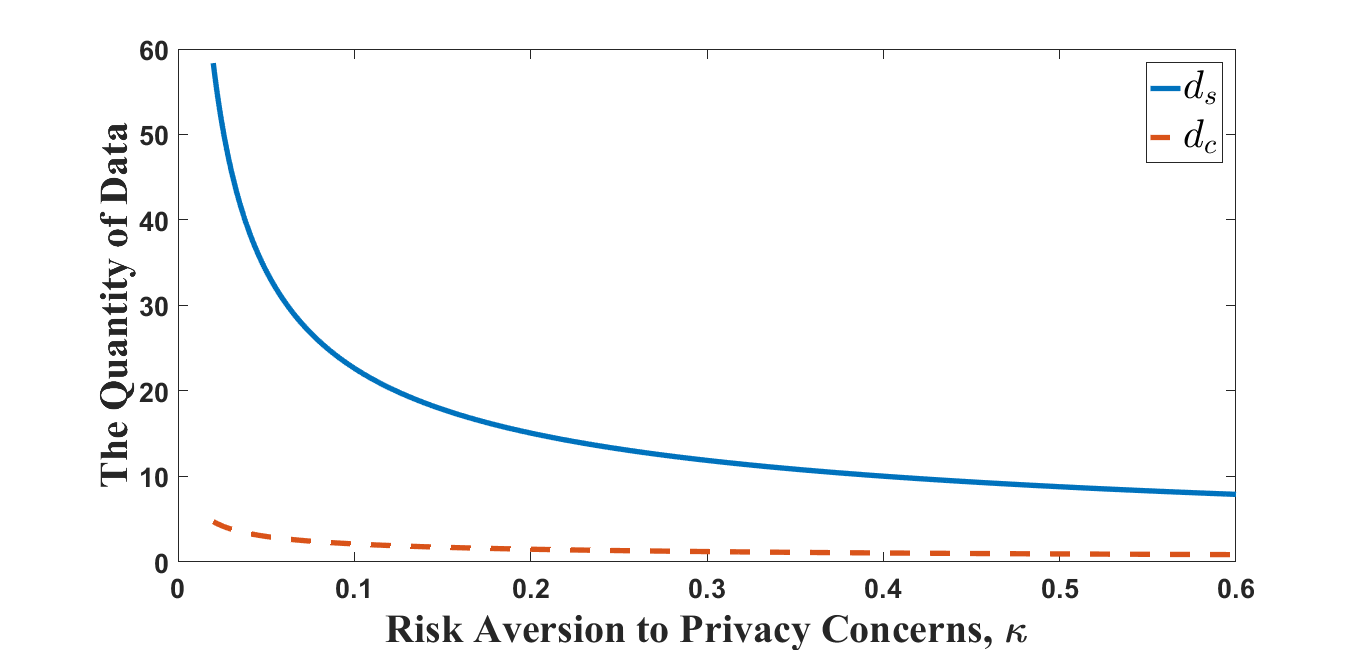}}}
\subfigure[Difference in labor allocation]{
\includegraphics[scale=0.18]{{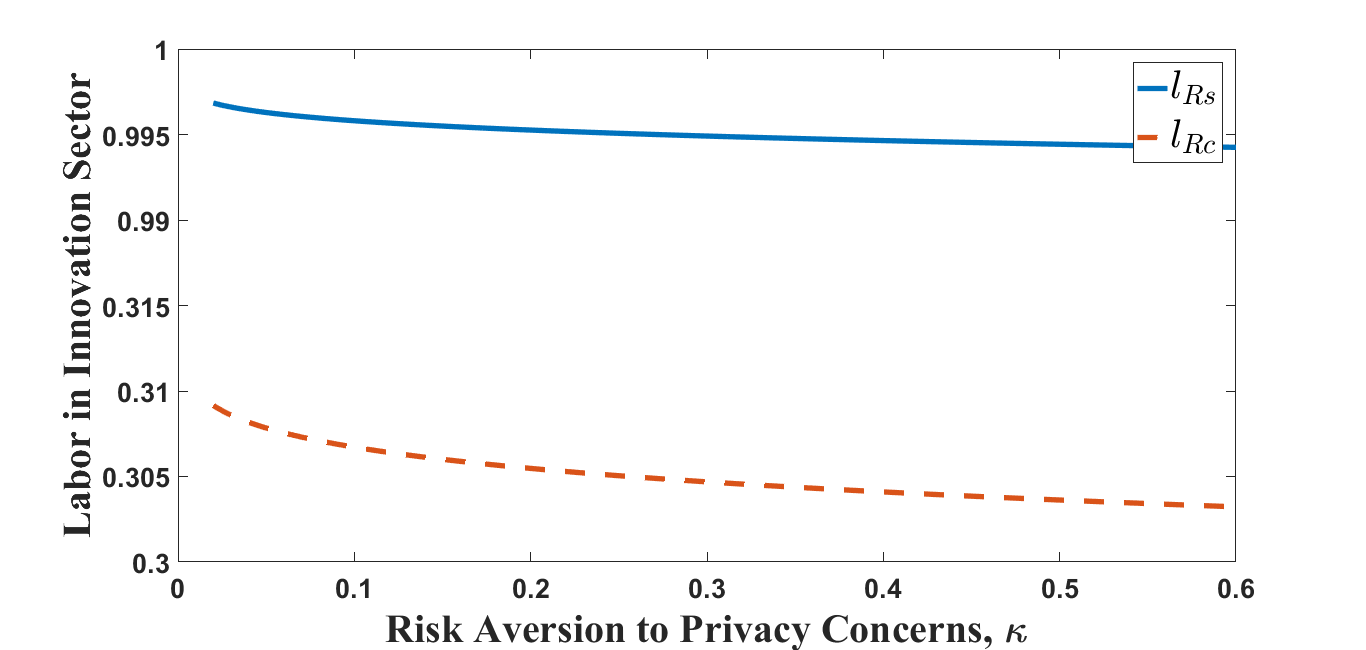}}}
\caption{\textbf{Different allocation with constant return of production and  different $\kappa$ }}
\label{fig:constant_kappa}
\end{center}
\qquad Notes: The figure depicts three key variables with different $\kappa$ when the production is constant return. The result is similar to Figure \ref{fig:diff_dc_kappa}.   
\end{figure}

Then, we have
\begin{equation}\nonumber
    \tilde g_{Ns}^1(d) = \frac{\rho}{\xi \left(\frac{1}{\gamma-1}+\theta\right)}(\kappa d^2 -\theta),
\end{equation}
and 
\begin{equation}\nonumber
    \tilde g_{Ns}^2(d) = \varepsilon L \left[{\frac{\frac{1}{1-\theta}\frac{1-\xi}{\xi}(\kappa d^2 -\theta)}{1+\frac{1}{1-\theta}\frac{1-\xi}{\xi}(\kappa d^2 -\theta)}}\right]^{1-\xi }d^\xi.
\end{equation}
The distortion is similar with Proposition \ref{propMis}. In Figure \ref{fig:constant_theta}-\ref{fig:constant_kappa}, we show that the distortion in the decentralized economy is similar to that in Proposition \ref{propMis} and this alternative assumption does not affect other results and numerical simulations.

\subsection{Cumulative Privacy Concern}
\label{cumulativeAPP}

The social planner problem is
\begin{equation}\nonumber
\max _{\left\{l_E(t), d(v,t)\right\}}  \int_0^\infty e^{-\rho t} L\left(\ln{c(t)}-\int_0^{N(t)} \frac{\kappa d(v,t)^2}{2} \mathrm{d}v \right) \mathrm{d} t.
\end{equation}
Since the price, outputs and profits are equal for all incumbent firms, we simplify the notations by $d(t)=d(v,t)$

s.t.,
\begin{align}
c(t) &=Y(t) / L, \nonumber\\
Y(t) &=N(t)^{\frac{1}{\gamma-1}}l_E(t)L^{1+\eta}d(t)^{\eta}, \nonumber\\
\dot{N}(t) &=\varepsilon l_R(t)^{1-\xi} L d(t)^{\xi} N(t), \nonumber\\
1 &= l_E(t) + l_R(t),\nonumber\\
d(t) &\le g(c(t)).\nonumber
\end{align}

Define the following Hamiltonian:
\begin{equation}\nonumber
    \mathcal{H} (l_E(t),d(t),N(t),\lambda (t))=\ln{\left[ N(t)^{\frac{1}{\gamma-1}} l_E(t) L^{\eta} d(t)^{\eta} \right]}-\frac{\kappa N(t) d(t)^2}{2}+\varepsilon \lambda (t) l_R(t)^{1-\xi} L d(t)^\xi N(t).
\end{equation}

We then have following FOC:
\begin{equation}
    \frac{\partial \mathcal{H}}{\partial l_E(t)}=\frac{1}{l_E(t)}-\lambda (t)\varepsilon (1-\xi) l_R(t)^{-\xi} L d(t)^\xi N(t)=0,
    \label{cum:HlE}
\end{equation}
\begin{equation}
    \frac{\partial \mathcal{H}}{\partial d(t)}=\frac{\eta }{d(t)}- \kappa N(t)d(t)+\lambda (t)\varepsilon \xi l_R(t)^{1-\xi } L d(t)^{\xi-1}N(t)=0,
    \label{cum:Hd}
\end{equation}
\begin{equation}\nonumber
    \frac{\partial \mathcal{H}}{\partial N(t)}=-\dot \lambda (t)+\rho \lambda (t)=\frac{1}{\gamma -1}\frac{1}{N(t)}-\frac{\kappa d(t)^2}{2}+\varepsilon \lambda (t){l_R(t)^{1-\xi}} L d(t)^{\xi},
\end{equation}

Consider (\ref{cum:HlE}), we have
\begin{equation}
    \frac{l_R(t)}{l_E(t)}=(1-\xi)\lambda(t)\dot N(t).
    \label{cum:lRlE}
\end{equation}

Plug (\ref{cum:lRlE}) into (\ref{cum:Hd}), then
\begin{align}
    \kappa N(t) d(t)^2 &= \eta+\xi\lambda(t)\dot N(t) \nonumber \\
    &= \eta+\frac{\xi}{1-\xi}\frac{l_R(t)}{l_E(t)}.
    \label{cum:dataUstility}
\end{align}

Along the BGP, $l_E(t)$ and $l_R(t)$ both remain constant. As a result, from (\ref{cum:dataUstility}), we have $-\dot{N}(t)/N(t)=2\dot{d}(t)/d(t)$. Plug (\ref{cum:dataUstility}) into (\ref{newentrants:p}), then 
\begin{equation}\nonumber
    \frac{\dot{N}(t)}{N(t)}=\varepsilon l_R(t)^{1-\xi} L \left[\frac{\eta+\frac{\xi}{1-\xi}\frac{l_R(t)}{l_E(t)}}{\kappa N(t)}\right ]^{\xi/2}.
\end{equation}
With increasing $N(t)$ and constant $l_R(t)$ (and $l_E(t)$) along the BGP, the growth rate of $N(t)$ gradually declines towards zero. This case is similar to the semi-endogenous growth model with constant population. As a result, the employment of the innovation sector denoted by $l_R(t)$ also vanishes gradually.

\subsection{Additional Privacy Concerns with Multiple Uses}
\label{additionalPrivacy}

The social planner problem is
\begin{equation}\nonumber
\max _{\left\{l_E(t), d(t)\right\}}  \int_0^\infty e^{-\rho t} L\left[\ln{c(t)}- (1+\alpha)\frac{\kappa d(t)^2}{2} \right] \mathrm{d} t, \end{equation}
s.t.,
\begin{align}
c(t) &=Y(t) / L, \nonumber\\
Y(t) &=N(t)^{\frac{1}{\gamma-1}}l_E(t)L^{1+\eta}d(t)^{\eta}, \nonumber\\
\dot{N}(t) &=\varepsilon l_R(t)^{1-\xi} L d(t)^{\xi} N(t), \nonumber\\
1 &= l_E(t) + l_R(t),\nonumber\\
d(t) &\le g(c(t)).\nonumber
\end{align}

Define the following Hamiltonian:
\begin{equation}\nonumber
    \mathcal{H} (l_E(t),d(t),N(t),\lambda (t))=\ln{\left[ N(t)^{\frac{1}{\gamma-1}} l_E(t) L^{\eta} d(t)^{\eta} \right]}-\frac{(1+\alpha) \kappa d(t)^2}{2}+\varepsilon \lambda (t) l_R(t)^{1-\xi} L d(t)^\xi N(t).
\end{equation}

We then have following FOC:
\begin{equation}\nonumber
    \frac{\partial \mathcal{H}}{\partial l_E(t)}=\frac{1}{l_E(t)}-\lambda (t)\varepsilon (1-\xi) l_R(t)^{-\xi} L d(t)^\xi N(t)=0,
\end{equation}
\begin{equation}\nonumber
    \frac{\partial \mathcal{H}}{\partial d(t)}=\frac{\eta }{d(t)}- (1+\alpha) \kappa d(t)+\lambda (t)\varepsilon \xi l_R(t)^{1-\xi } L d(t)^{\xi-1}N(t)=0,
\end{equation}
\begin{equation}\nonumber
    \frac{\partial \mathcal{H}}{\partial N(t)}=-\dot \lambda (t)+\rho \lambda (t)=\frac{1}{\gamma -1}\frac{1}{N(t)}+\varepsilon \lambda (t){l_R(t)^{1-\xi}} L d(t)^{\xi},
\end{equation}

The proof process is similar to Appendix \ref{App:OptimalAllocation}, then we have the following BGP:
\begin{equation}\nonumber
    l_R = \frac{\frac{1-\xi}{\xi}((1+\alpha) \kappa d^2 -\eta)}{1+\frac{1-\xi}{\xi}((1+\alpha)\kappa d^2 -\eta)}.
\end{equation} 
The balanced growth rate of varieties and quantity data per capita are determined by the following two equations:
\begin{equation}\nonumber
    g_N^1(d) = \frac{(\gamma-1)\rho}{\xi}((1+\alpha)\kappa d^2 -\eta),
\end{equation}

\begin{equation}\nonumber
    g_N^2(d) = \varepsilon L \left[{\frac{\frac{1-\xi}{\xi}((1+\alpha) \kappa d^2 -\eta)}{1+\frac{1-\xi}{\xi}((1+\alpha) \kappa d^2 -\eta)}}\right]^{1-\xi }d^{\xi }.
\end{equation}

\begin{figure}[H]
\begin{center}  
\subfigure[Difference in growth rate of varieties]{
\includegraphics[scale=0.22]{{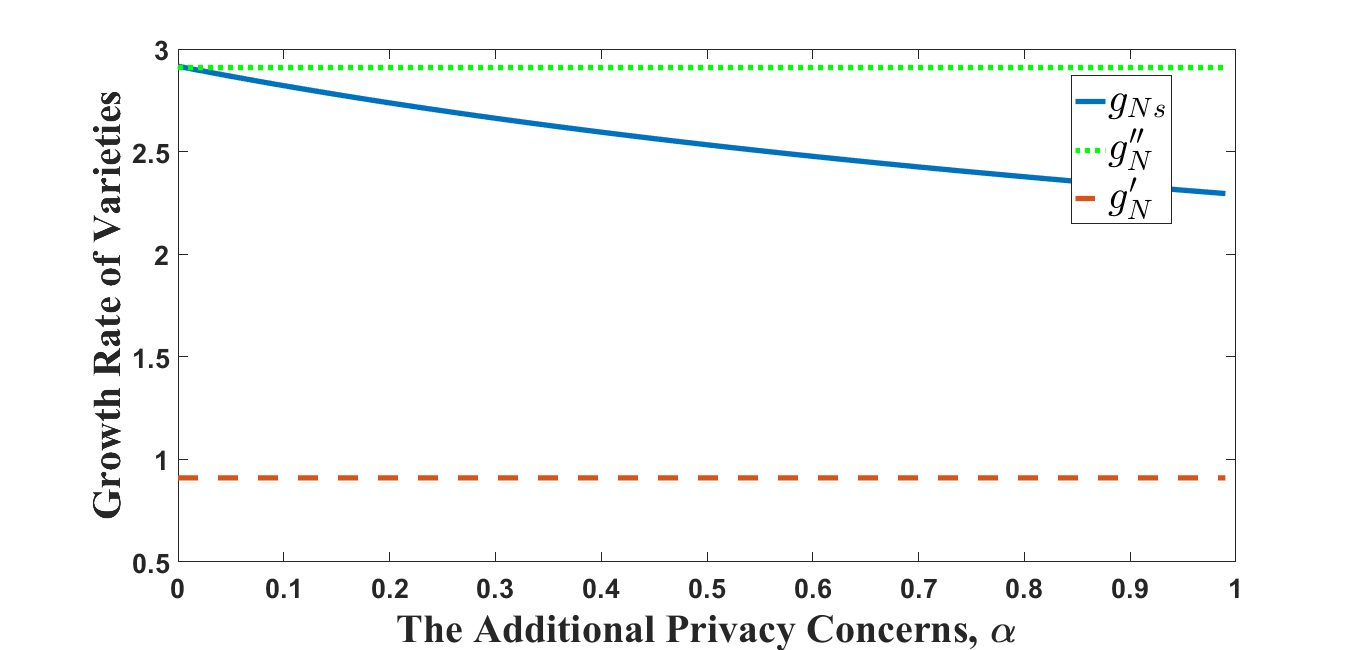}}}
\subfigure[Difference in the quantity of data]{
\includegraphics[scale=0.22]{{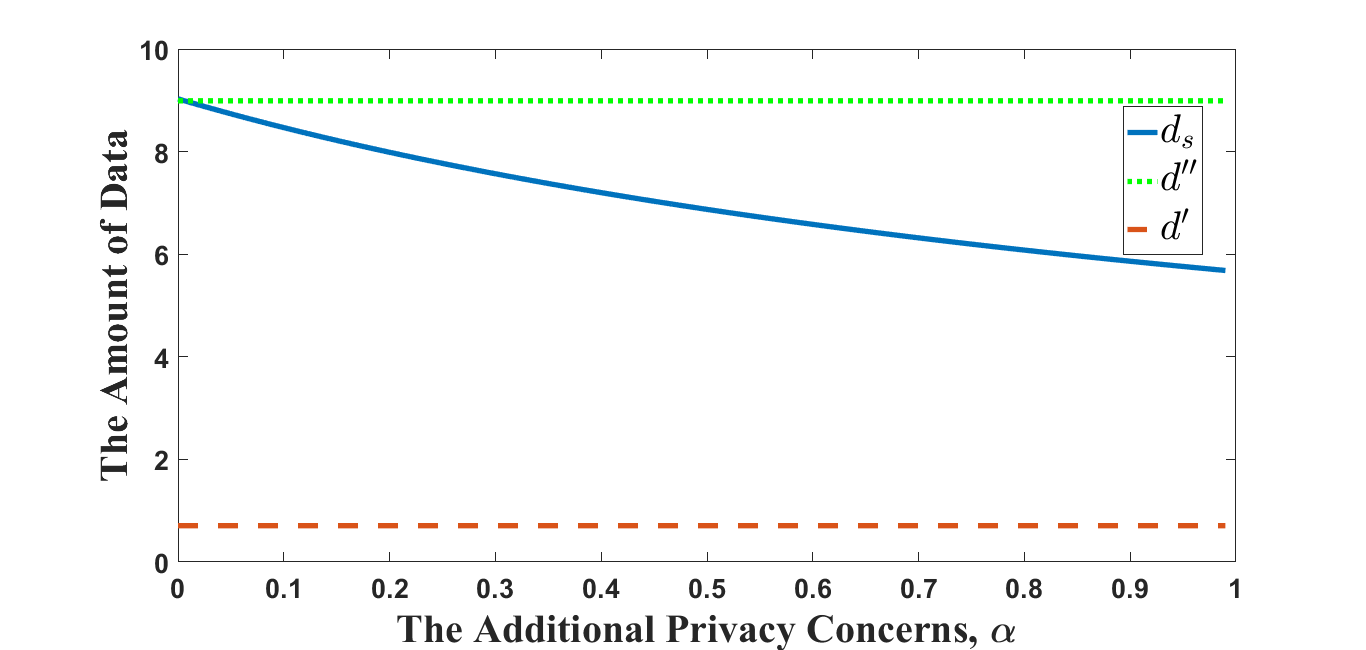}}}
\subfigure[Difference in labor allocation]{
\includegraphics[scale=0.22]{{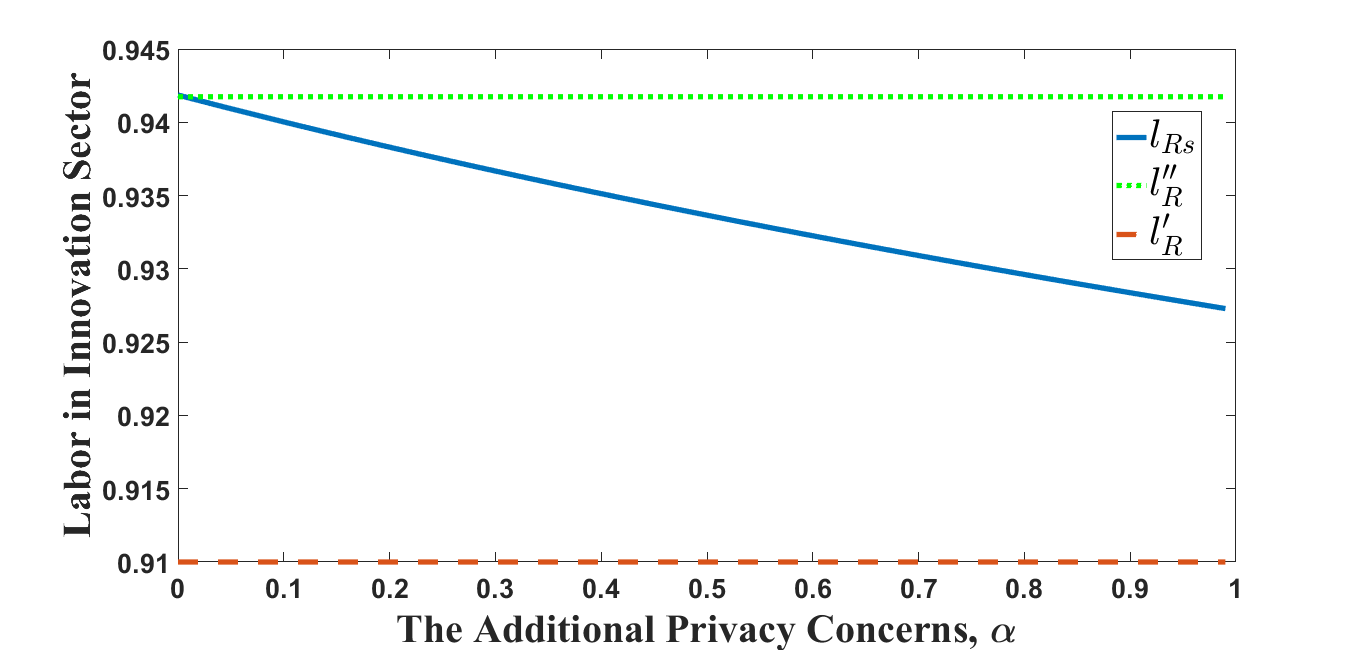}}}
\caption{\textbf{Difference allocation among three cases with different $\alpha$ }}
\label{fig:additional privacy}
\end{center}
\qquad Notes: The figure depicts different allocations with different $\alpha$. The full line represents the optimal allocation with multiple uses and additional privacy concerns, the dashed line represents the economy where data only enter the production sector, and the dotted line represents the economy where data only enter the innovation sector. Parameters are given in Table \ref{tab:parameterValues}. $g_{Ns}$, $d_s$ and $l_{Rs}$ are computed according to Appendix \ref{additionalPrivacy} and display negative correlation with $\alpha$ decreases. The other two cases are computed according to Appendix \ref{App:Only} and Appendix \ref{App:OnlyInno}. Table \ref{tab:additional privacy} further lists some specific values.
\end{figure}

\begin{table}[H]
\caption{\textbf{The Numerical Example from Figure \ref{fig:additional privacy}}}
\begin{center}
\begin{tabular*}{\hsize}{@{\extracolsep{\fill}}l c c c}
    \toprule
    Model & $g_N$ & $d$ & $l_R$ \\
    \midrule
    Social Planner ($\alpha$=0)$^*$ & 2.9160 & 9.0277 & 0.9419 \\
    Social Planner ($\alpha$=0.002)$^*$ & 2.9140 & 9.0157 & 0.9418 \\
    Social Planner ($\alpha$=0.004)$^*$ & 2.9120 & 9.0037 & 0.9418 \\
    Social Planner ($\alpha$=0.006)$^*$ & 2.9100 & 8.9918 & 0.9417 \\
    Social Planner ($\alpha$=0.008) & 2.9080 & 8.9799 & 0.9417 \\
    Social Planner ($\alpha$=0.010) & 2.9060 & 8.9680 & 0.9417 \\
    Only in Production (SP) & 0.9100 & 0.7071 & 0.9100  \\
    Only in Innovation (SP) & 2.9097 & 8.9903 & 0.9417  \\
    \bottomrule
    \label{tab:additional privacy}   
    \end{tabular*} 
\end{center}
\qquad Notes: The table reports three key variables for different allocations from Figure \ref{fig:additional privacy}. When $\alpha$ is closed to 0, compared with only used in the innovation sector, sharing data with production sector may increase the quantity of data and the growth rate. We mark this situation with an asterisk. When $\alpha$ is greater than or equal to 0.008, with vertical nonrivalry, more serious privacy concerns lead to a reduction of the quantity of data and the growth rate. 
\end{table}

\end{appendices}

\end{document}